\documentclass[sigconf, nonacm]{acmart}

\AtBeginDocument{%
  }

\usepackage{enumitem}
\usepackage{xspace}
\usepackage{graphicx,graphbox}

\usepackage{pifont}

\usepackage{tabularx} 

\usepackage{colortbl}   %

\usepackage{tikz}
\newcommand{\cdotcolor}[2][0.7ex]{%
  \tikz[baseline=-0.7ex]\draw[#2, fill=#2] (0,0) circle (#1);%
}

\usepackage{soul}

\usepackage{subcaption}

\newcommand{\SystemName}{PointAloud\xspace}

\newcommand{\rem}[1]{}

\copyrightyear{2026}
\acmYear{2026}
\setcopyright{cc}
\setcctype{by}
\acmConference[CHI '26]{Proceedings of the 2026 CHI Conference on Human Factors in Computing Systems}{April 13--17, 2026}{Barcelona, Spain}
\acmBooktitle{Proceedings of the 2026 CHI Conference on Human Factors in Computing Systems (CHI '26), April 13--17, 2026, Barcelona, Spain}
\acmPrice{}
\acmDOI{10.1145/3772318.3790797}
\acmISBN{979-8-4007-2278-3/2026/04}

\begin{document}

\title{PointAloud: An Interaction Suite for AI-Supported Pointer-Centric Think-Aloud Computing}

\author{Frederic Gmeiner}
\affiliation{
  \institution{Autodesk Research \&\\Carnegie Mellon University}
  \city{Toronto}
  \state{ON}
  \country{Canada} 
}
\email{gmeiner@cmu.edu}
\authornote{Work done as an intern researcher at Autodesk Research.}

\author{John Thompson}
\affiliation{
  \institution{Autodesk Research}
  \city{Atlanta}
  \state{GA}
  \country{USA} 
}
\email{john.thompson@autodesk.com}

\author{George Fitzmaurice}
\affiliation{
  \institution{Autodesk Research}
  \city{Toronto}
  \state{ON}
  \country{Canada} 
}
\email{george.fitzmaurice@autodesk.com}

\author{Justin Matejka}
\affiliation{
  \institution{Autodesk Research}
  \city{Toronto}
  \state{ON}
  \country{Canada} 
}
\email{justin.matejka@autodesk.com}

\renewcommand{\shortauthors}{Gmeiner et al.}

\begin{abstract}
Think-Aloud Computing, a method for capturing users’ verbalized thoughts during software tasks, allows eliciting rich contextual insights into evolving intentions, struggles, and decision-making processes of users in real-time. 
However, existing approaches face practical challenges: users often lack awareness of what is captured by the system, are not effectively encouraged to speak, and miss or are interrupted by system feedback.
Additionally, thinking aloud should feel worthwhile for users due to the gained contextual AI assistance. 
To better support and harness Think-Aloud Computing, we introduce PointAloud, a suite of novel AI-driven pointer-centric interactions for in-the-moment verbalization encouragement, low-distraction system feedback, and contextually rich work process documentation alongside proactive AI assistance. 
Our user study with 12 participants provides insights into the value of pointer-centric think-aloud computing for work process documentation and human-AI co-creation. 
We conclude by discussing the broader implications of our findings and design considerations for pointer-centric and AI-supported Think-Aloud Computing workflows. 
\end{abstract}

\ccsdesc[500]{Human-centered computing~Interaction paradigms}

\keywords{think-aloud computing, work process documentation, human-AI interaction, pointer interactions, context-aware support}

\begin{teaserfigure}
  \includegraphics[width=0.99\textwidth]{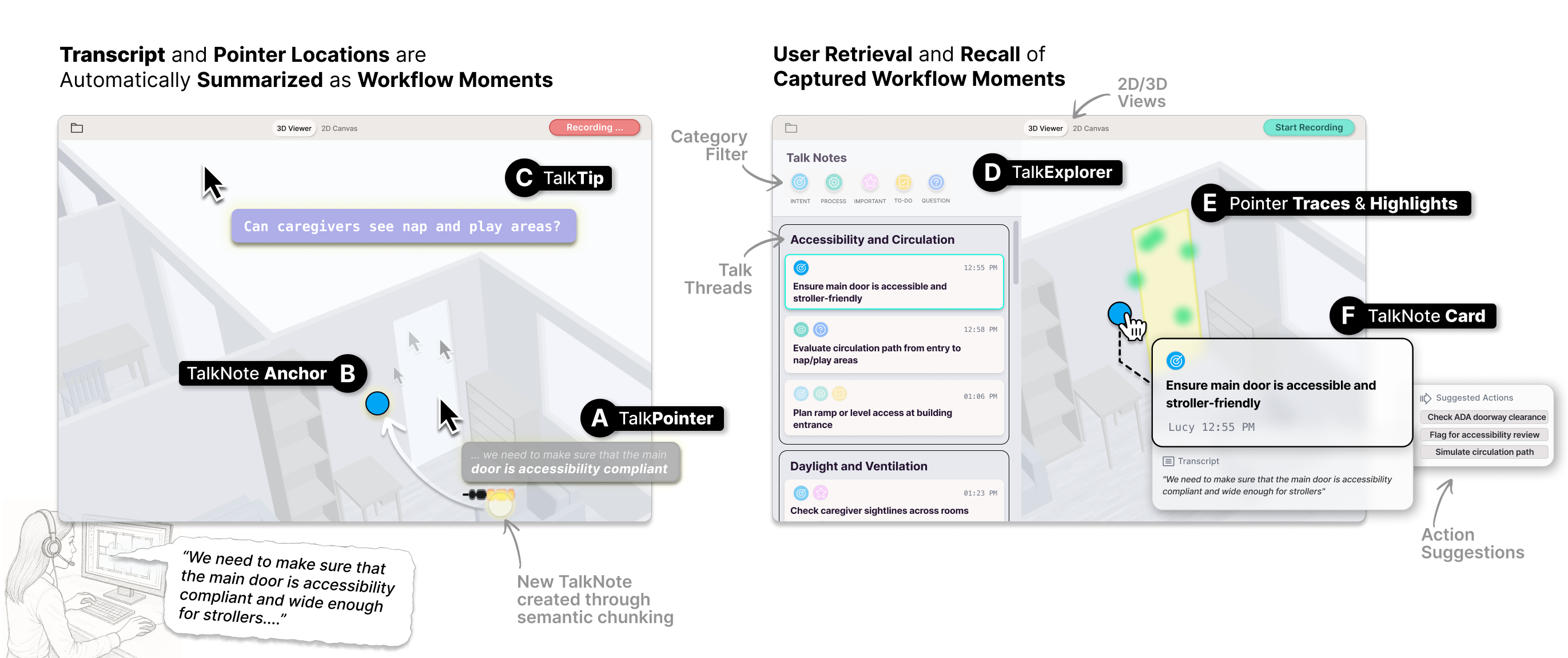}
\caption{ PointAloud allows users to (1) automatically capture their think-aloud verbalizations and pointer locations while working on architectural software tasks in 2D and 3D; with (A) TalkPointer providing pointer-adjacent low-distraction real-time feedback on the capture process and indicating when a new TalkNote gets created, the TalkNote is (B) contextually anchored in the design canvas; Additionally, (C) TalkTips provide short proactive system suggestions in response to users' activities. (2) To retrieve captured moments, (D) TalkExplorer provides a topically-clustered list view with filter options; when selecting TalkNotes, (E) captured pointer traces and relevant design elements are highlighted in the canvas, along with the TalkNote's (F) card, which features transcript, summary, process labels, and system-suggested follow-up actions.     
}
  \Description{The figure displays a user interface for the PointAloud system, highlighting how users can capture think-aloud verbalizations and pointer interactions while engaging in 2D and 3D architectural design. It features multiple labeled components, including TalkPointer for capturing pointer movements, TalkNotes for annotating insights, and TalkExplorer, which organizes notes in a list format with filter options. The visual elements indicate areas of focus in the design, including real-time feedback and suggested actions related to user input.}
  \label{fig:teaser}
\end{teaserfigure}

\maketitle

\section{Introduction}

While digital design tools support the \textit{production} of creative work, they offer little support for documenting the user's \textit{processes} that generate them. 
In domains such as architectural planning or engineering design, critical decisions often unfold tacitly through iterative cycles of exploration and reflection-in-action, shaped by evolving ideas, constraints, and partial understandings \cite{pirolli_sensemaking_2005, kolko_sensemaking_2010, schon_reflective_1983}. 
Despite their centrality to professional practice, these situated decision-making processes are difficult to capture, revisit, or communicate \cite{horner_design_2006,gruber_design_1991}. 
However, externalizing and documenting such fleeting thoughts and rationales is valuable: it can support individual reflection \cite{reymen_research_2003, schon_reflective_1983, dove_argument_2016}, preserve design rationales for later retrieval \cite{horner_design_2006,gruber_design_1991}, facilitate communication with collaborators \cite{eckert_role_2003, hisarciklilar_annotation_2009}, and even provide rich contextual data for more adaptive AI-driven design support \cite{gmeiner_exploring_2023}.

Building on this need, prior work has started to unpack \textbf{Think-Aloud Computing} \cite{krosnick_thinkaloud_2021} as a technique to capture knowledge in situ by encouraging users to verbalize their reasoning while working. 
However, previous approaches and interfaces for think-aloud computing face important challenges:

\begin{enumerate}
    \item Users often lack awareness of what is being captured;
    \item Users may not be effectively encouraged to verbalize their thoughts in the moment;
    \item Users’ thought verbalization does not receive real-time responses from the system, or such system feedback can feel either too disruptive or too subtle to notice; and
    \item The additional effort from users to verbalize their thoughts should lead to worthwhile design assistance from the system.
\end{enumerate}

In this paper, we introduce \textbf{PointAloud}, a suite of novel AI-driven pointer-centric interactions that address these challenges by embedding think-aloud support directly into the visual and interactive fabric of design tools. 
These interaction techniques support capturing, representing, and responding to users’ verbalizations in real-time as they work on a design task.

A core concept of the suite is \textbf{\texttt{Talk}Pointer}: Augmenting the space around the user's mouse pointer with a dynamic display to provide real-time feedback and system responses without causing workflow distractions. 
This way, users can receive low-distraction in-the-moment feedback from the system in response to their verbalization and screen actions.
 
Another core part of PointAloud is \textbf{\texttt{Talk}Notes}: real-time, semantically parsed annotations that automatically transform users’ spoken thoughts into lightweight ``sticky notes'' anchored directly to relevant areas of the workspace. 
When a user later selects a \texttt{Talk}Note, the system highlights the note together with the documented pointer activity and related design elements from the moment of verbalization, resituating the user’s reasoning in its original context.

In this way, PointAloud bridges the gap between ephemeral verbalization and persistent design knowledge, supporting both in-the-moment sensemaking and longer-term documentation of process.
At the same time, by externalizing and capturing users’ evolving intentions, concerns, and rationales, such interactions open \textbf{new opportunities for more context-aware forms of human–AI co-creation}, where AI support incorporates awareness of how the user’s thinking and reasoning develops over time, instead of responding only to momentary cues.

To further probe the potential of these interaction techniques, we developed the \textbf{PointAloud System}: a CAD application for inspecting and annotating architectural floor plans in both 2D and 3D. 
The system allows us to test the PointAloud interaction suite within the concrete design application of architectural planning.
Under the hood, the system leverages real-time transcription and a large language model (LLM) to segment, semantically cluster, summarize, categorize, and visually contextualize verbalized thoughts to create \texttt{Talk}Notes that can be reviewed later. 
The system extends beyond capture by providing tools for grouping related \texttt{Talk}Notes into thematic clusters, resurfacing notes when similar contexts arise in the design process, and offering dynamic actions for users to follow up within the workspace. 

To explore the benefits and limitations of PointAloud, we conducted a \textbf{user study} with 12 professionals. 
The study comprised various architectural planning-related tasks that participants completed using the PointAloud prototype system. 
In comparative tasks, the participants worked with a reduced feature set of the prototype as a baseline, mimicking a conventional text-based live transcription interface. 
The baseline interface, without components embedded around the cursor or canvas, offers a low-distraction, minimal think-aloud condition.

Our findings indicate that users perceived stronger support for making use of their spoken thoughts and for keeping track of their design process with PointAloud compared to the baseline.
Participants especially appreciated how PointAloud helped externalize fleeting ideas into automatically structured and contextualized \texttt{Talk}Notes, resurface earlier reasoning during later recap, and receive low-distraction, workflow-embedded AI suggestions.

In summation, this paper makes three main contributions:

\begin{enumerate}
    \item \textbf{\SystemName:} a suite of pointer-centric interaction techniques for AI-supported think-aloud computing, design workflow documentation, and context-aware human-AI co-creation;

    \item \textbf{User study insights} on how PointAloud interactions support concurrent thought verbalization, work process documentation, and human-AI co-creation, by using a PointAloud-based CAD prototype system as a technical probe;

    \item \textbf{Design considerations} for future pointer-centric and AI-supported think-aloud computing interactions, such as strategies for incentivizing verbalization, designing pointer-ambient displays, enabling more process-aware human–AI co-creation, and embedding documentation seamlessly within users’ ongoing workflows.

\end{enumerate}

Our work focuses on pointer-centric think-aloud interactions for supporting the creative activities of architectural designers, which we examine in depth throughout this paper. 
At the same time, the design patterns embodied in PointAloud offer an applicable foundation that researchers and practitioners can adapt across diverse software workflows. 
In doing so, PointAloud introduces new interaction techniques for documenting work processes and enabling richer forms of human-AI co-creation.

\section{Related Work} \label{sec:related-work}

\subsection{Pointer-Centric Display Techniques} \label{sec:related-work-pointer}

In traditional “point-and-click” WIMP-style interfaces,\textbf{ pointer-adjacent tool tips} or \textbf{right-click context menus} are often used to surface options relevant to the hovered/selected/dragged object or GUI element~\cite{myers_pick_2024a, ashdown2005hybridcursor}. 
Beyond such static pop-ups, research has also proposed dynamic and multimodal tool tips that adapt to user state or task context to provide pointer-adjacent multimedia assistance for learning feature-rich software applications~\cite{grossman_toolclips_2010, masson_supercharging_2022, fourney_queryfeature_2011}. 
In recent years, many commercial software applications, such as Adobe Photoshop or Autodesk Fusion, have also integrated animated tool tips that demonstrate feature usage to improve their learnability.

From a cognitive perspective, these forms of pointer-proximate augmentation \textbf{reduce mental load} by minimizing the cost of attention switching while supporting parallel information processing~\cite{wickens_attention_2021, ayres_splitattention_2005, mayer_principles_2014, paas_implications_2014}.
In WIMP-style and pointer-centric interfaces, a common assumption is that \textbf{users’ visual attention is close to their pointer}. 
Eye–pointer alignment studies provide empirical support for this assumption, and even suggest that the cursor can serve as an attention proxy with accuracy comparable to gaze tracking~\cite{huang_user_2012, demsar_quantifying_2017, liebling_gaze_2014}.

While a large body of work has explored improving the usability or accuracy of pointers as \textit{input} techniques (for example to improve accessibility~\cite{wobbrock_angle_2009}, gestures~\cite{dover_improving_2016}, editing details~\cite{payne_expansion_2016} or 3D pointing~\cite{zhou_indepth_2022}), little work has further explored augmenting the space around the pointer as a low-distraction \textit{output} technique.

Prior work in telepresence research has suggested conveying the presence, intention, and identity of collaborators in real-time, remote software workflows through \textbf{displaying remote users' mouse cursors}~\cite{gutwin_workspace_1996}. 
In recent years, the emergence of commercial collaborative shared canvas tools such as Miro or Figma has allowed users to see each other's cursors on the same canvas when collaborating in real-time, including chat bubbles pinned to collaborators'  cursors~\cite{miro_miro_2024, figma_cursor_2025}.
Similarly, recent research systems like \textit{“Pointer Assistant”}~\cite{prasongpongchai_talk_2025} propose LLM-driven pointers to represent agents in GUI workflows to guide users’ focus and enhance human-AI co-creation.

Our work builds upon these strands of work of pointer-centric display techniques by contributing a technique to “stick” dynamic visual feedback and suggestions adjacent to the pointer based on the user's captured verbalization and pointer movements during their design workflows. 
Through this, we aim to \textbf{(1) provide a low-distraction display} of real-time think-aloud verbalization feedback and proactive system responses while also \textbf{(2) capturing users' pointer locations during workflows} for enabling integrated process documentation and context-rich user models for improved human-AI co-creation.

\subsection{Context-Aware Support Systems} \label{sec:related-work-context-aware}

A second strand of research has explored how systems can automatically surface timely and relevant guidance, suggestions, or feedback by modeling the user’s task context and goals. 
Such \textbf{context-aware support systems} have been proposed to help users navigate complex, feature-rich software environments that often overwhelm novices~\cite{kiani_would_2020, mahmud_learning_2020}. 
For example, \textit{DiscoverySpace} suggests task-level action macros in Photoshop, allowing novices to maintain confidence and discover features more effectively~\cite{fraser_discoveryspace_2016a}. 
In addition to surfacing commands, several systems have explored how to provide contextualized tutorials and examples. 
\textit{TutoriVR} extends 2D video tutorials with embedded 3D contextual aids to support VR painting~\cite{thoravikumaravel_tutorivr_2019}. 
\textit{RePlay} gathers activity across applications to suggest video tutorials, helping users reduce time spent on web search~\cite{fraser_replay_2019}. 
Similarly, \textit{Shöwn} dynamically presents conceptual examples during comic drawing~\cite{ngoon_shown_2021}. 
Recent systems also integrate lightweight AI-generated commentary, such as \textit{FeedQUAC}, which provides persona-driven ambient feedback to support reflection and inspiration in creative workflows~\cite{long_feedquac_2025}.

Another related body of work highlights the role of \textbf{ambient or peripheral displays}~\cite{pousman_taxonomy_2006} for surfacing contextual knowledge and resources in the background of the primary user activity. 
For instance, \textit{Ambient Help} presents automatic, context-sensitive videos and help resources on a secondary display to support opportunistic learning without harming productivity~\cite{matejka_ambient_2011a}. 
Similarly, \textit{SidePoint} integrates a peripheral knowledge panel to surface concise, relevant knowledge items alongside the slide presentation authoring activities~\cite{liu_sidepoint_2013a}. 
Another example is \textit{InterWeave}, which embeds contextual search suggestions directly into users’ note-taking environments, helping participants connect new information to their emergent structures of knowledge~\cite{palani_interweave_2022}. 
Together, these works show how presenting ambient contextual support, such as side panels and peripheral displays, can reduce interruption costs and embed guidance into users’ ongoing, primary workflows.

Finally, research on \textbf{mixed-initiative systems} has articulated how proactive system feedback can be balanced with the user's notion of control. 
Early principles emphasize the coupling of automated services with direct manipulation in ways that preserve user agency; exemplified by Bayesian user models for inferring users’ goals and needs from observed actions and queries~\cite{horvitz_principles_1999a, horvitz_lumiere_2013}. 
Subsequent domain-specific systems have extended these ideas, such as \textit{Soloist}, which leverages deep-learning audio processing to generate customizable tutorials and provide real-time feedback for guitar learning~\cite{wang_soloist_2021}. 

Across these streams, context-aware support systems demonstrate different strategies for helping users: inferring goals and surfacing relevant functionality, providing tutorials and examples in situ, embedding knowledge in peripheral displays, and balancing initiative between the system and the user. 
Building atop this prior work, we contribute novel interaction techniques for proactive, AI-driven assistance grounded in multimodal user context models elicited through capturing users' concurrent think-aloud verbalization and pointer activities.

\subsection{Workflow Capture, Documentation, and Retrieval Systems} \label{sec:related-work-capture}
A long-standing line of work in HCI has examined how digital systems can capture, document, and later retrieve knowledge about work processes, design decisions, and task histories. 
For example, research has explored systems that \textbf{capture and document workflows} to create rich histories toward understanding and reuse. 
\textit{Chronicle}~\cite{grossman_chronicle_2010} records the complete creation history of graphical documents by linking its content to the steps, tools, and settings used, thereby turning the resulting replayable document into a rich archive and learning resource. 
Similarly, \textit{Co-notate}~\cite{rasmussen_conotate_2019b} explores real-time annotations that combine audio, video, and textual traces to capture evolving situational knowledge during collaborative design activities. 
More recently, \textit{Meta-Manager} allows automatically collecting and organizing provenance information in software development by tracking code changes, including AI-generated or copy-pasted fragments, enabling developers to later answer questions about unfamiliar code bases and design rationale~\cite{horvath_metamanager_2024}. 
Other approaches augment physical fabrication activities by capturing multimodal sensor data about users, tools, and expertise to enable personalized feedback and contextual analysis of making processes~\cite{gong_instrumenting_2019}.

Another strand has investigated \textbf{documenting work processes through demonstration} and automated tutorial generation. 
For example, \textit{Grabler et al.}~\cite{grabler_generating_2009} introduced a system that automatically generates step-by-step tutorials from demonstrations in photo manipulation software. 
\textit{Torta}~\cite{mysore_torta_2017} extends this approach by combining operating-system-wide activity tracing with screencasts to generate mixed-media tutorials. 
Crowd-powered systems such as \textit{StreamWiki}~\cite{lu_streamwiki_2018} leverage audiences of live knowledge-sharing streams to collaboratively generate archival documentation in real time. 

Systems have also explored \textbf{capturing the contextual “why”} (rationale) behind work artifacts~\cite{horner_design_2006,gruber_design_1991}. 
For example, \textit{Callisto} links conversational chat traces with computational notebook elements to surface the rationale behind data analytic workflows~\cite{wang_callisto_2020}, while other systems capture and link online discussion threads to the code under debate~\cite{park_postliterate_2018}. 
These approaches highlight the importance of recording not only what was done but also \textit{the reasoning behind design decisions}.

Complementary work has addressed the challenge of \textbf{retrieving captured process information} for reflection and sensemaking through novel interfaces. 
For example, \textit{Delta} allows visualizing and comparing alternative software workflows at multiple levels of granularity to help users understand trade-offs~\cite{kong_delta_2012}. 
Similarly, \textit{Tesseract} introduces spatial querying of design recordings through a Worlds-in-Miniature interface, enabling expressive search of past multimodal design activities~\cite{karthikmahadevan_tesseract_2023}. 
In the context of online meetings, \textit{TalkTraces}~\cite{chandrasegaran_talktraces_2019} provides real-time capture and visualization of relevant discussed content, while \textit{MeetingVis}~\cite{shi_meetingvis_2018a} uses visual narratives to support remembering meeting content and context. 

Other systems have explored \textbf{implicit multimodal process documentation} by capturing users' speech and gaze activities. For example, \textit{GAVIN} combines gaze with voice input to anchor voice notes to text passages~\cite{khan_gavin_2021}, while follow-up work explores gaze and speech more broadly for implicit multimodal interaction design~\cite{khan_integrating_2022}. 

Similarly, recent research by \textit{Krosnick et al.} has proposed the concept of \textbf{Think-Aloud Computing}~\cite{krosnick_thinkaloud_2021}: using concurrent verbalization as a lightweight but rich mechanism for documenting users' work processes. 
The authors extend the traditional think-aloud protocol into everyday computing, prompting users to speak while working and contextualizing this speech with system state. 
Their findings show that think-aloud computing captures subtle design intent, rationale, and process knowledge that traditional documentation often misses, while requiring comparable effort. 
By situating knowledge capture in natural verbalization rather than post hoc reporting, think-aloud computing represents a promising direction for low-effort and rich workflow documentation. 

Building atop this rich body of work on capturing, documenting, and retrieving workflows, our contribution lies in extending the vision of Think-Aloud Computing by contributing AI-supported interaction techniques that are tightly integrated into pointer-centric, think-aloud design workflows, as detailed in the next section.

\section{Exploring Interaction Design Principles for Supporting Think-Aloud Computing}

Our work is grounded in the concept of \textit{Think-Aloud Computing}~\cite{krosnick_thinkaloud_2021}---a low-effort method for systems to elicit users’ knowledge in situ by encouraging them to verbalize their reasoning while engaging in software tasks. 
Such verbalizations can surface rich, contextual insights into users’ evolving intentions, struggles, and decision-making processes in real time, which are otherwise difficult to capture by only observing their interactions with graphical user interfaces. 

Especially in the context of design-related activities, such as in architectural planning or engineering design, critical decisions often unfold tacitly through cycles of exploration and reflection-in-action, shaped by evolving ideas, constraints, and partial understandings~\cite{pirolli_sensemaking_2005, kolko_sensemaking_2010, schon_reflective_1983}.
These situated processes are notoriously difficult to document for later revisitation or communication~\cite{horner_design_2006,gruber_design_1991}. 
Think-aloud computing, however, has the potential to externalize and record such fleeting rationales and can thus support self-reflection~\cite{reymen_research_2003, schon_reflective_1983, dove_argument_2016}, preserve design justifications for future retrieval~\cite{horner_design_2006,gruber_design_1991}, and facilitate collaboration~\cite{eckert_role_2003, hisarciklilar_annotation_2009}. 

Beyond documentation, capturing verbalized thoughts \textbf{also creates opportunities for more effective human–AI co-creation} \cite{gmeiner_exploring_2023, gmeiner_exploring_2025}. 
Users’ spoken reasoning can provide context-rich data and serve as a foundation for AI systems to deliver more aligned and situated assistance. 
At the same time, prior approaches to think-aloud computing interfaces face several challenges: (1) users often lack awareness of what is actually being captured, (2) they may not feel sufficiently encouraged to articulate their thoughts in the moment, (3) system responses to verbalizations are frequently either absent, overly disruptive, or too subtle to notice, and (4) the additional effort of speaking aloud must ultimately translate into tangible and worthwhile assistance for the user.

These challenges raised several guiding questions for our design exploration:
\begin{enumerate}
    \item \textit{How might we \textbf{design systems} that \textbf{actively incentivize and sustain thinking aloud}?}
    \item \textit{What complementary \textbf{contextual data}, alongside verbalizations, should be captured to \textbf{represent user processes more holistically}?}
    \item \textit{How can \textbf{in-the-moment feedback} be delivered \textbf{without interrupting} or distracting users’ workflows?}
    \item \textit{In what ways can \textbf{captured process data} be harnessed to \textbf{foster} more effective \textbf{human–AI co-creation}?}
\end{enumerate}

To address these questions, we engaged in a set of formative activities:  
(1) a literature review on workflow capture, documentation, and retrieval systems, as well as on context-aware support and low-distraction feedback techniques \textit{(see Section \ref{sec:related-work})}; and  
(2) a four-week ideation phase involving iterative cycles of sketching and refining interaction concepts, supported by daily check-ins of the first and second authors and weekly meetings with the broader research team, where feedback was also gathered from a larger group of HCI researchers. 
Across this period, we generated over 15 distinct design concepts and iteratively distilled them into a smaller set of directions, all documented on a shared Miro board.
Through this process, we distilled our explorations into four overarching \textit{design principles}, detailed below:

\subsection{Design Principles}

Based on our guiding questions and formative activities, we developed the following interaction design principles for AI-supported think-aloud computing in the context of design-related tasks:

\begin{itemize}[font=\bfseries,
  align=left]
    \item[DP1] \textbf{Subtle yet perceivable feedback on user context capture.}
    To increase users’ awareness and trust in the capture process, the system should provide real-time visual feedback on both verbalization activity and live transcription. 
    Such feedback must be lightweight: Noticeable enough to reassure users about what is being captured, but subtle enough to avoid interrupting the primary design workflow. 
    This ensures that users feel confident their thinking is being recorded without suffering unnecessary distraction.

    \item[DP2] \textbf{Automatically document the user's design process as structured and contextually rich moments.} \\
    While the user works and verbalizes, the system should automatically detect and document distinct moments of the design process. 
    Each documented moment should combine the user’s verbalization with contextual process information \textit{(such as pointer activity, screenshots, and linked design elements)} so that captured moments form a richer record of the workflow. 
    In doing so, the system creates a durable memory of the process that supports later continuity, reflection, and recall, while also providing a data foundation (user models) for more context-aligned AI-suggestions, enabling more effective human–AI co-creation.

    \item[DP3] \textbf{Provide diverse mechanisms to represent and retrieve documented moments.} \\
    To support sensemaking and provenance, the system should support users in flexibly revisiting and working with previously documented moments. 
    Captured moments should be represented and accessible in multiple complementary forms — such as in-context overlays in the 2D/3D design workspace or thematically clustered in side-panel lists. 
    Retrieval should be further supported through categorization and filtering mechanisms, allowing users to navigate thematic aspects of their workflow.

    \item[DP4] \textbf{Offer proactive, context-aware prompts and follow-up actions to the user.} \\
    To encourage verbalization and foster human-AI co-creation, the system should proactively respond to users’ speech with context-aware prompts and suggested actions. 
    For example, when silence occurs, it may nudge users to elaborate on relevant topics or dynamically resurface past notes that connect to the current situation. 
    The system should also suggest situation-specific follow-up actions, enabling designers to fluidly extend and act upon prior lines of thought as part of their design process.
\end{itemize}

\begin{figure*}[t]
\includegraphics[width=1\linewidth]{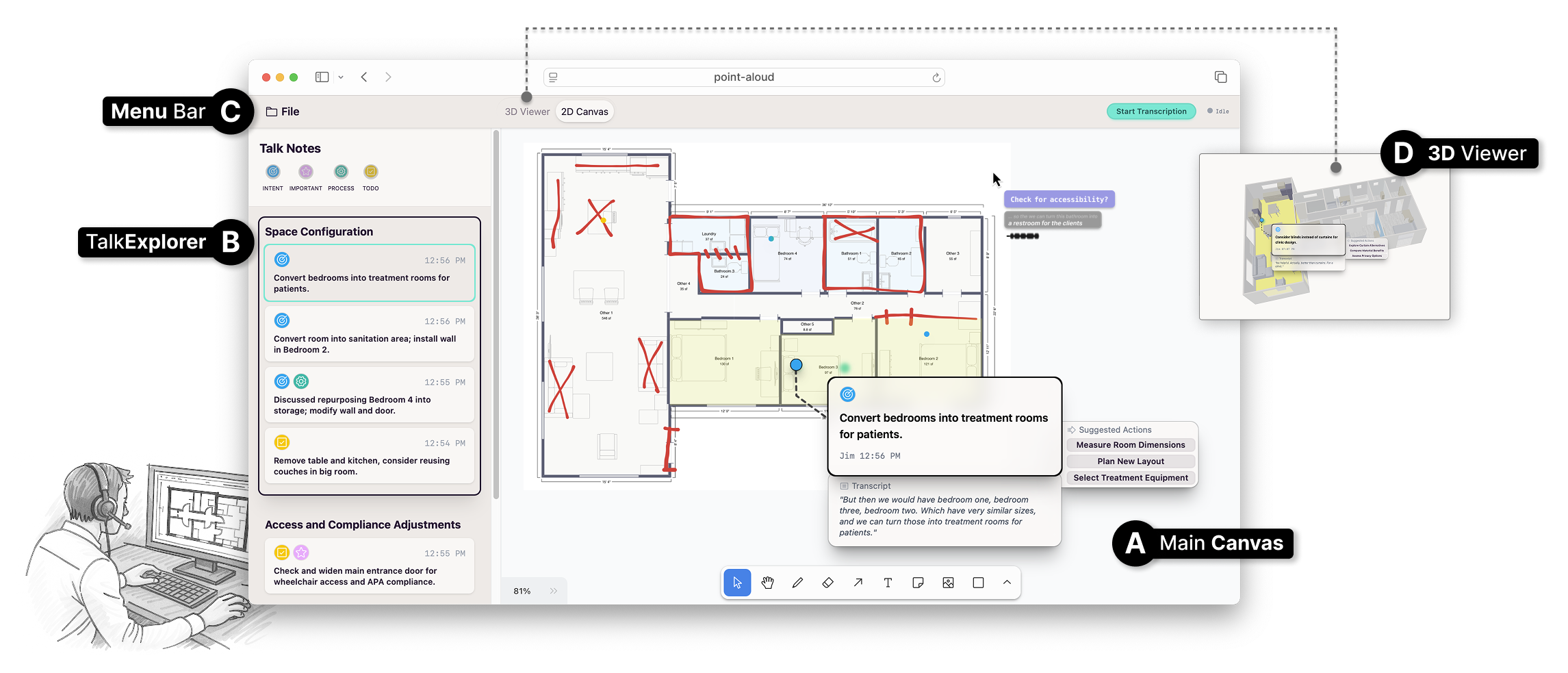}
\caption{Screenshot of the \textit{PointAloud} system: (A) Main canvas with activated 2D view containing sketches on a floor plan with an unfolded \texttt{Talk}Note card and \texttt{Talk}Pointer next to the mouse cursor; (B) \texttt{Talk}Explorer sidebar for browsing and filtering of captured \texttt{Talk}Notes; (C) Menu bar with controls for starting/stopping transcription and switching between 2D and (D) 3D view.}
\label{fig:system-ui}
\Description{The figure displays a screenshot of the PointAloud system. It features a main canvas showing a 2D floor plan with annotations, a TalkExplorer sidebar for managing TalkNotes, and a menu bar at the top for various functions. An additional 3D viewer is also present, displaying a model of the space and interactive elements for user input.}
\end{figure*}

\section{\SystemName System: A Pointer-Centric AI-supported Think-Aloud Computing System}

Informed by our formative activities and the design principles for supporting Think-Aloud Computing described in the previous section, we designed a \textit{suite} of novel pointer-centric interaction techniques and adaptive user interface (UI) elements we coined \textit{PointAloud}. 

To probe the potential of this collection of techniques in a concrete design-related task context, we developed the \textbf{PointAloud System}: a GenAI-driven CAD application for inspecting and annotating architectural floor plans in 2D and reviewing corresponding 3D models. 
The prototype embeds the PointAloud interaction suite directly into the design workflow, allowing us to explore how pointer-centric think-aloud computing can support process documentation, reflection, and AI-assisted co-creation (Figure~\ref{fig:system-ui}).

In this section, we demonstrate the utility of \textit{PointAloud} by first illustrating its functionalities within brief use-case scenarios. 
We then follow with a detailed description of the proposed interface mechanisms and their implementation.

\subsection{Example Use Case Scenarios}

Lucy is an architect tasked with converting an existing apartment into a licensed childcare micro-center for toddlers. 
She opens the \textbf{Point\-Aloud} system to begin exploring the client-provided floor plan.

\subsubsection{\textbf{2D Use Case: Annotating the floor plan}} \label{sec:walkthrough-2d}\hfill

\noindent Lucy loads the provided 2D floor plan and 3D model of the apartment into the workspace. To capture her reasoning as she works, she presses the \textit{“Start Transcription”} button.

\vspace{0.5em}
\noindent \textbf{(1a) Talking Through Early Design Concerns:}
As she pans across the floor plan, she verbalizes her initial thoughts: \textit{“We need to make sure that the main door is accessibility compliant and wide enough for strollers.”} Her speech appears live as a small text element next to her mouse pointer \textbf{(\texttt{Talk}Text)}. 
A soft, bubble-style indicator grows as she continues talking, letting her see that the system is actively listening without pulling her out of the design task \textbf{(\texttt{Talk}Vis)}.

\noindent \textbf{(1b) Automatic \texttt{Talk}Note Creation:}
When Lucy shifts focus to another concern—\textit{“These two interior walls might be load-bearing; I’ll need to check structural drawings before removing them.”}—the system recognizes this as a new line of thought and generates a \textbf{\texttt{Talk}Note}. A small colored dot appears on the floor plan at the entrance she was pointing to, directly anchoring the \texttt{Talk}Note within the spatial design context.  
Each anchor's color corresponds to the \texttt{Talk}Note's process category label  (e.g., light blue for \textit{Design Intent}, magenta for \textit{Important}, dark blue for \textit{Questions}), giving Lucy a quick visual cue of the note’s type at a glance.

\vspace{0.5em}
\noindent \textbf{(2) Revisiting and Exploring \texttt{Talk}Notes:}
Hovering over the dot expands it into a \textbf{\texttt{Talk}Note Card}, containing the transcript snippet, an automatically generated summary (“Ensure main door is accessible and stroller-friendly”), and a category label (\textit{Design Intent}). 
Alongside the card, Lucy sees a faint yellow highlight over the entrance door and traces of her pointer movements during that utterance, situating the note in its original design context.
Lucy hovers over another previously created \texttt{Talk}Note about natural light and sees her earlier pointer traces near the living room windows. 
This helps her quickly recall why she considered converting the space into the main play area to maximize daylight.

\vspace{0.5em}
\noindent \textbf{(3) Receiving Proactive System Suggestions (\texttt{Talk}Tips):}
As she continues sketching and talking, a subtle prompt appears near her pointer: 
\texttt{“Can caregivers see nap and play areas?”} 
This is a system-generated \textbf{\texttt{Talk}Tip}. 
Lucy responds aloud: 
\textit{“That’s an important point—I’ll need to make sure caregivers can supervise both areas.”} 
The system records her response and creates a new \texttt{Talk}Note tagged as \textit{Important} and \textit{Design Intent}, anchoring it near the playroom partition wall.

\vspace{0.5em}
\noindent \textbf{(4) Surfacing Related Prior Notes (\texttt{Talk}Reminders):} 
While reflecting on possible layouts for the nap room, two earlier \texttt{Talk}Notes reappear on the floorplan, highlighted in red and briefly summarized next to their anchors. 
The system has detected that Lucy’s current verbalization relates to these earlier concerns, resurfacing them as \textbf{\texttt{Talk}Reminders}. 
Lucy states her appreciation for the prompt since it jogs her memory about a prior decision on daylight and supervision that is again relevant to the new arrangement. 

\vspace{0.5em}
\noindent \textbf{(5) Triggering Contextual Follow-Up Actions (\texttt{Talk}Note Action Suggestions):}
When Lucy revisits the accessibility note at the entrance, it contains a contextual action suggestion: \texttt{“Check ADA doorway clearance.”} 
She clicks the small button, which automatically overlays the required clearance radius onto the floor plan. 
This helps her visually confirm that the current doorway is too narrow and flags the element with a subtle red outline.  
For another \texttt{Talk}Note about daylight, the action menu offers to \texttt{“Simulate daylight exposure.”} 
Lucy accepts, and the workspace briefly shades the floor plan according to window orientation, helping her evaluate which areas will receive the most natural light.  
These contextual follow-up actions extend \texttt{Talk}Notes from passive memory cues into active design supports, giving Lucy lightweight but targeted tools right when they are relevant.
After continuing for 20 minutes, Lucy stops the transcription and saves the project, which now contains 55 \texttt{Talk}Notes.

\subsubsection{\textbf{3D Use Case: Preparing for a client meeting}}
\label{sec:walkthrough-3d}
{\raggedright
The next day, Lucy re-opens the project in \textbf{PointAloud} to prepare for an upcoming meeting with her clients.  
In the \textbf{\texttt{Talk}Explorer} side panel, she sees all of her previously captured \texttt{Talk}Notes automatically clustered by theme, such as \textit{Accessibility \& Circulation}, \textit{Daylight \& Ventilation}, and \textit{Supervision \& Safety}.  
}
When hovering over a note in the side panel, its corresponding anchor reappears in the 2D/3D workspace, showing the highlighted design elements and her pointer traces from the moment of verbalization. 
This allows Lucy to quickly re-situate her earlier reasoning in the design context.  

To organize her discussion points, Lucy filters the \texttt{Talk}Explorer to show only notes labeled as \textit{Design Intent}, \textit{Important}, \textit{Questions}, and \textit{To-Dos}.  
She then uses these filtered notes to draft her meeting document, capturing early design decisions, unresolved questions, and key concerns such as supervision sightlines and acoustic separation in the nap room.  
This way, her spontaneous thoughts from the exploration session are transformed into a structured, filterable work process documentation to draw upon as she crafts a focused agenda for client discussion.

\subsection{Interface Features}

In this subsection, we describe the main interaction features of the \textit{PointAloud Interaction Suite}. 
Together, these features demonstrate how pointer-centric feedback and persistent contextual representations can support designers in capturing, revisiting, and extending their thinking while working on design tasks.
We developed the \textit{PointAloud System} to instantiate these interactions and interfaces with architectural planning as a proxy design task.
The prototype provides two spatial planning tools: a 2D floor plan tool and 3D model inspector tool.
Both tools provide interaction mechanisms to support pointer-centric think-aloud computing.

\subsubsection{\textbf{System Interface Overview}}

The interface of the \textit{PointAloud System} (Figure~\ref{fig:system-ui}) consists of three main components:  
\begin{enumerate}
    \item \textbf{Main Canvas:} The central workspace for sketching in a 2D floor plan view and navigating corresponding 3D models.  
    \item \textbf{TalkExplorer Sidebar:} A panel for browsing, clustering, and filtering captured \texttt{Talk}Notes, supporting organization and review of verbalized ideas.  
    \item \textbf{Menu bar:} Controls for starting or stopping transcription, switching between 2D and 3D views, and accessing file operations.  
\end{enumerate}

\subsubsection{\textbf{\texttt{Talk}Pointer: Dynamic Feedback Anchored to the Cursor (DP1, DP4)}}

\begin{figure}[h!]
\includegraphics[width=0.7\linewidth]{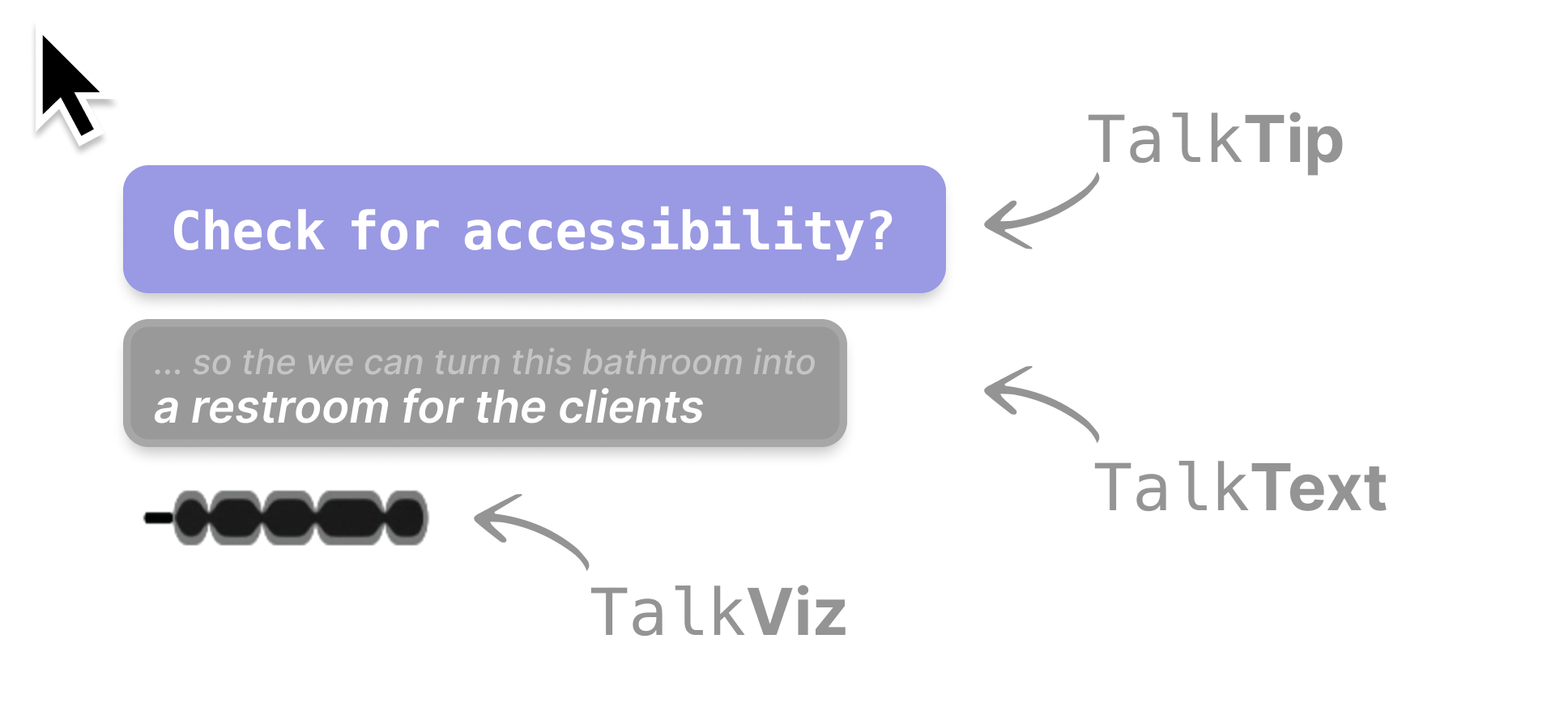}
\caption{Pointer-adjacent \texttt{Talk}Pointer display comprising \texttt{Talk}Tip, \texttt{Talk}Text, and \texttt{Talk}Viz }
\Description{The figure shows a user interface element labeled "TalkPointer" with a tooltip titled "Check for accessibility?" visible at the top. Below, there are three distinct components labeled "TalkTip," "TalkText," and "TalkViz," each connected to the main tooltip by lines, indicating their relationship to the displayed content. A cursor is positioned near the tooltip, suggesting interactivity.}
\label{fig:feature-talkpointer}
\Description{The figure shows a user interface element labeled "TalkPointer" with a tooltip titled "Check for accessibility?" visible at the top. Below, there are three distinct components labeled "TalkTip," "TalkText," and "TalkViz," each connected to the main tooltip by lines, indicating their relationship to the displayed content. A cursor is positioned near the tooltip, suggesting interactivity.}
\end{figure}

To provide lightweight but continuous feedback on the think-aloud process, the system features the \textbf{\texttt{Talk}Pointer}: a dynamic display anchored right of the user’s mouse cursor (Figure \ref{fig:feature-talkpointer}). 
By visually linking proactive system messages and feedback about the user’s speech activity to their current pointer location, \texttt{Talk}Pointer provides process-related cues and reassures users that their verbalizations are being captured and contextualized without requiring them to divert attention from the design task.

\noindent \texttt{Talk}Pointer integrates three complementary components:

\begin{figure}[H]
  \begin{minipage}[t]{3cm}
    \includegraphics[align=t,width=\columnwidth]{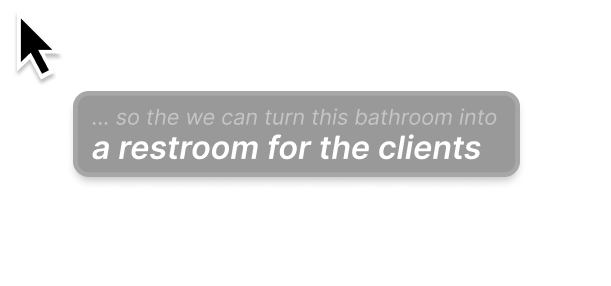}
    \caption{}
    \label{fig:feature-talktext}
    \Description{The figure displays a speech transcription overlay interface labeled "TalkText." It shows a text bubble containing partially transcribed dialogue, with words visually represented in a sequential format, indicating a real-time capture of spoken language. The focus on the text emphasizes immediate feedback regarding ongoing speech recognition.}
  \end{minipage}
   \begin{minipage}[t]{\dimexpr\columnwidth-3.2cm\relax}
          \textbf{(1) \texttt{Talk}Text} (DP1): A short transcription overlay that streams the user's most recently captured words in real time, providing immediate feedback on the system's ongoing speech transcription.  
  \end{minipage}
\end{figure}

\begin{figure}[H]
  \begin{minipage}[t]{3cm}
    \includegraphics[align=t,width=\columnwidth]{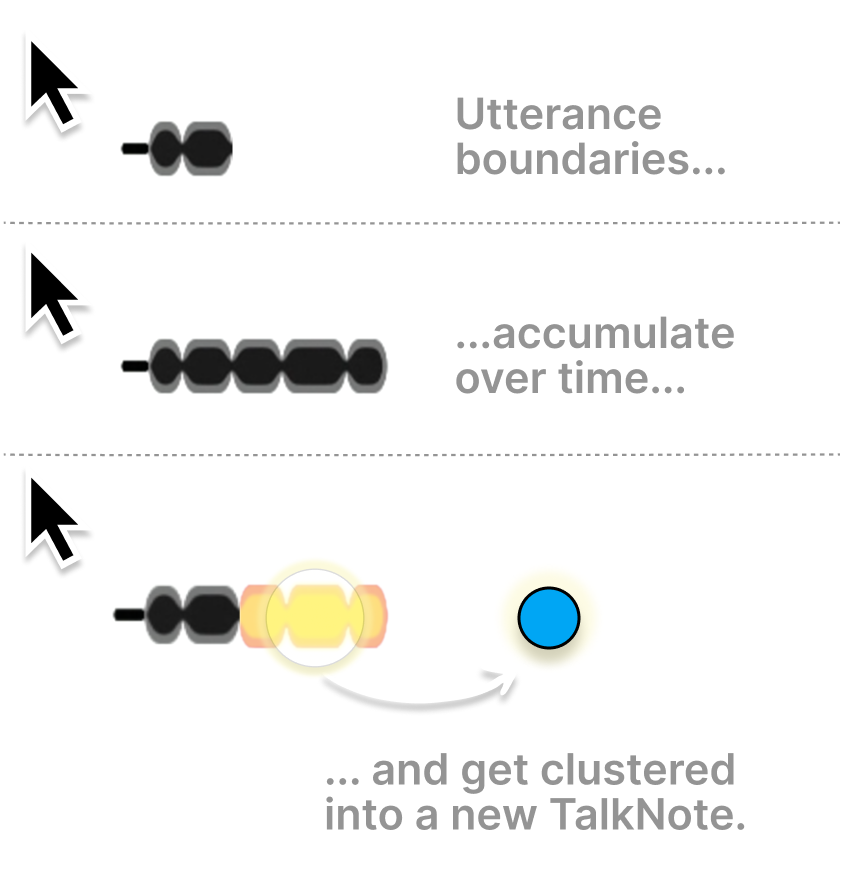}
    \caption{}
    \label{fig:feature-talkviz}
    \Description{The figure displays a series of visual indicators that represent utterance boundaries and how speech segments accumulate over time. It illustrates a process where these segments cluster together into a new element, labeled as a TalkNote. Arrows and icons are used to guide the viewer through the sequence of actions involving speech segmentation.}
  \end{minipage}
   \begin{minipage}[t]{\dimexpr\columnwidth-3.2cm\relax}
          \textbf{(2) \texttt{Talk}Viz} (DP1): Visual indicators that signal utterance boundaries and chunking operations, allowing users to see how/when their speech has been segmented and clustered into new \texttt{Talk}Notes. 
  \end{minipage}
\end{figure}

\vspace{-1em}

\begin{figure}[H]
  \begin{minipage}[t]{3cm}
    \includegraphics[align=t,width=\columnwidth]{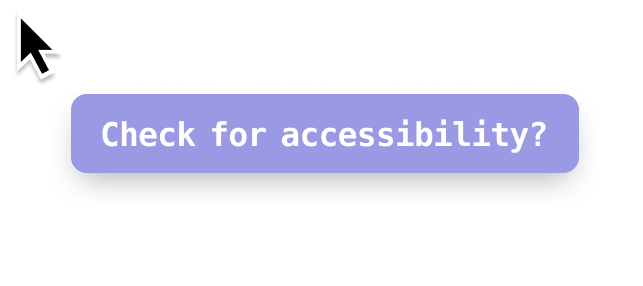}
    \caption{}
    \label{fig:feature-talktip}
    \Description{The figure displays a user interface component labeled "Check for accessibility?" within a rounded rectangle that has a light purple background. The prompt is positioned at the top left and is intended to encourage users to consider accessibility features in their interactions. A cursor is visible, indicating potential user engagement with the interface element.}
  \end{minipage}
   \begin{minipage}[t]{\dimexpr\columnwidth-3.2cm\relax}
          \textbf{(3) \texttt{Talk}Tip} (DP1, DP4): Brief, context-sensitive prompts that appear both during pauses and in response to users’ speech. 
          They encourage continued verbalization, surface relevant considerations, or pose open-ended questions to support deeper reflection.
  \end{minipage}
\end{figure}

\begin{figure}[h!]
\includegraphics[width=0.8\linewidth]{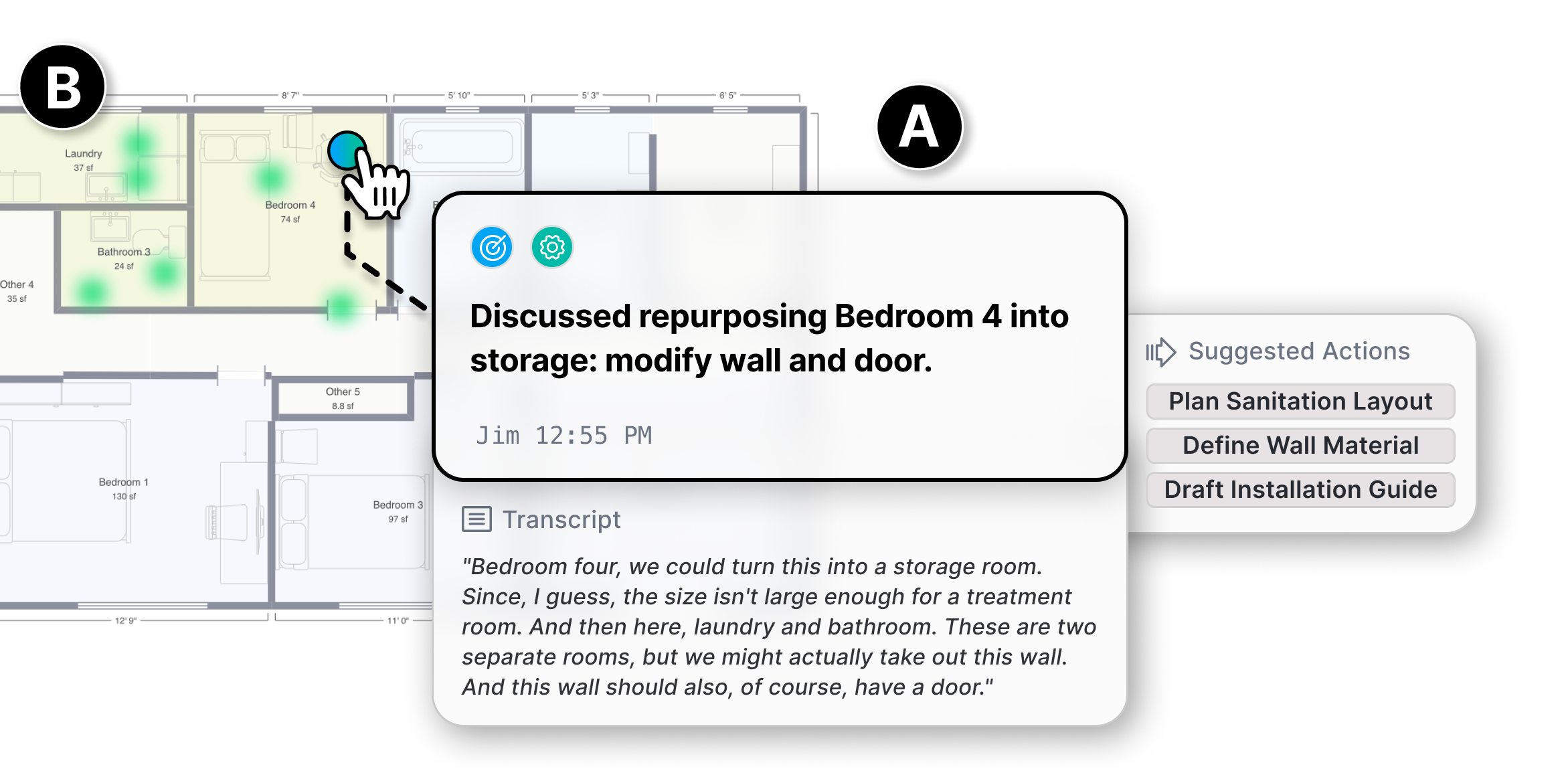}
\caption{An unfolded TalkNote with two key components: (A) structured note content combining the user’s transcript, system-generated summary, process labels, and suggested follow-up actions; and (B) spatial anchoring that situates the note at the location where the user was pointing during verbalization, complemented by pointer traces (green dots) and design element highlights (yellow overlays).}
\label{fig:feature-talknotecard}
\Description{The figure displays an unfolded TalkNote interface featuring two primary components. Section A shows a structured note that contains the user's transcript, along with system-generated summaries and suggested actions, while section B illustrates a spatial layout of a room with green point traces indicating where the user pointed during the discussion, supplemented by yellow highlights on relevant design elements.}
\end{figure}

\subsubsection{\textbf{\texttt{Talk}Notes: Contextualized Representations of Design Moments (DP2, DP3, DP4)}} 

\SystemName captures segments of speech as \textbf{\texttt{Talk}Notes}: persistent, structured units that combine transcribed text with contextual metadata. 
\texttt{Talk}Notes are generated automatically when the system detects a shift in topic or intent, allowing designers to externalize their reasoning without manual effort.  
Each \texttt{Talk}Note contains the original transcript, a concise generated summary, and process category labels (Figure \ref{fig:feature-talknotecard}).

\begin{figure}[H]
  \begin{minipage}[t]{3cm}
    \includegraphics[align=t,width=\columnwidth]{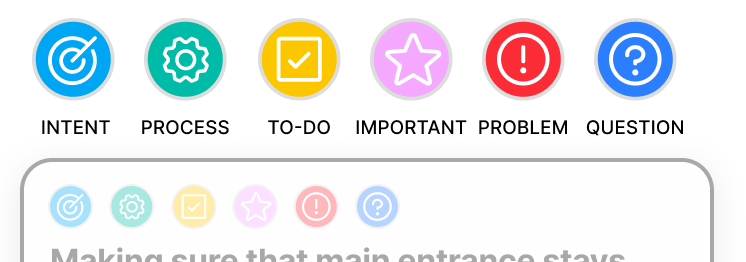}
    \caption{}
    \label{fig:feature-processLabels}
    \Description{The figure displays a series of colored circular icons, each representing a distinct category for organizing design reasoning in TalkNotes. The icons are labeled as Intent, Process, To-Do, Important, Problem, and Question, with corresponding colors that differentiate each category. Below the icons, there is a textual element that appears to provide context or instructions related to a design task.}
  \end{minipage}
   \begin{minipage}[t]{\dimexpr\columnwidth-3.2cm\relax}
          \textbf{Process Label Categories} (DP2): Each TalkNote is automatically assigned one or more categories, reflecting different kinds of design reasoning: \textit{Design Intent} (high-level goals or rationales), \textit{Process} (operations, tools, or workflow steps), \textit{ToDo} (tasks to complete later), \textit{Important} (flagged critical information), \textit{Problem} (issues or obstacles), and \textit{Question} (open uncertainties). 
  
  \end{minipage}
\end{figure}

Beyond textual content, \texttt{Talk}Notes are anchored to the design workspace through multiple forms of contextualization:

\begin{figure}[H]
  \begin{minipage}[t]{3cm}
    \includegraphics[align=t,width=\columnwidth]{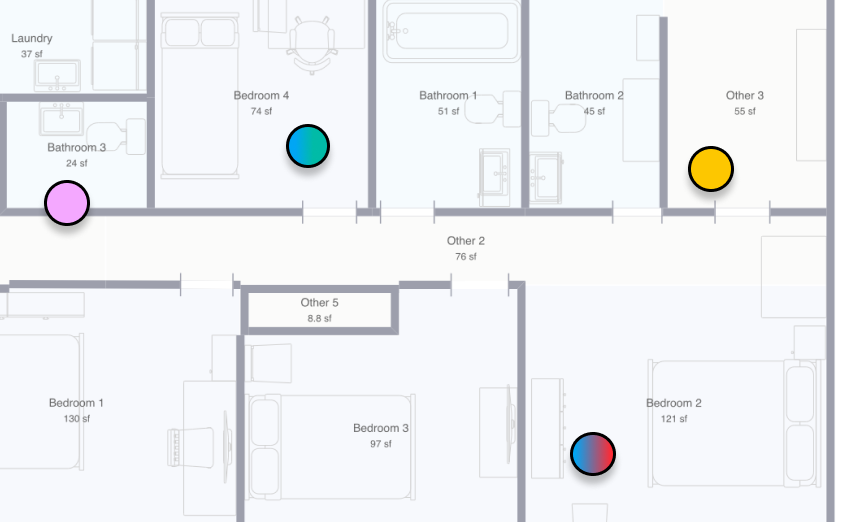}
    \caption{}
    \label{fig:feature-anchors}
    \Description{The figure displays a floor plan of a space with labeled rooms, such as "Room 1" and "Room 2." Colored circular markers indicate specific locations where user notes appear as overlays, symbolizing spatial anchors corresponding to user interaction during verbalization. The layout shows the relationship between the notes and the physical arrangement of the spaces.}
  \end{minipage}
   \begin{minipage}[t]{\dimexpr\columnwidth-3.2cm\relax}
          \textbf{Spatial Anchors} (DP3): Notes appear as overlays on the canvas at the location where the user was pointing during verbalization (as 2D/3D location).  
  \end{minipage}
\end{figure}

\begin{figure}[H]
  \begin{minipage}[t]{3cm}
    \includegraphics[align=t,width=\columnwidth]{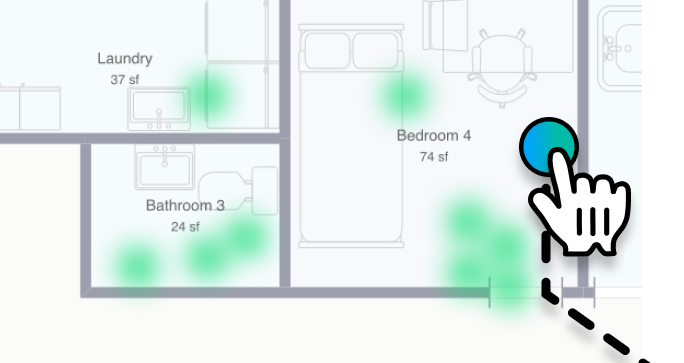}
    \caption{}
    \label{fig:feature-traces}
    \Description{The figure shows a stylized floor plan of a building, with labeled rooms such as "Laundry," "Bedroom," and "Bathroom." Overlaying the floor plan are green dots indicating points of cursor movement, with a hand icon and a curved line representing the path taken by the cursor. The overall design illustrates the tracing of cursor paths in relation to specific areas within the layout.}
  \end{minipage}
   \begin{minipage}[t]{\dimexpr\columnwidth-3.2cm\relax}
          \textbf{Pointer Traces} (DP2, DP3): Visual paths of cursor movement during speech are stored and shown as overlays, providing contextual grounding.  
  \end{minipage}
\end{figure}

\begin{figure}[H]
  \begin{minipage}[t]{3cm}
    \includegraphics[align=t,width=\columnwidth]{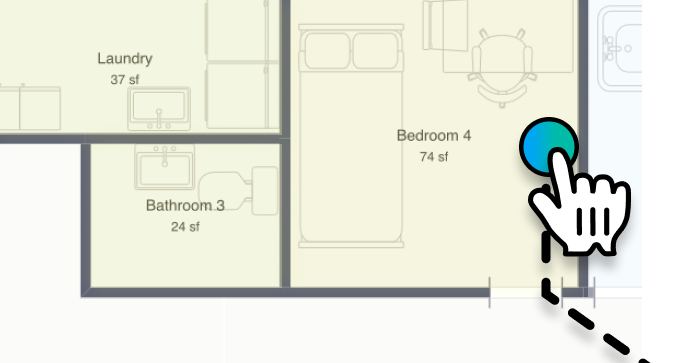}
    \caption{}
    \label{fig:feature-highlights}
    \Description{The figure displays a floor plan of a residential layout, indicating various rooms such as a laundry area, bathroom, and bedrooms. A hand cursor symbol highlights or points to a specific location on the floor plan, suggesting an interaction or reference to an architectural element within the design. Dotted lines extend from the cursor to visually link these elements to additional contextual information.}
  \end{minipage}
   \begin{minipage}[t]{\dimexpr\columnwidth-3.2cm\relax}
          \textbf{Design Element Highlights} (DP2, DP3): Relevant architectural elements referenced during users' speech are visually linked to the note.  
  \end{minipage}
\end{figure}

\begin{figure}[H]
  \begin{minipage}[t]{3cm}
    \includegraphics[align=t,width=\columnwidth]{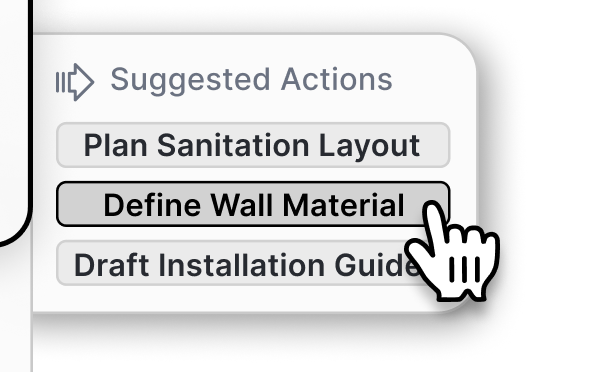}
    \caption{}
    \label{fig:feature-actionSuggestions}
    \Description{The figure displays a user interface element featuring a menu labeled "Suggested Actions" with three selectable items: "Plan Sanitation Layout," "Define Wall Material," and "Draft Installation Guide." A hand cursor hovers over the "Define Wall Material" option, indicating user interaction.}
  \end{minipage}
   \begin{minipage}[t]{\dimexpr\columnwidth-3.2cm\relax}
          \textbf{Action Suggestions} (DP4): Based on the \texttt{Talk}Note's captured context, the system dynamically generates a UI button menu with follow-up system actions for users to trigger\footnotemark.
  \end{minipage}
\end{figure}
\footnotetext{Action suggestion buttons are generated per \texttt{Talk}Note but remain non-functional in our prototype, serving solely as probes.}

To make these representations lightweight and accessible, each \texttt{Talk}Note appears initially as a small anchor on the canvas. 
Hovering over or selecting the anchor expands it into a \textbf{\texttt{Talk}Note Card}, showing the transcript snippet, summary, and category labels (Figure \ref{fig:feature-talknotecard} A). 
This provides designers with a quick way to revisit and reflect on earlier reasoning in its original design context.

\begin{figure}[h!]
\includegraphics[width=0.8\linewidth]{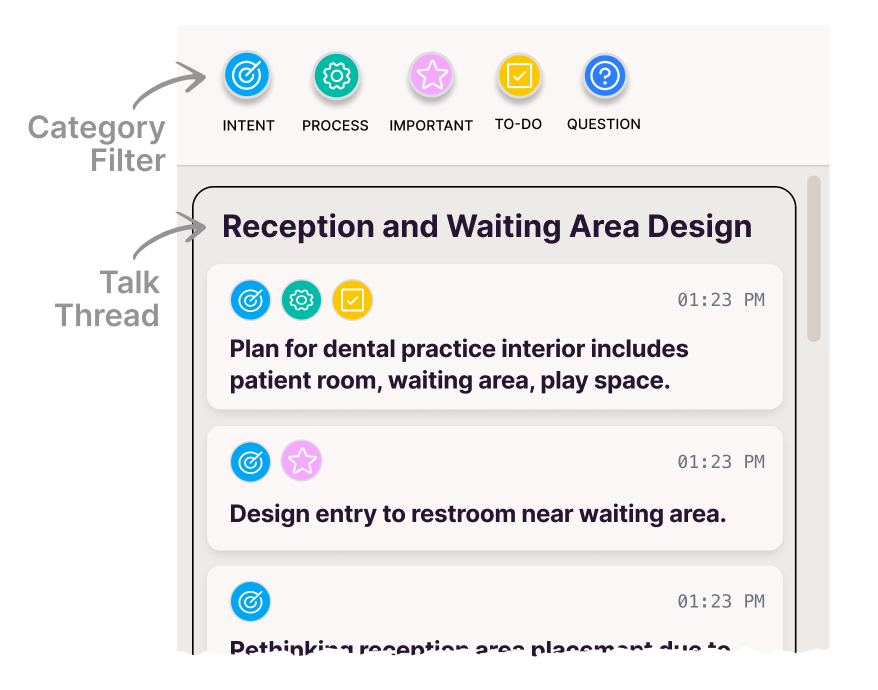}
\caption{TalkExplorer sidebar with filter options and TalkNotes topically clustered into TalkThreads.}
\label{fig:feature-talkexplorer}
\Description{The figure shows a graphical user interface of the TalkExplorer sidebar. It includes a category filter with options labeled "Intent," "Process," "Important," "To-Do," and "Question" at the top, and below it, a section labeled "Reception and Waiting Area Design" that contains multiple TalkNotes organized in a threaded format, each showing a timestamp and brief text.}
\end{figure}

\subsubsection{\textbf{\texttt{Talk}Explorer: Browsing, Clustering, and Filtering Notes} (DP3)} 

For more deliberate review and organization, \SystemName provides the \textbf{\texttt{Talk}Explorer}: a sidebar list view for navigating captured \texttt{Talk}Notes (Figure \ref{fig:feature-talkexplorer}). 
Notes are dynamically organized into \textbf{\texttt{Talk}Threads}, which cluster related ideas based on semantic similarity and temporal proximity. 
Users can filter displayed \texttt{Talk}Notes to focus on specific categories (e.g., \textit{Design Intent}, \textit{Question}, \textit{To-Do}). 

Selecting a note in the \texttt{Talk}Explorer resurfaces its contextual anchors: the corresponding pointer traces and design elements reappear within the main canvas, situating the designer in the original context of their thought and supporting reflection or continuation of interrupted reasoning.

\subsubsection{\textbf{\texttt{Talk}Reminders}: Resurfacing Related Prior Notes (DP4)}

In addition to manual browsing, the system proactively resurfaces relevant prior notes through \textit{\texttt{Talk}Reminders}.

\begin{figure}[H]
  \begin{minipage}[t]{3cm}
    \includegraphics[align=t,width=\columnwidth]{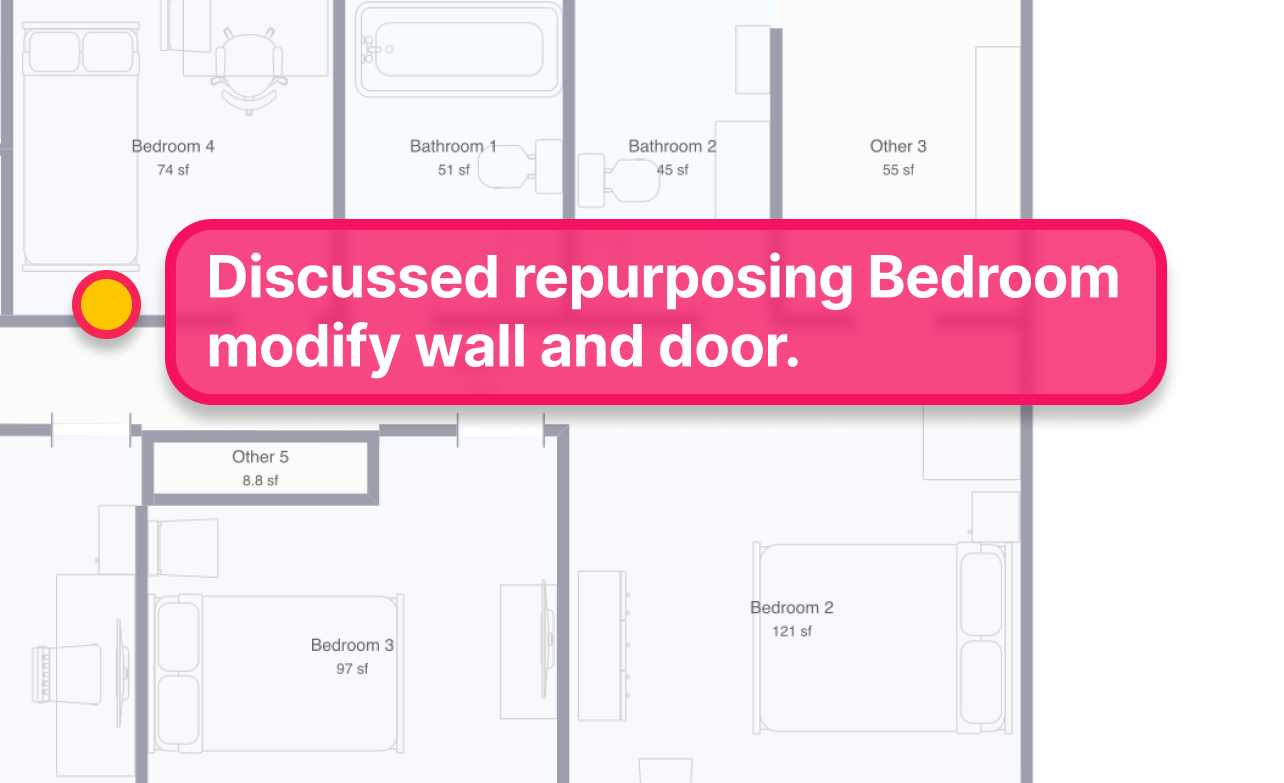}
    \caption{}
    \label{fig:feature-talkreminder}
    \Description{The figure shows a section of a digital canvas displaying a floor plan, with multiple rooms labeled. An annotation appears prominently in a pink box, highlighting a previous discussion about repurposing a bedroom wall and door. The visual elements aim to recall earlier user inputs related to the current context.}
  \end{minipage}
   \begin{minipage}[t]{\dimexpr\columnwidth-3.2cm\relax}
          When the user’s current verbalization relates to earlier concerns, previously created \texttt{Talk}Notes briefly reappear on the canvas, highlighted and summarized next to their anchors. 
This lightweight mechanism helps jog memory and connect ongoing reasoning with past decisions without requiring explicit search or navigation.  
  \end{minipage}
\end{figure}

\subsection{Implementation Details}

We developed \SystemName as a web-based prototype, following a 
client–server architecture. 
The front end was implemented in \textit{React} with \textit{Three.js} for 3D rendering and \textit{TLDraw} for the 2D sketching canvas.
The back end, built with \textit{Node.js} and \textit{Express}, manages data flow between transcription services \textit{(Deepgram Nova-3) }and large language models (LLMs) APIs \textit{(GPT-4o, Gemini 2.5 Pro)}. For further implementation details, see Appendix Section \ref{sec:appendix-implementation-details}.

\section{User Study}

To better understand possible benefits and limitations of pointer-centric AI-supported think-aloud computing interactions, we conducted a within-subject remote user study aimed at providing insights into these research questions:
\begin{itemize}[font=\bfseries, align=left, labelwidth=2.5em, leftmargin=3em]
    \item[RQ1a] \textit{What are the key differences between working with real-time text-based transcription only and \SystemName? }
    \item[RQ1b] \textit{Does \SystemName incentivize people to think aloud more? }
    \item[RQ2] \textit{How do people think aloud and work with \SystemName?  }
    \item[RQ3] \textit{What are users’ perceived benefits and challenges for working with \SystemName?}
 \end{itemize}

These research questions aim to understand different aspects of the PointAloud interaction suite, collectively unpacking hypotheses on how users respond to these four aspects: \textit{think-aloud computing}, \textit{pointer-adjacent ambient displays}, \textit{pointer-attention context}, and \textit{AI-driven support features}. 
Questions \textit{RQ1a} and \textit{RQ2a} largely explore the effect of pointer-adjacent ambient displays on think-aloud computing.
While \textit{RQ2} equally touches on each of these interaction suite concepts to understand how they interplay to support the CAD task scenario.
Finally, \textit{RQ3} tends to lean toward insights on AI-driven support as participants would tend to associate those features as the benefits of the system.

For RQ1a and RQ1b, we decided to compare \SystemName interactions
with a modified (limited) version of the \SystemName system, where instead of the \texttt{Talk}Pointer and \texttt{Talk}Note features, the live user's transcription appears in the left side panel as text (optionally see Appendix Figure~\ref{fig:text-based-screenshot}). 
This limited, low distraction system (baseline) aims to mimic the contemporary form of live transcription system UI (for example, in Zoom or Teams) -- the baseline provides support for think-aloud capture, without the additional embedded user interfaces of \SystemName.

\begin{figure}[h!]
  \includegraphics[width=0.5\linewidth]{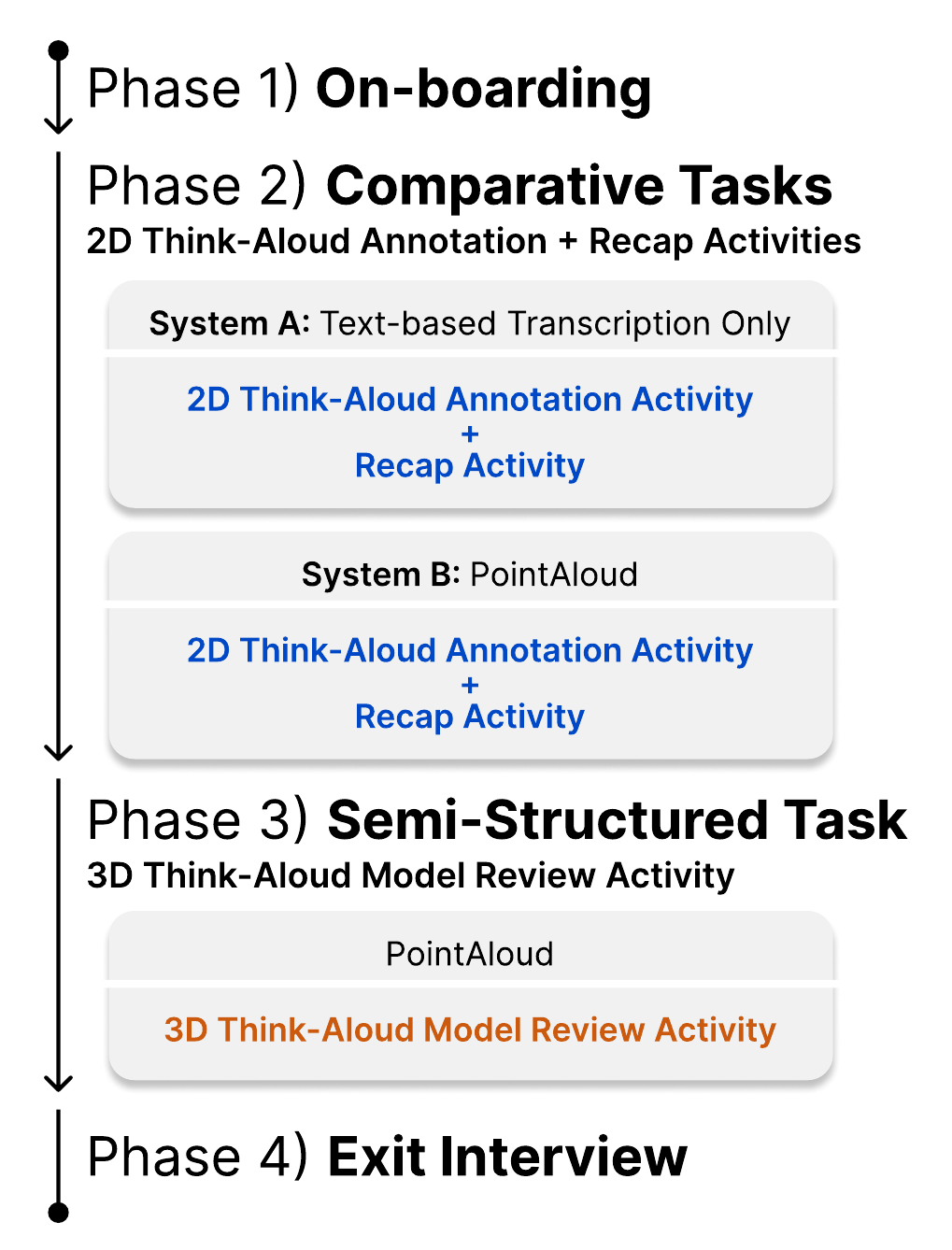}
  \caption{Process diagram of the four-phase study procedure.}
  \label{fig:study-procedure}  
  \Description{The figure illustrates a process diagram outlining the phases of a study procedure. It is organized into four main phases: On-Boarding, Comparative Tasks, Semi-Structured Task, and Exit Interview, with specific activities and systems included under the comparative tasks. Arrows indicate the flow from one phase to the next, emphasizing the sequential nature of the study.}
\end{figure}

 \subsection{Participants}

We recruited 12 participants \textit{(7 self-identified as female and 5 as male, aged 23–49 years, $M=30.8$, $SD=6.7$)} with professional backgrounds in architecture and interior design. 
Participants had between 2 and 19 years of professional experience \textit{($M=6.9$, $SD=4.5$)}, see Appendix Table~\ref{tab:participants}. 
Recruitment was conducted via snowball sampling and professional online user forums. 
Eligibility was determined through a screening form, requiring a minimum of two years of professional experience in architecture or interior design, as well as at least two years of CAD experience and fluent English-level proficiency. 
All participants signed an IRB-approved consent form prior to the first study session and were compensated with a \$125 gift card upon completion.

\subsection{Procedure and Tasks} \label{sec:study-procedure}

The 90-minute-long remote study was structured into four phases:

\textbf{Phase 1) On-Boarding (20–30 min):}
At the beginning of the session, the facilitator welcomed participants, introduced the study, and ensured that consent was signed. 
Participants then completed a short pre-task survey and watched a three-minute tutorial video demonstrating PointAloud’s core functionalities. 
Afterward, they performed a five-minute guided hands-on trial to familiarize themselves with the prototype interface and features.

\textbf{Phase 2) Comparative Tasks (2 × 20 min):}  
In the comparative phase, participants completed two design tasks, alternating between a baseline version of the prototype (limited to live text-based transcription) and the full version of \SystemName.  
Each task was paired with a distinct design brief (similar to the task described in \ref{sec:walkthrough-2d}), with the order of briefs counterbalanced across participants to mitigate learning effects.  
These design briefs were intentionally framed to elicit early-stage design reasoning, requiring participants to repurpose an apartment layout while balancing functional requirements with spatial constraints and emerging client needs.  
For each task, participants first worked for ten minutes on a \textbf{2D floor plan annotation activity} while verbalizing their thoughts aloud, followed by a five-minute \textbf{recap activity} in which they summarized their early design decisions, open questions, and issues to flag as if preparing for a client meeting.  
After each task, participants completed an attitudinal post-task questionnaire.

\textbf{Phase 3) Semi-structured Task (5 min):}  
In the third phase, participants were asked to continue working on their previous design brief, this time by reviewing a spatial model of the floor in the 3D model view (similar to the task described in \ref{sec:walkthrough-3d}).  
They could only view and explore a model of the apartment (without making changes or annotations) to analyze the existing space and reflect on interior design considerations.  
Participants worked for five minutes using the full prototype system while continuing to think aloud.  
This task was designed to encourage spontaneous engagement with the prototype’s features in a spatial 3D interface.

\textbf{Phase 4) Exit Interview (15 min):}
In the final phase, participants took part in a semi-structured interview. They reflected on their experience with \SystemName, comparing it to their usual tools and workflows and discussing its effectiveness in supporting and harnessing think-aloud practices. Participants provided feedback on strengths, limitations, and opportunities for integrating such a system into professional design work.

\subsection{Collected Data, Measures, and Analysis}

Across the study, we collected the following data:
\begin{itemize}
\item Video, screen, audio recordings, and machine-generated transcripts of all task sessions (phase 2 and 3)
\item System interaction log data of all task sessions (phase 2 and 3)
\item Post-task surveys data (phase 2)
\item Audio recordings and machine-generated transcripts of the exit interviews (phase 4)
\end{itemize}

To compare the baseline (text-based transcript-only system) and \SystemName (Q1a), we analyzed the post-task surveys from phase 2 that probed participants’ perceived thinking-aloud support, task support (including relevance of system suggestions), process awareness, and cognitive load/distraction on a 6-point Likert scale.
We applied the Wilcoxon signed-rank test to assess statistical significance and calculated 95\% confidence intervals for mean differences via bootstrapping with 10,000 replications using R~\cite{rcoreteam_language_2024}. 
This approach has been suggested for similar data and studies~\cite{zhu_assessing_2018, masson_statslator_2023}.

To assess to what extent \SystemName incentivized participants to think aloud (Q1b), we compared participants’ words-per-minute (WPM) across the baseline and the \SystemName condition \footnote{Transcripts were tokenized using spaCy~\cite{honnibal_spacy_2020}. We counted tokens that were not punctuation or whitespace and contained at least one alphabetic character. Common English contractions were merged and treated as single words.}.
We then conducted paired comparisons using the Wilcoxon signed-rank test and paired t-test to assess differences in WPM between conditions. 

To answer how people think aloud and work with \SystemName (Q2), we conducted a video interaction analysis~\cite{baumer_comparing_2011} of the video recordings collected in the PointAloud-supported activities in phase 2 and 3.
Using a custom-built coding tool, we reviewed the video recordings alongside the system interaction log files for each session and coded participants' interactions, such as their verbal responses to system suggestions. 
From this data, we then created timeline visualizations for each session using ggplot2~\cite{wickham_ggplot2_2016}.

Finally, to investigate users’ perceived benefits and challenges of working with \SystemName (Q3), we conducted a reflexive thematic analysis~\cite{braun_reflecting_2019} of the interview transcripts. 
We followed an iterative inductive coding process and generated themes through affinity diagramming using Dovetail~\cite{dovetail_customer_2025}.

\begin{figure*}[h!]
  \centering
  \includegraphics[width=1\linewidth]{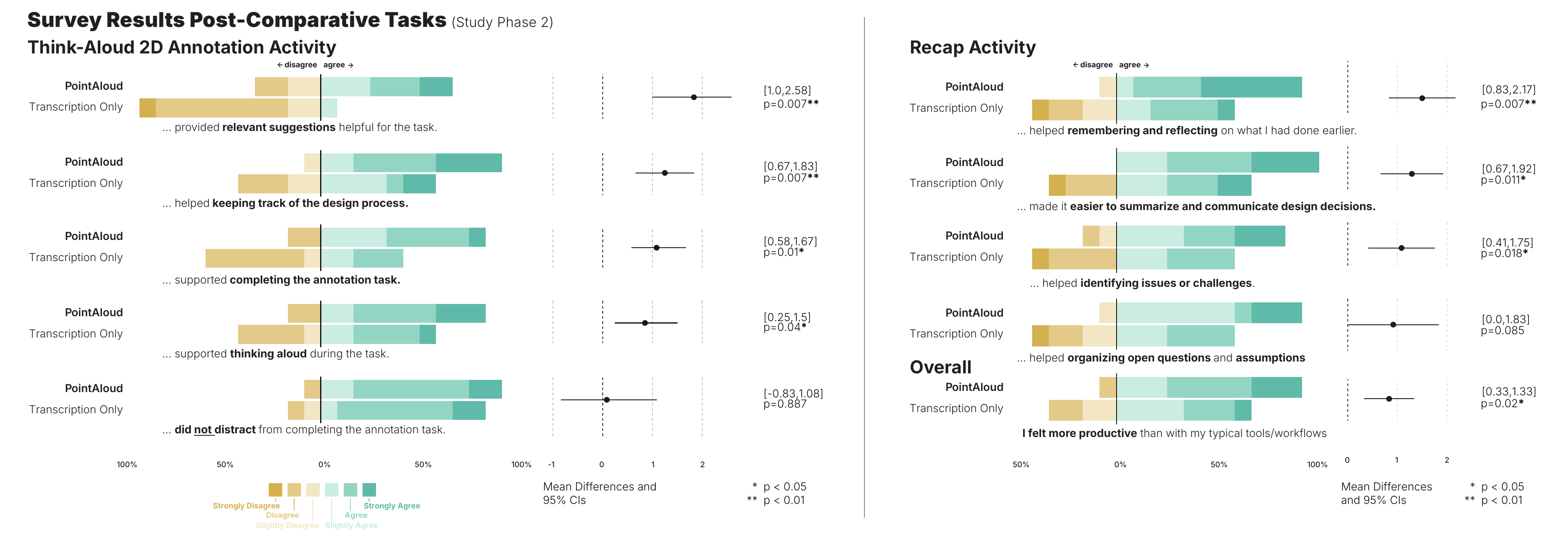}
  \caption{Participants’ responses when rating the 6-point Likert statements for annotation and recap activities completed with PointAloud and text-based live transcription only (baseline), ranked from largest to smallest effects; Dots show the mean difference of PointAloud compared to Text-based Transcription Only; Bars are the 95\% CIs calculated with the studentized bootstrap method.}
  
  \Description{The figure presents the results of a survey where participants rated their experiences on a Likert scale regarding two transcription methods: PointAloud and text-based transcription only. Each statement is represented by a horizontal bar indicating the proportion of agreement and disagreement, with dots illustrating the mean differences between the two methods along with 95 percent confidence intervals. Statistical significance is denoted with asterisks.}
  \label{fig:survey-results}
\end{figure*}

\section{Study Findings} \label{sec:findings}

\subsection{Key differences: Live Text-based Transcription vs \SystemName (RQ1a, RQ1b)} \label{sec:findings-RQ1}

In the two comparative tasks (phase 2), participants worked on apartment redesign briefs that asked them to annotate a 2D floor plan and articulate early zoning ideas while thinking aloud. 
They then completed a recap activity, summarizing key design decisions, open questions, and concerns as if preparing for a client meeting using the annotated floor plan and the think-aloud text-based transcript or \texttt{Talk}Notes, respectively. 
Here, we report on the observed differences in participants’ processes when working with conventional live text-based transcription (baseline) versus \SystemName.  
Optionally, we included screenshots in Appendix \ref{apdx:task-outcomes} to give an impression of the variety of participant-created floor plans using PointAloud.

\subsubsection{\textbf{Questionnaire}} \label{sec:findings-RQ1-Q}

In the comparative post-task surveys, participants consistently rated \SystemName higher than the baseline text-based transcription system across both task phases (Figure~\ref{fig:survey-results}).
During the \textbf{floorplan annotation activity}, participants reported a significant improvement in \textbf{relevance of system suggestions} (\textit{“…provided relevant suggestions helpful for the task”}, $MD = 1.83$, $p=0.007$), \textbf{process awareness} (\textit{“…helped keep track of the design process”}, $MD = 1.25$, $p=0.007$), and \textbf{task support} (\textit{“…supported completing the annotation task”}, $MD = 1.08$, $p=0.01$). They also noted stronger \textbf{thinking-aloud support} (\textit{“…supported thinking aloud during the task”}, $MD = 0.83$, $p=0.04$), while ratings for \textbf{cognitive load / distraction} showed no significant difference ($MD = 0.08$, $p=0.887$).
While the baseline lacks the \SystemName features, it provides an opportunity to meaningfully compare the cognitive load and distraction between the two conditions. Overall, it is expected that participants would prefer \SystemName for these tasks compared to the baseline, however additional benefits from \SystemName did not increase effort or attention when compared to the baseline.

During the \textbf{client recap activity}, PointAloud was valued for aiding \textbf{memory and reflection} (\textit{“…helped remembering and reflecting on what I had done earlier”}, $MD = 1.50$, $p=0.007$), \textbf{summarization and communication} (\textit{“…made it easier to summarize and communicate design decisions”}, $MD = 1.25$, $p=0.011$), and \textbf{issue identification} (\textit{“…helped identifying issues or challenges”}, $MD = 1.08$, $p=0.018$). Participants also perceived improved \textbf{organization of open questions and assumptions}, though this effect was marginal ($MD = 0.92$, $p=0.085$).
During the brief recap activity, participants preferred the AI-supported documentation features of \SystemName compared to the text-based baseline. On reflection, the baseline might be improved with AI support such as chat or generative search powered by Retrieval-Augmented Generation (RAG)~\cite{lewis_retrievalaugmented_2020}. Future work should take this into consideration when evaluating think-aloud systems.

Across both phases, participants expressed feeling \textbf{more productive than with their typical tools and workflows} when using PointAloud ($MD = 0.83$, $p=0.02$).

\subsubsection{\textbf{Observations}} \label{sec:findings-RQ1-O}

\noindent\textbf{Differences During Think-Aloud Activity.}  
With the text-based transcription, participants spoke aloud while their speech appeared in the left side panel as text.
However, the transcript was never actively visited and used during this task activity. 
In contrast, with PointAloud, participants’ utterance stream was displayed via \texttt{Talk}Text and \texttt{Talk}Viz near their cursor while periodically forming anchored \texttt{Talk}Notes on the canvas and side panel. 
While \texttt{Talk}Tips frequently appeared next to the cursor, participants’ responses varied widely. 
Some rarely engaged with them, while others actively used them to elaborate their reasoning and to guide ongoing design decisions. 
Similarly, \texttt{Talk}Notes themselves functioned differently across participants—ranging from passive documentation to active use as a springboard for reflection and further exploration. 
These varied engagement patterns with \texttt{Talk}Notes and \texttt{Talk}Tips during think-aloud are analyzed in more detail in Section~\ref{sec:findings-RQ2}.

\vspace{0.5em}
\noindent\textbf{No Difference in Extent of Verbalization (RQ1b).}  
For the \textit{annotation} think-aloud task activities, we compared words-per-minute (WPM) across the two conditions to assess whether \SystemName incentivized more verbalization. Differences were participant-specific: some (e.g., P07, P10) spoke considerably more with \SystemName, while others (e.g., P04, P11) spoke less (for further details see also Appendix Table~\ref{tab:wpm}). 
On average, the difference across conditions was negligible (Mean = 1.23 WPM, SD = 17.46). 
Neither a paired t-test ($t = -0.245, p = .81$) nor a Wilcoxon signed-rank test ($W = 34, p = .73$) indicated a significant effect. 
This suggests that \SystemName did not uniformly increase the quantity of verbalization, though it shifted how participants engaged with their speech as design material.
However, because participants were explicitly instructed to think aloud in both conditions, our study design does not isolate whether PointAloud itself prompted additional verbalization.

\begin{figure}[h!]
  \includegraphics[width=0.9\linewidth]{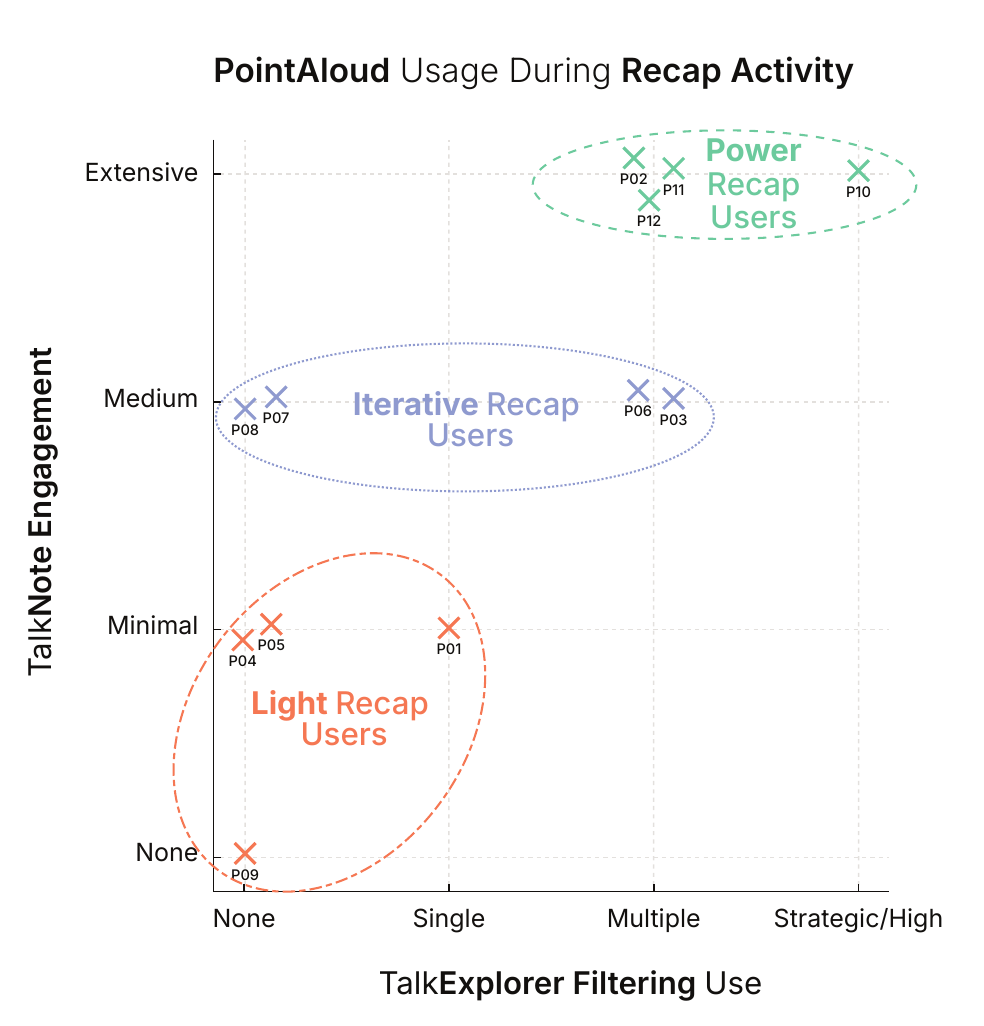}
  \caption{Patterns of \texttt{Talk}Note engagement and \texttt{Talk}Explorer filtering use during the recap activity, distinguishing \textit{Light Users}, \textit{Iterative Users}, and \textit{Power Users}.}
  \label{fig:talknote-usage}  
  \Description{The figure presents a scatter plot with two axes: TalkNote Engagement on the vertical axis and TalkExplorer Filtering Use on the horizontal axis. Three distinct user groups—Light Users, Iterative Users, and Power Users—are represented by ellipses and are positioned throughout the plot, indicating varying levels of engagement and filtering use during recap activities.}
\end{figure}

\vspace{0.5em}
\noindent\textbf{Extreme Differences During Recap Activity.}  
When completing their \textit{recap activity} in which participants summarized their earlier design process as if preparing for a client meeting, engagement diverged strongly between conditions. 
With text-based transcription alone, no participant made substantive use of the transcript (only P07 briefly skimmed the transcript for a few seconds). 
In contrast, \texttt{Talk}Notes in \SystemName provided an externalized memory that participants engaged with to varying degrees. 
As shown in Figure~\ref{fig:talknote-usage}, \textbf{we identified three usage patterns during the recap activity} with PointAloud in regard to users' engagement with \texttt{Talk}Notes and the \texttt{Talk}Explorer's category filters:  
\begin{itemize}
    \item \textit{\textbf{Light Recap Users}} (e.g., P01, P04, P05, P09) who made minimal or no use of PointAloud (\texttt{Talk}Notes and category filter).  
    
    \item \textit{\textbf{Iterative Recap Users}} (e.g., P03, P06, P07, P08) who selectively revisited and filtered \texttt{Talk}Notes to organize their summaries.  
    
    \item \textit{\textbf{Power Recap Users}} (e.g., P02, P10, P11, P12) who engaged extensively, strategically filtering and cross-referencing notes to construct detailed recaps.  
\end{itemize}
These differences highlight how \SystemName not only captured verbalizations but also supported many users in retrospectively reflecting on their design process and structuring their documented ideas for communicating their process.

\begin{table*}[h!]
\centering
\caption{Overview of the results of the PointAloud-supported think-aloud activities (2D annotation and 3D review activities combined); reporting session duration and process statistics for \texttt{Talk}Notes and \texttt{Talk}Tips interactions. 
Columns with conditional coloring indicate relative intensity within the measure. Final column indicates user engagement pattern: \textit{Note Explorer} (\cdotcolor{green!40}), \textit{Tip-driven Elaborator} (\cdotcolor{blue!40}), \textit{Heavy Integrator} (\cdotcolor{green!40}+\cdotcolor{blue!40}), \textit{Documentation-only User} (see Section \ref{sec:findings-RQ2}). \textit{* Note: In P02’s first activity, \texttt{Talk}Note creation was periodically interrupted by a browser plug-in conflict.}}
\begin{tabularx}{0.6\textwidth}{p{0.4cm}p{0.7cm}p{0.7cm}p{0.7cm}p{0.7cm}p{0.7cm}p{0.7cm}p{4cm}}
\hline
\rotatebox{55}{\textbf{ID}} 
& \rotatebox{55}{\begin{tabular}[c]{@{}l@{}}\textbf{Activity Duration}\\(mm:ss)\end{tabular}}
& \rotatebox{55}{\begin{tabular}[c]{@{}l@{}}\# \textbf{TalkNotes}\\ \textbf{Created}\end{tabular}}
& \rotatebox{55}{\begin{tabular}[c]{@{}l@{}}\# \textbf{TalkNotes}\\ \textbf{Merged}\end{tabular}}
& \rotatebox{55}{\begin{tabular}[c]{@{}l@{}}\# \textbf{User-Checked}\\ \textbf{TalkNotes}\end{tabular}}
& \rotatebox{55}{\begin{tabular}[c]{@{}l@{}}\# \textbf{TalkTips}\\ \textbf{Shown}\end{tabular}}
& \rotatebox{55}{\begin{tabular}[c]{@{}l@{}}\# \textbf{TalkTip}\\ \textbf{Responses}\end{tabular}}
& \rotatebox{55}{\begin{tabular}[c]{@{}l@{}}\textbf{Engagement}\\ \textbf{Pattern}\end{tabular}} \\
\hline
P01 & 15:18 & 28 & 13 & \cellcolor{green!40} 11 & 63 & \cellcolor{blue!3} 2 & \cdotcolor{green!40} Note Explorer  \\
P02* & 14:29 & 35 & 13 & 0  & 56 & \cellcolor{blue!5} 3 & Documentation-only\\
P03 & 15:44 & 40 & 26 & 0  & 83 & \cellcolor{blue!2} 1 & Documentation-only\\
P04 & 15:23 & 57 & 27 & \cellcolor{green!21} 6  & 94 & 0 & \cdotcolor{green!40} Note Explorer  \\
P05 & 14:07 & 55 & 30 & 0  & 91 & \cellcolor{blue!3} 2 & Documentation-only\\
P06 & 15:22 & 55 & 27 & \cellcolor{green!14} 4  & 73 & \cellcolor{blue!40} 24 & \cdotcolor{green!40}+\cdotcolor{blue!40} Heavy Integrator \\
P07 & 15:06 & 60 & 32 & 0  & 70 & \cellcolor{blue!5} 3 & Documentation-only\\
P08 & 14:56 & 58 & 41 & 0  & 76 & \cellcolor{blue!2} 1 & Documentation-only\\
P09 & 15:19 & 42 & 24 & 0  & 92 & \cellcolor{blue!2} 1 & Documentation-only\\
P10 & 15:23 & 63 & 36 & \cellcolor{green!21} 6  & 78 & \cellcolor{blue!20} 12 & \cdotcolor{green!40}+\cdotcolor{blue!40} Heavy Integrator\\
P11 & 13:37 & 46 & 22 & 0  & 62 & \cellcolor{blue!2} 1 & Documentation-only\\
P12 & 15:20 & 65 & 40 & \cellcolor{green!3} 1  & 64 & \cellcolor{blue!11} 7 & \cdotcolor{blue!40} Tip-driven Elaborator\\
\hline
\textit{Mean} & \textit{15:01} & \textit{50.3} & \textit{27.6} & \textit{2.3} & \textit{75.2} & \textit{4.8} & \\
\bottomrule
\end{tabularx}
\Description{The table presents a summary of results from PointAloud-supported think-aloud activities, detailing various metrics for individual participants identified by ID. It includes columns for activity duration, the number of TalkNotes created and merged, user engagement metrics, and the engagement pattern represented with specific symbols. Conditional coloring highlights the intensity of the values within certain measures, facilitating easy comparison across participants.}
\label{tab:think-aloud-stats}
\end{table*}

\subsection{Observed Think-Aloud Workflows Using PointAloud (RQ2)} \label{sec:findings-RQ2}

During the two PointAloud-supported think-aloud activities, participants repurposed an apartment for childcare by first annotating the floor plan in 2D (\textit{phase 2}), then reviewing and reflecting on their design in the 3D viewer (\textit{phase 3}).

To analyze how participants interacted with PointAloud’s features throughout the tasks, we aggregated system and user interaction event counts of both think-aloud activities from phase 1 and phase 2 (see Table ~\ref{tab:think-aloud-stats}) and visually compared all sessions' timeline visualizations (optionally, see Figure ~\ref{fig:timelines-all} in the Appendix).
Overall, \textbf{we identified four different engagement patterns during the think-aloud activities} across participant sessions, according to their interactions with the \texttt{Talk}Notes and \texttt{Talk}Tip features during the think-aloud planning activity--- 
\textit{Note Explorers} (\cdotcolor{green!40}), \textit{Tip-driven Elaborators} (\cdotcolor{blue!40}), \textit{Heavy Integrators} (\cdotcolor{green!40}+\cdotcolor{blue!40}), and \textit{Documentation-only Users} (see Figure \ref{fig:timelines-examples})---which we unpack below:

\begin{figure*}[h!]
\includegraphics[width=1\linewidth]{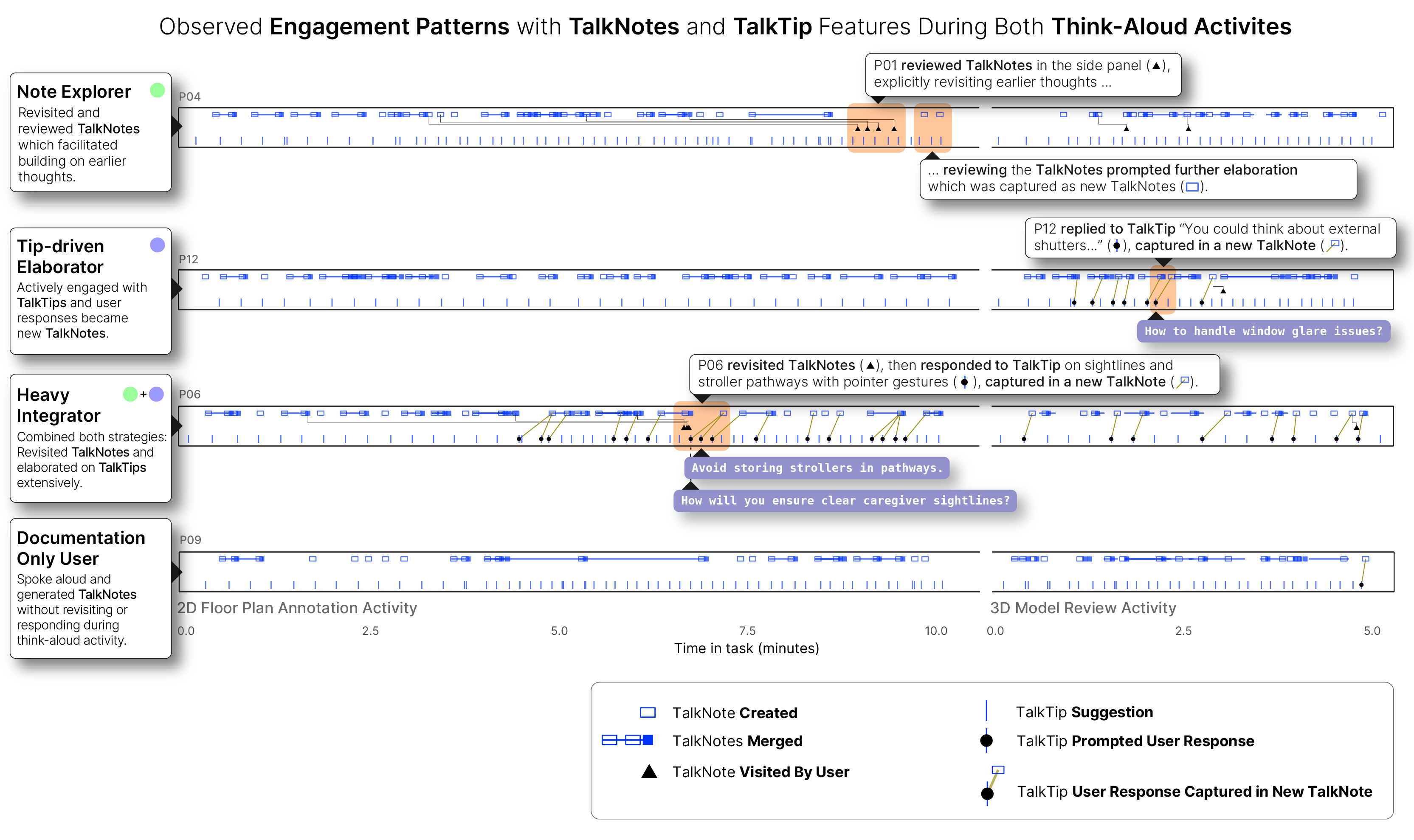}
\caption{Timeline visualizations of interaction events, illustrating four engagement patterns with \texttt{Talk}Notes and \texttt{Talk}Tips during the 2D floor plan annotation and 3D model review activities: \textit{Note Explorer}, \textit{Tip-driven Elaborator}, \textit{Heavy Integrator}, and \textit{Documentation-only User}.}
\label{fig:timelines-examples}
\Description{The figure presents a timeline visualization of interaction events categorized into four distinct user engagement patterns: Note Explorer, Tip-driven Elaborator, Heavy Integrator, and Documentation Only User. Each pattern is depicted along a horizontal timeline showing the duration of two activities—2D floor plan annotation and 3D model review—with specific symbols representing actions such as TalkNotes creation, merging, and responses to \texttt{Talk}Tip suggestions. Notable user behaviors and interactions are highlighted, illustrating how users engaged with the features during each task.}
\end{figure*}

\paragraph{\textbf{Varied Engagement Patterns with \texttt{Talk}Notes: From Background Documentation to Active Think-Aloud Support}} 
\texttt{Talk}Notes were generated automatically as participants verbalized, and the system created between 28 and 65 \texttt{Talk}Notes per session ($M=50.3, SD=11.9$) across both 2D/3D activities (see sequences of squares 
\textcolor{blue}{\ding{112}}/\textcolor{blue}{\ding{110}} in Figure~\ref{fig:timelines-examples} and Figure \ref{fig:timelines-all}). 
However, participants varied in whether they \textit{engaged with these notes during the think-aloud activities}: 
Most participants refrained from exploring the captured notes, while some occasionally revisited them by clicking/hovering (see sequences of triangles \ding{115} in Figure \ref{fig:timelines-examples}, Figure \ref{fig:timelines-all}). 
For example, P01, P04, P06, and P10 repeatedly inspected their \texttt{Talk}Notes in the 2D and 3D activities (between 11 and 4 inspected \texttt{Talk}Notes), while most others left the accumulating notes untouched. 
In the case of P04, after annotating the floor plan for about eight minutes, the participant paused to scroll through the \texttt{Talk}Notes in the side panel, explicitly revisiting earlier thoughts (\textit{“just reviewing my notes here…”}). This review prompted further elaboration, as they considered follow-up questions about zoning, occupancy, sound, and parking (see Figure \ref{fig:timelines-examples}).
Overall, these observations highlight two broad workflows: \textit{Docu\-men\-tation-only users} who treated \texttt{Talk}Notes primarily as an \textit{automatic background capture for documentation purposes} while thinking aloud, and those who acted as \textit{note explorers} (\cdotcolor{green!40}), actively integrating \texttt{Talk}Notes into their reasoning process while thinking aloud.
The variability in engagement patterns hints toward user preferences in how participants designed while thinking aloud. Identifying the rationale behind their preferences was not part of our initial hypotheses. These varied engagement patterns emerged from questions on how users would think aloud and work with PointAloud (RQ2). However, the identification of these patterns provides food for thought for further explorations into think-aloud computing and work process documentation.

\paragraph{\textbf{Varied Response Patterns to \texttt{Talk}Tips: From Ignoring to Guided Exploration}}
\texttt{Talk}Tips are brief, pointer-attached prompts delivered both during pauses and in response to users’ verbalizations.
Across both 2D/3D tasks, participants saw between 56 and 94 \texttt{Talk}Tips ($M=75.2, SD=12.8$; see Table~\ref{tab:think-aloud-stats}).
Response behavior varied widely—from no responses (e.g., P04) to sustained engagement (e.g., P06 with 24 responses)—and often shifted with task context.

For example, in the 3D model review, P12 read and replied to multiple prompts while scanning the main room.
When prompted about window glare (\texttt{Talk}Tip: \texttt{“How to handle window glare issues?”}), P12 immediately proposed actionable options: \textit{“You could think about external shutters, which will regulate temperature as well...”} (see Figure \ref{fig:timelines-examples} and optionally Appendix Figure \ref{fig:appendix_task3_P12}).
This pattern—brief system cue, situated inspection, concise verbal elaboration—recurred as P12 moved through adjacent areas of the plan.

An even more intensive workflow appeared with P06, who repeatedly used \texttt{Talk}Tips to steer their reasoning in rapid succession.
Around minute~7, after a short review of their \texttt{Talk}Notes (\textit{“looking back at everything I’ve said…”}), a prompt surfaced: \texttt{“How will you ensure clear caregiver sightlines?”}
P06 responded while pointer-gesturing over the plan, articulating a rationale for room layout and corridor openness.
Moments later, a safety prompt (\texttt{“Avoid storing strollers in pathways”}) led P06 to clarify the circulation zones verbally (\textit{“strollers to the corner, away from the main path”}), which the system captured as a new \texttt{Talk}Note summarizing the design intent as \textit{“Caregiver access prioritized; strollers redirected from pathways to corners.”} (see Figure \ref{fig:timelines-examples}, optionally Appendix Figure \ref{fig:appendix_task2_P06_2d}).
Here, the sequence of \texttt{Talk}Tips, pointer-driven think-aloud activity, and \texttt{Talk}Notes formed a tight loop of prompt→reflection→decision capture.

Taken together, we observed two characteristic workflows in response to \texttt{Talk}Tips: \textit{tip-driven elaborators} (\cdotcolor{blue!40}), who frequently replied and extended their thinking aloud, and \textit{documentation-only users}, who largely ignored system prompts.
A third pattern, \textit{heavy integrators} (\cdotcolor{green!40}+\cdotcolor{blue!40}), cycled between reviewing \texttt{Talk}Notes and answering \texttt{Talk}Tips to progress their design reasoning, using prompts to probe trade-offs and then consolidating decisions into new notes.

\paragraph{\textbf{Contrasting \texttt{Talk}Tip Engagement Across 2D Annotation and 3D Review Activities.}}  
Participants responded to more \texttt{Talk}Tips in the 2D annotation task activity (phase 2) than in the shorter 3D review task (phase 3) when looking at absolute counts ($M=2.8$ vs.\ $M=2.0$ responses per session).  
However, after normalizing for task duration, the rate of engagement was slightly higher in 3D ($M=0.39$ responses/min) compared to 2D ($M=0.27$ responses/min), though with a skewed distribution: in 2D most participants engaged a little, while in 3D a few engaged intensively but many did not respond at all (see also Appendix Figure \ref{fig:timelines-all}).  
Interview reflections provide a lens on this discrepancy: several participants reported feeling more attentive to \texttt{Talk}Tips in the 3D review task, since they were not simultaneously sketching and could devote more cognitive bandwidth to responding.  
As P02 explained, \textit{``Having to do the annotations [in 2D] distracted me from noticing those [TalkTip] bubbles. 
But in the 3D space, because the annotation was taken away completely, those things were more noticeable. 
So I was looking at the wall and then I saw the [TalkTip]} \texttt{'Hey, think about the structure. Load bearing.'} \textit{And I was like 'Oh, yeah. Let's think about the load bearing.' 
So that was really helpful [...] and only noticeable when I didn't have to worry about annotation.''} 

Together, these findings reveal a tension between observed behavior and perceived experience: while absolute logs show similar engagement in 2D annotation task and 3D review task, participants experienced \texttt{Talk}Tips as more salient in 3D without additional annotations.

\paragraph{\textbf{TalkReminders as Subtle but Ineffective Cues.}}  
\texttt{Talk}Reminders were designed to proactively resurface relevant prior notes: when a participant’s current verbalization related to earlier concerns, previously created \texttt{Talk}Notes briefly reappeared on the canvas, highlighted and summarized next to their anchors.  
Although such reminders appeared in the logs (between 26 and 48 times during participant sessions), our video analysis did not reveal participants visibly responding to them.
Participants either ignored or did not consciously register these resurfaced notes, suggesting that the reminders were too subtle to shape behavior or that participants noticed them but did not perceive them as actionable in the moment.
For example, P12 desired to have notes re-surfaced based on their current process: \textit{``I wanted to [...] go back to my notes [...] to a certain thought [...] especially when you're on a fast [...] iterative processes.''} However, they found browsing through \texttt{Talk}Notes to sidetrack their train-of-thought: \textit{``instead of, like, helping me find exactly what I was going for, I was just [...] met with all of these other issues and [Talk]notes.''} Despite a preference to have relevant \texttt{Talk}Notes re-surfaced, this testimonial indicates that \texttt{Talk}Reminders went unseen by P12.

\subsection{Users’ Perceived Benefits and Challenges For Working With PointAloud (RQ3)} \label{sec:findings-RQ3}

We clustered the perceived benefits and challenges for users working with PointAloud into three themes: \textit{Supporting Thinking Aloud}, \textit{Supporting Design Process}, and \textit{Supporting Human-AI Co-Creation}. 

\subsubsection{\textbf{Supporting Thinking Aloud}} \label{findings:supporting-thinking-aloud} \hfill

\textbf{Users appreciated thinking aloud for fostering a more reflective practice and how the system helped them stay in the flow.}
Overall, participants saw value in verbalizing their thoughts during the tasks, with several framing verbalization as a welcomed reflective design practice:
\textit{“I don’t usually […] talk out loud, [but] this kind of speaking [my thoughts] helps me to reflect on my own thinking”} (P03).
Many participants also reported that speaking while working would lead to a more organized design process, as P04 described:
\textit{“It forces you to organize your thoughts as you’re going through the space.”} (P04)

Participants also described that verbalization was experienced as focus-enhancing rather than intrusive and that real-time capture allowed them to keep moving and stay in the "flow" without stopping to take notes, with \texttt{Talk}Notes acting as lightweight “traces” of thinking, as P12 described: 
\textit{“[It was] capturing these little snippets of thought as I was pretty much […] free flowing. The fact that the notes are being taken automatically, it's a cool thing. 
That might […] lead you to focus more actually and not lose your train of thought, which is quite precious."}

\paragraph{\textbf{Users felt motivated by the system's ambient nature that created a sense of being "listened to.'}}
Participants highlighted how the unobtrusive ambient transcription display of the \texttt{Talk}Pointer and \texttt{Talk}Notes created a sense of being “listened to,” which motivated them to keep verbalizing, as P11 explained: 
\textit{“This tool really helps me to stay focused [...] it's transcribing what I was talking about [...] I don't feel like it was distracting [...] because I was treating it as some ambient activity [and] it's not occupying that much space of the screen, so you don't have to really pay attention to what is going on there. But I do have a sense, `oh, something is happening there.' And I know that this computer is, like, listening to me. So [...] it motivated me to keep talking.”}

Others compared the system to a conversational partner that helped articulate and surface ideas that might otherwise remain unspoken: 
\textit{“It makes sense because it’s like working with another person [...] you have to express [your thoughts] out loud [because otherwise] some of the ideas never come to the surface.”} (P10)

Participants also reflected on how the \texttt{Talk}Text feature influenced their verbalization.  
For several, seeing the live transcript next to the cursor prompted clearer articulation. 
Others also mentioned that having the live transcript feed next to their cursor would reduce context switching by not having to
\textit{“look on the other side of the screen to see the transcription.”} (P01)

\paragraph{\textbf{While \texttt{Talk}Text signaled verbalization activity, participants largely ignored it while speaking.}}

Overall, participants felt motivated by the system to think aloud, but many also  reported that while \texttt{Talk}Text signaled activity, they rarely paid attention to the transcribed words and activity visualization, as P04 reflected:  \textit{“It gave a cue that it was capturing what I was saying [...] but because it wasn’t readily identifiable [...] I kind of ignored it.”} 
For a few, \texttt{Talk}Text also introduced anxiety around accuracy and the risk of being misunderstood.  
Together, these reflections highlight how \texttt{Talk}Text shaped awareness of being recorded and encouraged clarity, while also revealing challenges around limited engagement with the displayed text and worries about potential transcription errors.

\paragraph{\textbf{Thinking aloud also introduced a learning curve and felt impractical for some.}}
At the same time, a few participants also described a learning curve and initial hesitations around thinking aloud. 
For example, P06 mentioned: 
\textit{“Thinking out loud is a little bit of a vulnerable thing to do [...] I was afraid a little bit that [it] would judge you [...] but then I saw [...] it’s beneficial to think out loud.”}
Lastly, P08 also doubted the practicality of thinking  outside the study setting: 
\textit{“Thinking aloud doesn’t really work in an office environment because you cannot [...] talk out loud [near] other people [and] if I’m working alone, I’m not gonna talk aloud [...] it takes energy [...] it wouldn’t be as practical and functional, but it can work in some settings.”}

\subsubsection{\textbf{Supporting Design Process}} \label{findings:supporting-design-process} \hfill 

\noindent\textbf{Users valued the tool's engaging support for externalizing, structuring and resurfacing their fleeting thoughts.}
Participants emphasized how PointAloud made their design process feel more engaging, deliberate and manageable. By externalizing fleeting thoughts, the tool helped structure what might otherwise remain chaotic, as P11 put it: 
\textit{“I've never imagined designing something this way [...] 
Because this makes [the] design process [...] much more engaging for me as a designer. 
Also [...] one struggle for me while doing design is [...] thinking about this complex problem from multiple dimensions. You sometimes [get lost] in your own thinking. 
And this tool is really helpful, in terms of documenting what is still a chaos in my head, but it's somehow jogging down everything in a very efficient way. 
And what is even cooler is that it helps you to kind of organize your thinking”} (P11).

The system was also described as a promising way to archive and resurface thoughts for later recall, as P05 explained:  
\textit{“It's like your thoughts are not going in the way [...] it's just putting them in a book or a file [...] that will help you when you keep thinking and designing, and you come back to this zone later.”} %

Many participants valued how the system captured their situational reasoning and made design intent accessible for later reference. 
In particular, \texttt{Talk}Notes were seen as more effective than having the raw text transcript, as P02 reflected: 
\textit{“I didn't really use the [text only] transcript that much. 
But [...] the talk notes were actually quite helpful. So I did forget a lot, [...] So recording that and going back to those talk notes was actually quite helpful to record what my sort of thought process was.”}

\paragraph{\textbf{Users saw benefits in linking \texttt{Talk}Notes to 2D/3D spatial contexts.}}
Participants highlighted the value of linking \texttt{Talk}Notes to specific elements in the floor plan or 3D model, which supported later recall and continuity.  
Many imagined \texttt{Talk}Notes as a new form of note-taking, combining verbalization with spatial, pointer-centric context, as P04 put it:  
\textit{“It is an interesting idea [...] verbalizing your thought process for the design and how it's highlighting the different areas.”}
Frequently, participants valued the novelty and convenience of automated transcription and cursor location to place notes directly in relevant spaces by \textit{“moving my cursor and just talking, it would add the notes to the area that I'm talking about.”} (P08)  
Linking notes with floor plans and 3D views preserved context, making it easier to retrieve details and make sense of past work:
\textit{"Another thing that is helpful for me is the visual connection between the floor plans, the 3D view and the notes. [...] Sometimes we remember we said something, but we don't remember the exact details of the context [...] And [it would be impossible documenting my process in detail how] this tool can provide."} (P11)

Interestingly, some participants found notes easier to interpret in 3D, where design discussions like wall changes were more clearly tied to spatial context than in 2D, as mentioned by P02:
\textit{"This is actually very helpful in 3D. Because, [...] it's a lot easier for me to understand or see these points and where they're related to in the 3D model compared to the 2D canvas."}

\paragraph{\textbf{Participants highlighted challenges of automatic pointer-attention alignment.}}
A recurring reported friction was that the pointer would not always represent where participants’ attention was directed, as P04 reflected:   
\textit{“The tricky part is [...] when I’m going through the thought process, I’m moving the mouse a lot as I am thinking [...] so it may be difficult to pinpoint my notes to where they’re relevant in that particular space.”}  
P05 also mentioned that their awareness of the pointer capture could feel distracting if misaligned, worrying it might create inaccurate \texttt{Talk}Notes:  
\textit{“Because I know it links what I say to where the mouse is, it sometimes distracts me when I maybe forget [to move] my mouse pointer to here [...] and I'm talking about another space. [This] would create confusion.”}
This recurring misalignment calls to attention one of the pitfalls of using pointer location data as a proxy for user attention, future work could consider ways to filter out irrelevant pointer location data based on the ongoing user context. For example, what the user is talking about should align with what they are pointing at.

\paragraph{\textbf{Users valued \texttt{Talk}Note grouping and category filtering for easily navigating their documented process.}}
Participants emphasized that the automatic grouping of \texttt{Talk}Notes into \texttt{Talk}Threads and their semantic classification into process categories supported clearer retrieval of their design process, as P06 reflected:
\textit{“I think something that helped me [was] that it classified [the TalkNotes] so I was able to check: 'What is the process that I went through?' [...] So it doesn't overwhelm you when you filter that. I usually like when things are sorted out [and] grouped together. I like the fact that it does it for you.”} (P06)
Categorization and filtering were also seen as valuable for navigating large sets of notes and directly revisiting specific moments:  
\textit{“I wanted to [...] check which was one of the problem areas that I kinda forgot [...] it was really nice to be able to click on 'problem' and see exactly what I needed.”} (P12)  
At the same time, participants expressed a desire for customization, noting that predefined labels might not always match their personal workflow, and also for the ability to search for \texttt{Talk}Notes and arrange them more flexibly outside of the list view.

\paragraph{\textbf{Participants saw great potential in \texttt{Talk}Notes for also supporting \underline{human–-human} collaboration.}}
Participants also frequently envisioned the system not only as a personal tool but also as a shared resource for collaborative design. 
They described how \texttt{Talk}Notes could allow teammates to follow each other’s reasoning without requiring extensive explanation.  
P10 expanded on how passing along notes with design intent was seen as particularly valuable:
\textit{“I like that there’s added data that you can pass along. So it’s not just your first pass on the layout design, but also your thoughts before making it. So it could provide some insights to the next person [...] That’s good information because sometimes [...] you don’t know who else designed this or what they were thinking of. That’s very helpful [...] like a preview [...] from another designer's perspective.”}

Participants also noted potential information boundaries for certain collaborations. 
While \texttt{Talk}Notes were valuable for internal teams, external consultants might require filtered or tailored views, as P02 elaborated:   
\textit{“Internally, it could work well. [But] working with separate consultants [...] might be a bit confusing. They might not want all of that information. So if there was an ability to maybe give cues and filter [...] that might be quite helpful."}

\subsubsection{\textbf{Supporting Human-AI Co-Creation}} \hfill \label{findings:supporting-human-AI-co-creation}

\noindent\textbf{\textit{From a ``thought recorder'' to co-creative AI support.}}
Participants described how PointAloud’s context-aware suggestions shifted the system from a passive recorder of verbalized thoughts to a virtual collaborator that could help externalize ideas and seed design moves. 
As P10 noted, \textit{``There are times where you actually need some feedback [...] it’s helpful for the design process [...] you are able to, like, bounce ideas out with someone else.''} 
Others emphasized how novel it felt for a design tool to actively engage, as P06 reflected: \textit{``I have never seen a design tool that [...] tries to understand what I’m saying, and that was really cool to watch, actually.''} 
Similarly, P05 highlighted the practical value of the \texttt{Talk}Notes' AI-generated Action Suggestions, observing that \textit{``You can see what the AI got from your speech and it gives you more suggestions [...] Maybe these are links for some furniture, similar projects, or some permit or code pages [...] That’s always so helpful.''}

\paragraph{\textbf{\texttt{Talk}Tips can aid designers' thinking without taking over.}}
When aligned with participants’ current focus, \texttt{Talk}Tips acted as lightweight thinking supports—\-providing nudges, references, and reminders that helped them sustain momentum without feeling overridden. 
For instance, P06 described, \textit{``the prompts were very impactful [...] it made me think about what I actually need to think about. Because I often deviate, but it kept putting me back on track. So that was really cool [...] and I felt like whenever I would stop talking, the prompt would come up. And that was something I liked.''}

\paragraph{\textbf{Overall mixed feelings about proactive system suggestions: value vs.\ distraction when misaligned.}}
While many valued the system's proactive nudges, a few participants also described distraction when timing, persistence, or placement did not match their current focus. 
P09 summarized this tension: \textit{``The prompts [were] distracting [...] because it kept on saying something that I was not thinking about at that point. But it might also be helpful to read because it was prompting me to think about the lighting quality in this space. So it is distracting, but at the same time, needed [...] I have mixed feelings about that.''} 
Some struggled to notice or process \texttt{Talk}Tip suggestions while working or found their transient presentation too brief, as P04 reflected, \textit{``I didn't really read it while I was working. My eyes did notice this bar piece that was showing up [...] but I didn't really comprehend the text that was being written.''} 
Participants also flagged cursor-adjacent pop-ups as interruptive, as 
P12 shared: \textit{"Talking out loud reinforces and is helpful [...] but not necessarily the notes popping by my cursor [...] I found that rather distracting [...] breaking my train of thought."}

\section{Discussion}

In the following sections, we reflect on the lessons from designing, implementing, and studying \textit{PointAloud}, focusing on how pointer-centric interactions can \textit{incentivize Think-Aloud Computing}, \textit{provide real-time feedback through pointer-ambient displays}, enable richer forms of \textit{process-aware human–AI co-creation}, and support \textit{design process documentation}. 
We also discuss how the PointAloud interaction suite contributes a transferable approach for other domains, such as writing in a text editor interface and visual data analysis via a computational notebook. 
Each subsection highlights challenges, design considerations, and open questions for advancing pointer-centric think-aloud computing into everyday workflows.

\subsection{Incentivizing Think-Aloud Computing}

Building on the original notion of Think-Aloud Computing~\cite{krosnick_thinkaloud_2021}, a central question of our work is whether interaction techniques such as PointAloud meaningfully incentivize people to verbalize their thoughts during design tasks.
Our study provides a nuanced answer.
Quantitative analysis showed no significant difference in words-per-minute (WPM) between PointAloud and the baseline live transcription condition (see Section~\ref{sec:findings-RQ1-O}).
This suggests that PointAloud did not uniformly increase the amount of verbalization.
However, because participants were explicitly instructed to think aloud in both conditions, our study design does not isolate whether PointAloud itself prompted additional articulation.
Future dedicated studies are needed to assess this effect better.

While the volume of speech did not change, qualitative findings indicate that PointAloud influenced how participants experienced and valued verbalization assisted by PointAloud (Section~\ref{sec:findings-RQ3}).
Many highlighted benefits such as a sense of being ``listened to,'' greater clarity of articulation, and the usefulness of externalizing fleeting thoughts for later reflection and recap.
In some cases, participants also engaged actively with \texttt{Talk}Tips in rapid succession, producing extended chains of verbalized reasoning that were immediately documented as \texttt{Talk}Notes (Section~\ref{sec:findings-RQ2}).
Thus, while overall speech quantity remained stable between conditions, PointAloud appeared to incentivize specific forms of verbalization. Feedback from participants indicates a change from a monologue to a dialogue with the system. The participants thought aloud, and the system responded via real-time documentation (\texttt{Talk}Notes) and AI-generated suggestions (\texttt{Talk}Note Actions, \texttt{Talk}Tips).

At the same time, thinking aloud also presented a novel practice for most participants.
Participants quickly adapted and saw value in thinking aloud, while some described initial hesitation, and some questioned its practicality in office settings where speaking aloud can feel socially inappropriate (Section~\ref{findings:supporting-thinking-aloud}).
This echoes challenges discussed in prior work on integrating think-aloud methods into everyday computing contexts~\cite{krosnick_thinkaloud_2021}, underscoring that adoption requires not only functional features but also attention to comfort and situational fit.

In other domains, such as writing, the need to incentivize users to think aloud may be greater, because verbally expressing thoughts while simultaneously writing text may be unfamiliar or cognitively demanding. 
However, other parts of the writing process may feel more natural, such as verbalizing fleeting ideas during the ideation phase of writing. Also, more problem-solving-oriented tasks, such as visual data analysis, might lend themselves to more frequent verbalization, particularly when analysts ask questions aloud about the data.

\textbf{Design Considerations.} First, incentivizing users to think aloud requires anchoring verbalizations to visible and useful outcomes, such as immediate feedback, design suggestions, or process documentation. 
Second, systems must make the benefits of speaking aloud transparent to users in the moment.
For example, PointAloud visibly demonstrates how the user's verbalizations (\texttt{Talk}Text and \texttt{Talk}Vis) translate into actionable \texttt{Talk}Notes or trigger relevant \texttt{Talk}Tips. 
Finally, designers should account for the learnability of think-aloud computing: it may require gradual onboarding, scaffolds that ease initial discomfort, and longitudinal use before its full value may be realized.

\textbf{Open Questions.} Future research should examine in which task contexts think-aloud computing most effectively encourages verbalization.
Without instructing participants to verbalize their thoughts, do they think aloud solely to gain the benefits of the system?
How does sustained use over weeks or months shift both the quantity and quality of articulation? 
Addressing these questions will be critical to understanding whether interaction suites like PointAloud can sustainably embed think-aloud practices into everyday design workflows.

\subsection{Designing Pointer-Adjacent Ambient Displays for Real-time Feedback}

Our findings show that pointer-adjacent  ambient displays, such as \texttt{Talk}Text and \texttt{Talk}Vis, provided valuable reassurance that speech was being captured without distracting from design tasks.
Survey results revealed no increase in perceived distraction compared to baseline transcription ($p = 0.887$, see Section~\ref{sec:findings-RQ1-Q}), suggesting that embedding capture indicators around the pointer offers a low-friction way to build awareness and trust.
However, participants often reported only glancing at these elements rather than continuously monitoring them, indicating that such feedback must strike a balance: frequent enough to confirm capture, but lightweight and glanceable to avoid shifting attention toward evaluating transcription quality.

Other ambient mechanisms were less effective.
\texttt{Talk}Reminders were rarely acted upon (Section~\ref{sec:findings-RQ2}), likely because their subtle resurfacing did not call users to action.
By contrast, \texttt{Talk}Tips were more salient and sometimes prompted rapid cycles of reflection and decision-making, though their effectiveness varied with task context.
Notably, participants experienced \texttt{Talk}Tips as more noticeable in 3D model review tasks than in 2D annotation, a difference they attributed to having more cognitive bandwidth when not simultaneously sketching.
This underscores how task load mediates the salience of pointer-adjacent interfaces.

Overall, these observations extend prior work on peripheral and ambient displays (e.g., \textit{Ambient Help}~\cite{matejka_ambient_2011a}, \textit{SidePoint}~\cite{liu_sidepoint_2013a}, \textit{Feed\-QUAC}~\cite{long_feedquac_2025}), which emphasize embedding contextual information into users’ ongoing workflows.
PointAloud contributes to this space by demonstrating how pointer-adjacent cues can reassure users about process capture while also serving as proactive verbalization and thinking supports.

The pointer-adjacent interface featured in PointAloud would need to be reconsidered for other user interface paradigms. For example, in a text editor interface, used in both writing and data analysis, the user's attention is digitally represented at differing moments by the pointer and the text cursor. Identifying which of these simultaneous locations best represents the user's attention is an open question. However, capturing the user's focus may be improved in other domains that introduce selection interaction paradigms. In writing, users could select text and speak aloud about a desired change. Also, when working in visual data analysis, users could directly select data points of interest in interactive charts. We hypothesize that selection, coupled with think-aloud verbalizations, could increase clarity on the user's underlying intent.

\textbf{Design Considerations.} Ambient pointer displays should remain subtle and glanceable, updating frequently enough to provide reassurance without overloading the user. Proactive cues must be tuned to task context and cognitive load, while secondary cues such as \texttt{Talk}Reminders require more careful design to make them more noticeable and actionable.

\textbf{Open Questions.} Future work should investigate how to adapt the salience of pointer-ambient feedback dynamically to different activity phases and workload levels.
What forms of lightweight error correction (e.g., voice-based commands) can help users manage transcription without leaving the flow of work? 
And how can such displays scale to richer system-suggested actions that operate on the design itself (e.g., auto-complete) without becoming overwhelming?

\subsection{Towards Process-Aware Human-AI Co-Creation}

Beyond documentation, a central promise of think-aloud computing is to enable more context-aware forms of human–AI co-creation.
By capturing users’ ongoing verbalized reasoning, PointAloud provides AI systems with semantically rich data about design intent that can be used to deliver more aligned and situated assistance.
Our study highlights this potential: participants valued \texttt{Talk}Tips when they connected directly to their current reasoning, and described the system as shifting from a passive recorder toward an active collaborator (see Section~\ref{sec:findings-RQ3}).
At the same time, when prompts were misaligned with the task at hand, participants experienced them as distracting (Section~\ref{sec:findings-RQ3}).
This underscores that effective AI support requires more than task-domain knowledge—it must also be attuned to where users are in their process.

PointAloud contributes here by coupling speech capture with additional process signals such as pointer traces and design context, which provide further cues about what users are attending to in the workspace.
These multimodal traces offer opportunities for anticipating user needs, for example, by tailoring suggestions based on what element is being inspected or which concern has just been verbalized.
Related research on workflow capture and mixed-initiative systems has long emphasized the importance of balancing proactive support with user control~\cite{horvitz_principles_1999a,fraser_discoveryspace_2016a}, and our findings extend this by showing how verbalization data can anchor system initiative in the \textit{evolving reasoning} of the user.

An opportunity for improving human-AI co-creation lies in using think-aloud data not only for semantic context but also for inferring underlying cognitive and metacognitive user states.
Prior work in learning analytics has shown that self-regulated learning phases—such as planning, enacting, and evaluating—can be detected from think-aloud protocols~\cite{bannert_process_2014,borchers_using_2024}.
Applied to think-aloud computing design workflows, similar approaches could allow systems to detect whether a user is in early ideation, evaluating trade-offs, or is unfamiliar with a certain software feature, and to adapt proactive support accordingly.
Complementary work also suggests that cursor movements can act as behavioral indicators of users' attention and cognitive states~\cite{diasdasilva_wandering_2020}, further pointing to how multimodal traces might enrich models of user state.

Think-aloud computing has the opportunity to enrich the user context for AI support systems. In writing, a tight editing loop could be powered by pointer-centric think-aloud interactions where the user selects text, verbalizes intended changes, and the LLM-driven support generates recommendations based on that context. Furthermore, think-aloud data attached to sections of writing could preserve the writer's intention. 
This metadata about the user's intent provides rationale and commentary that could enrich downstream LLM-recommended changes to the text.
In visual data analysis, visualization recommendation has improved based on LLM-supported pipelines driven by natural language question and answering~\cite{mackinlay_2007_show,wu_ai4vis_2022}.
However, introducing PointAloud concepts could move data analysis toward mixed-initiative systems where the analysis process is an ongoing iteration based on user intention. In this new model, AI agents could operate in the background based on the user's spoken hypotheses and appraisals.

\textbf{Design Considerations.} To be effective supporters, AI agents must be not only context-aware but also \textit{process-aware}, responding differently to exploratory reasoning versus evaluative reflection.
Capturing speech alongside pointer and interaction signals offers a foundation for such adaptive support, but systems must remain transparent about how inferences are drawn to sustain trust.

\textbf{Open Questions.} Future work should examine how reliably think-aloud transcripts and multimodal traces can be used to infer metacognitive states in real-world design contexts.
How should AI agents calibrate the timing and content of suggestions to different phases of work without becoming intrusive? What ethical implications arise when systems infer or attempt to influence users’ cognitive states? Exploring these questions will be critical for realizing the full potential of think-aloud computing in human–AI co-creation.

\subsection{PointAloud Interfaces as a Solution for Design Process Documentation}

A core contribution of PointAloud lies in demonstrating a new approach for documenting design processes.
Participants valued the system for helping externalize, structure, and archive their fleeting reasoning in ways that traditional live transcription did not (Section~\ref{sec:findings-RQ3}).
\texttt{Talk}Notes enabled participants to resurface earlier decisions, link ideas directly to spatial context in 2D and 3D, and retrieve information through grouping and category filtering.
These features made design processes more tangible and navigable, supporting reflection and recap while reducing the risk of losing important rationales.
In this way, PointAloud instantiates a transferable method for capturing tacit decision-making processes that are often left undocumented.

Beyond individual use, the findings also point to the potential of PointAloud as a collaborative documentation tool.
Participants imagined \texttt{Talk}Notes as artifacts that could be shared with colleagues to explain their reasoning without extensive verbal walkthroughs (Section~\ref{sec:findings-RQ3}).
By situating documentation directly in the design workspace, PointAloud opens pathways for process knowledge to travel across users, tasks, and applications.
This aligns with prior systems for workflow capture and retrieval~\cite{grossman_chronicle_2010,rasmussen_conotate_2019b,wang_callisto_2020}, and extends them by linking verbalized rationales with spatial pointer context.
Such an approach could connect across diverse tools (e.g., CAD models, project management systems, BIM environments) to create richer, cross-application process histories.

Taken together, these findings suggest that PointAloud contributes a novel way of documenting and sharing design processes, moving beyond capturing \textit{“what was done”} toward preserving \textit{“why it was done.”} 
This dual emphasis on action and rationale reflects a broader need in professional design practice to surface tacit knowledge and support continuity across time and collaborators.

The research field of visual data analysis has a rich history in tracking user interactions and visualization provenance~\cite{ragan_vda_provenance_2016, xu_survey_vis_provenance_2020}. Think-aloud capture introduces low-effort user rationale that can be attached to the existing provenance user interfaces and interactions in visual data analysis research. Additionally, the PointAloud AI-driven method for documenting work processes may be helpful in organizing and structuring streams of data analysis and resulting insights for future recall.

\textbf{Design Considerations.} Systems for process documentation should tightly couple reasoning with context—whether spatial, temporal, or semantic—to make captured knowledge useful both for individual reflection and for collaborative coordination.
Lightweight retrieval mechanisms, such as grouping and filtering, can reduce overwhelm and support targeted reuse.

\textbf{Open Questions.} Future work should also investigate alternative representations of captured processes, such as chronological timelines that complement spatial anchoring.
How might such documentation scale to complex, multi-user projects, and what forms of representation best support reuse across heterogeneous tools and workflows? Addressing these questions will help extend pointer-centric documentation into broader ecosystems of collaborative design practice.

\section{Conclusion}

Think-Aloud Computing offers the potential to capture rich contextual insights into users’ evolving intentions, struggles, and decision-making in real time. Yet, existing approaches face challenges: users often lack awareness of what is being captured, are not sufficiently encouraged to verbalize their thoughts, and may miss or be disrupted by system feedback. 
Moreover, thinking aloud must feel worthwhile by yielding meaningful assistance. 
To address these challenges, we introduced \textit{PointAloud}, a suite of AI-driven pointer-centric interactions designed for in-the-moment verbalization encouragement, low-distraction system feedback, and contextually rich process documentation alongside proactive AI assistance. We instantiated these techniques in the \textit{PointAloud System}, a CAD application for annotating 2D floor plans and inspecting 3D architectural models, allowing us to explore pointer-centric think-aloud computing in a concrete design context.
Our user study with 12 participants demonstrates how pointer-centric think-aloud support can facilitate documentation and enrich human–AI co-creation. Building on these findings, we outline design considerations for future pointer-centric and AI-supported Think-Aloud Computing workflows, including strategies for incentivizing verbalization, designing pointer-ambient displays, enabling more process-aware human–AI co-creation, and embedding documentation seamlessly within users’ ongoing workflows. While our study focused on architectural design, the PointAloud interaction suite illustrates a transferable design pattern that could inform other AI-assisted creative and knowledge work scenarios, offering HCI researchers and practitioners a novel interaction approach for integrating AI into in-the-moment user reasoning, documenting work processes, and enabling richer forms of human-AI co-creation.

\begin{acks}
We thank all study participants and the reviewers for their constructive feedback on the paper. 
The prototype was implemented with assistance from GitHub Co\-pilot, with all generated code reviewed, edited, and tested by the authors. 
Figures \ref{fig:teaser} and \ref{fig:system-ui} contain image elements (illustrations of people) that were generated with ChatGPT. 
Several figures in this paper include screenshots of 2D floor plans and corresponding 3D models sourced from Polycam community content, used under the Creative Commons Attribution 4.0 (CC BY 4.0) license.
\end{acks}



\appendix

\newpage
\onecolumn

\section{Additional Materials}

\begin{table*}[h]
\caption{Overview of study participants.}
\begin{tabular}{lllll}
\toprule
\textbf{ID}  & \textbf{Age} & \textbf{Gender} & \textbf{Role} & 
\begin{tabular}[c]{@{}l@{}}\textbf{Years of} \\ \textbf{Prof. Exp.}\end{tabular}\\
\midrule
P01 & 31  & Female & Interior designer                                   & 2                                     \\
P02 & 27  & Male   & Architect                                           & 6                             \\
P03 & 26  & Male   & Assistant Architect                                 & 6                                     \\
P04 & 49  & Male   & Faculty / Facilities Director / Licensed Architect & 19                            \\
P05 & 34  & Male   & Head of Architecture Design Department              & 11                                    \\
P06 & 23  & Female & Construction Management Intern / Architect          & 6                             \\
P07 & 30  & Female & Interior Designer                                   & 3                                     \\
P08 & 30  & Female & Architectural Designer                              & 4                                     \\
P09 & 25  & Female & Real Estate Intern / Architect                      & 7                                     \\
P10 & 29  & Female & Exhibition Design Intern                            & 5                                     \\
P11 & 30  & Female & Interior Designer                                   & 5                                     \\
P12 & 35  & Male   & Artist Studio Designer                              & 9                             \\                                       
\bottomrule
\end{tabular}
\label{tab:participants}
\Description{The table presents an overview of twelve study participants, detailing their ID, age, gender, professional role, and years of professional experience. Each row corresponds to a different participant, showcasing a variety of roles within the architecture and design fields, alongside demographic information. The data illustrates a range of ages and levels of experience in the industry.}
\end{table*}

\begin{table}[h]
\centering
\caption{Per-participant words per minute (WPM) across baseline (Task A) and \SystemName (Task B).}
\small
\begin{tabular}{lccc}
\toprule
PID & \begin{tabular}[c]{@{}l@{}} Task A WPM \\ (Baseline)\end{tabular} & \begin{tabular}[c]{@{}l@{}} Task B WPM \\ (PointAloud) \end{tabular} & Diff (B--A) \\
\midrule
P01 & 70.5 & 55.7 & -14.8 \\
P02 & 79.5 & 85.1 & +5.6 \\
P03 & 65.4 & 76.3 & +10.9 \\
P04 & 110.8 & 82.4 & -28.4 \\
P05 & 74.9 & 70.3 & -4.6 \\
P06 & 100.2 & 96.1 & -4.1 \\
P07 & 73.7 & 105.4 & +31.7 \\
P08 & 49.5 & 65.1 & +15.6 \\
P09 & 34.6 & 40.8 & +6.2 \\
P10 & 61.9 & 82.6 & +20.7 \\
P11 & 102.7 & 81.9 & -20.8 \\
P12 & 87.1 & 83.9 & -3.2 \\
\midrule
Mean Diff & \multicolumn{3}{c}{+1.23 (SD = 17.46)} \\
\bottomrule
\end{tabular}
\Description{The table presents per-participant words per minute (WPM) data for two tasks: Task A (Baseline) and Task B (PointAloud). It includes participant identifiers (PID) alongside their respective WPM scores for each task, as well as the difference in WPM between Task B and Task A. A mean difference and standard deviation for the WPM differences are also provided at the bottom of the table.}
\label{tab:wpm}
\end{table}

\begin{figure}[H]
\centering
\includegraphics[width=1\linewidth]{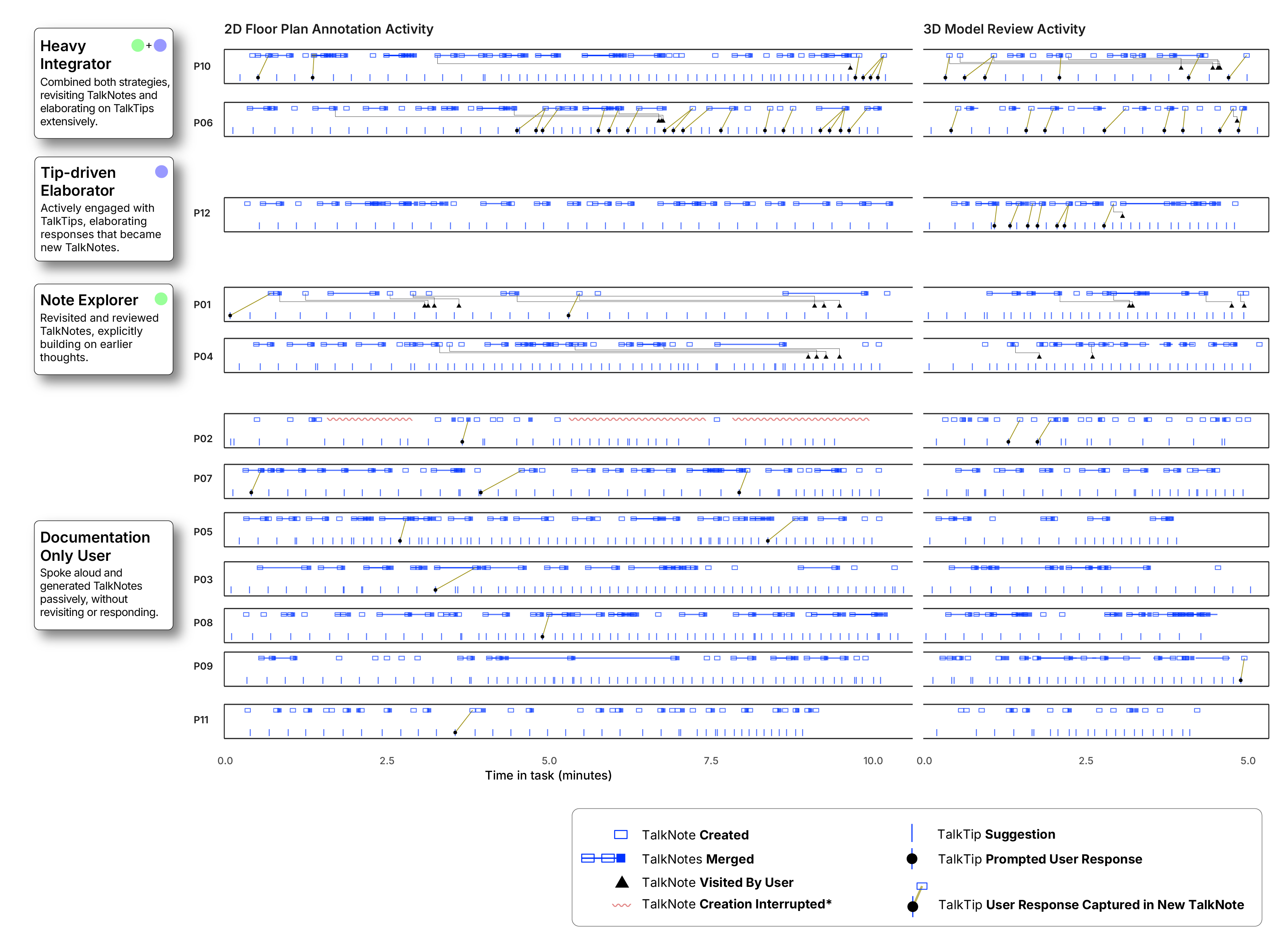}
\caption{Timeline visualizations of interaction events of all user sessions, clustered into four engagement patterns with TalkNotes and TalkTips during the 2D floor plan annotation and 3D model review activities: \textit{Note Explorer}, \textit{Tip-driven Elaborator}, \textit{Heavy Integrator}, and \textit{Documentation-only User}. \textit{*Note: In P02’s first activity, TalkNote creation was periodically interrupted by a browser plug-in conflict.}
  }
\label{fig:timelines-all}
\Description{The figure displays a timeline visualization of user interactions across two activities: 2D floor plan annotation and 3D model review. Each user's engagement is categorized into four patterns—Note Explorer, Tip-driven Elaborator, Heavy Integrator, and Documentation Only User—illustrated through varying symbols that denote different types of actions taken during the tasks. Interaction events are plotted along a timeline, showcasing how users engaged with TalkNotes and TalkTips throughout the sessions.}
\end{figure}

\begin{figure}[H]
\centering
\includegraphics[width=1\linewidth]{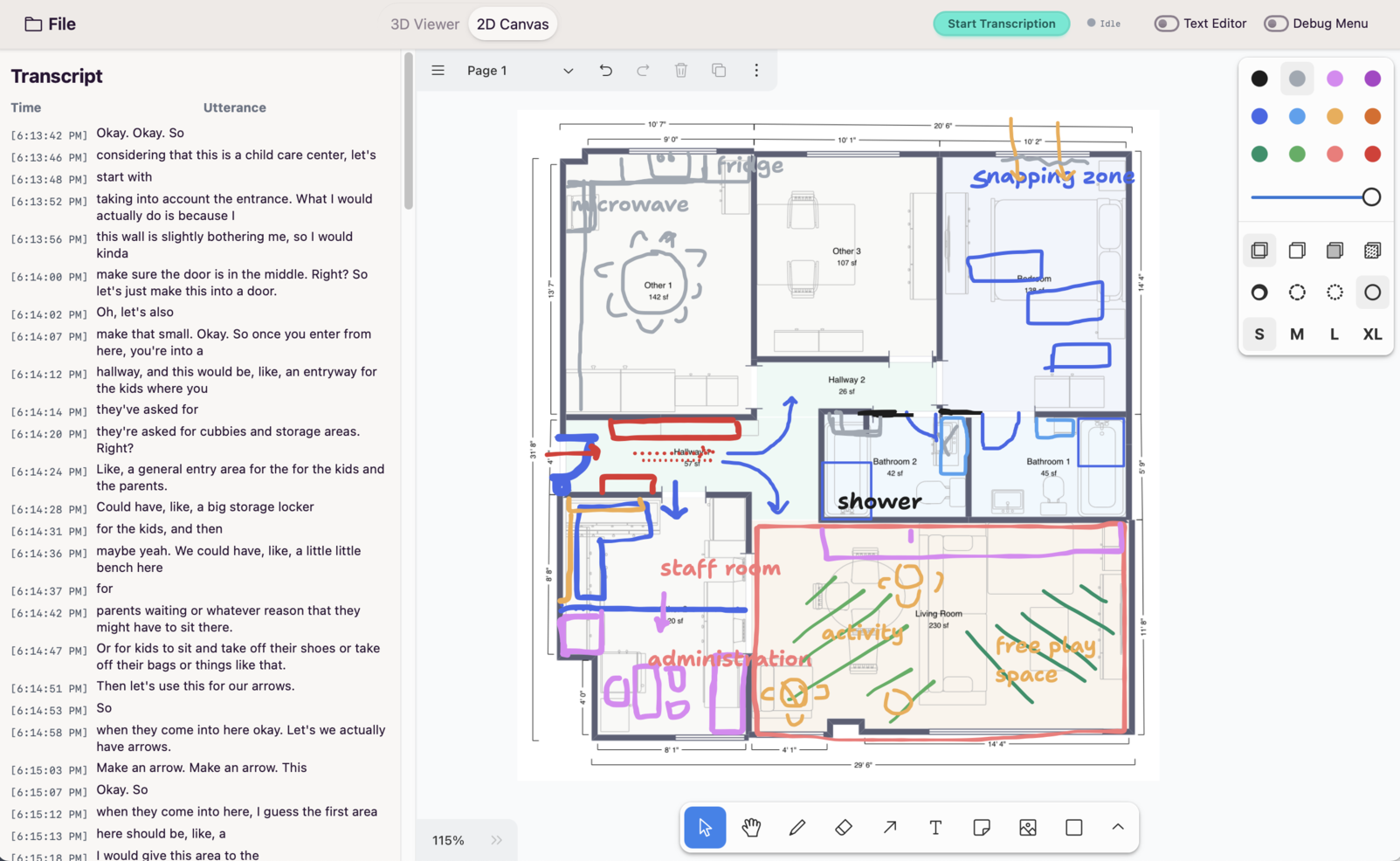}
\caption{Screenshot of the PointAloud system with the text-based live transcription in the left side panel serving as the baseline used by participants in the comparative tasks (phase 2). }
\Description{The figure shows a screenshot of the PointAloud system interface, featuring a floor plan of a child care layout with various labeled areas such as "staff room," "activity," and "free play space." Colors and arrows indicate design suggestions and movement paths discussed during a transcription session displayed in the left panel. Annotations in different colors provide context for the layout and proposed changes.}
\label{fig:text-based-screenshot}
\end{figure}

\subsection{Implementation Details} \label{sec:appendix-implementation-details}

\subsubsection{\textbf{Speech Transcription and Language Models}} 

User speech is captured via the browser’s microphone input and streamed to a commercial transcription service (\textit{Deepgram Nova-3}). 
Partial incoming transcription chunks are displayed live through \texttt{Talk}Text, while finalized chunks are sent to the back end for processing (see Section \ref{sec:implementation-semantic-chunking}). 
GPT-4o is used for: determining semantic chunking of incoming transcripts, \texttt{Talk}Note-related processing, and \texttt{Talk}Tip prompts.
Gemini 2.5 Pro is used for visually linking \texttt{Talk}Notes to scene elements.

\subsubsection{\textbf{Semantic Chunking and \texttt{Talk}Note Processing}} \label{sec:implementation-semantic-chunking}

Transcribed speech is segmented into \texttt{Talk}Notes through a semantic chunking pipeline powered by GPT-4o. Transcript fragments are buffered, and new fragments are evaluated in context to determine whether they continue the current topic or begin a new topic. Splits are avoided for filler words, minor topic shifts, or short clarifications, and triggered when a new idea, problem, or task is introduced.  

When a split is detected—or when the user pauses for more than eight seconds—the buffered text is promoted to a new \texttt{Talk}Note. Each \texttt{Talk}Note is then processed asynchronously with individual LLM prompts: it is summarized, assigned to one or more process labels, augmented with action suggestions, and checked for potential merging with the preceding note if the user resumes the same line of thought. In parallel, dynamic clustering routines group related \texttt{Talk}Notes into \texttt{Talk}Threads, enabling users to revisit connected reasoning across time.

\newpage

\subsubsection{\textbf{Multimodal Context Capture}} 

Beyond transcription, Point\-Aloud captures contextual signals from the design interaction. 
Cursor traces are logged and attached to the \texttt{Talk}\-Note as temporal-spatial metadata. 
For every \texttt{Talk}Note, the system reconstructs a visual overlay by rendering the relevant canvas screenshot and superimposing circular markers at the recorded cursor locations corresponding to each spoken fragment.
This composite image, along with a textual timeline of utterances, pointer coordinates, and the design scene, is then provided as input to the LLM (Gemini Pro 2.5) for identifying referenced regions in the design scene and highlighting specific floorplan areas. 
This multimodal capture ensures that \texttt{Talk}\-Notes can later re-situate users in the original design context by visually highlighting both pointer movement and referenced elements.

\subsubsection{\textbf{\texttt{Talk}Tip and \texttt{Talk}Note Suggestion Mechanisms}}

The system periodically generates \texttt{Talk}Tip candidate suggestions in the background by prompting GPT-4o with the current transcript and design brief, requesting concise, actionable tips in three categories: \textit{potential issue}, \textit{new idea}, and \textit{probing question}. 
These suggestions are stored and, at regular intervals, the system queries GPT-4o whether any tip is sufficiently relevant to interrupt the user. 

To support resurfacing contextually similar \texttt{Talk}Notes, the system uses the current transcript, previous \texttt{Talk}Notes, and the design brief to query GPT-4o to return the IDs of previously created \texttt{Talk}Notes that are highly relevant to the user's present focus. 
These notes are then visually highlighted in the interface, enabling users to quickly revisit and build upon earlier reasoning or decisions.

\subsection{PointAloud Annotation Task Outcome } \label{apdx:task-outcomes}
This section contains screenshots of the PointAloud system taken at the end of each participant's annotation task from our user study (see study phases 2 and 3 in Section \ref{sec:study-procedure}).

\begin{figure}[H]
  \centering
  \begin{subfigure}{0.99\linewidth}
    \centering
    \includegraphics[width=\linewidth]{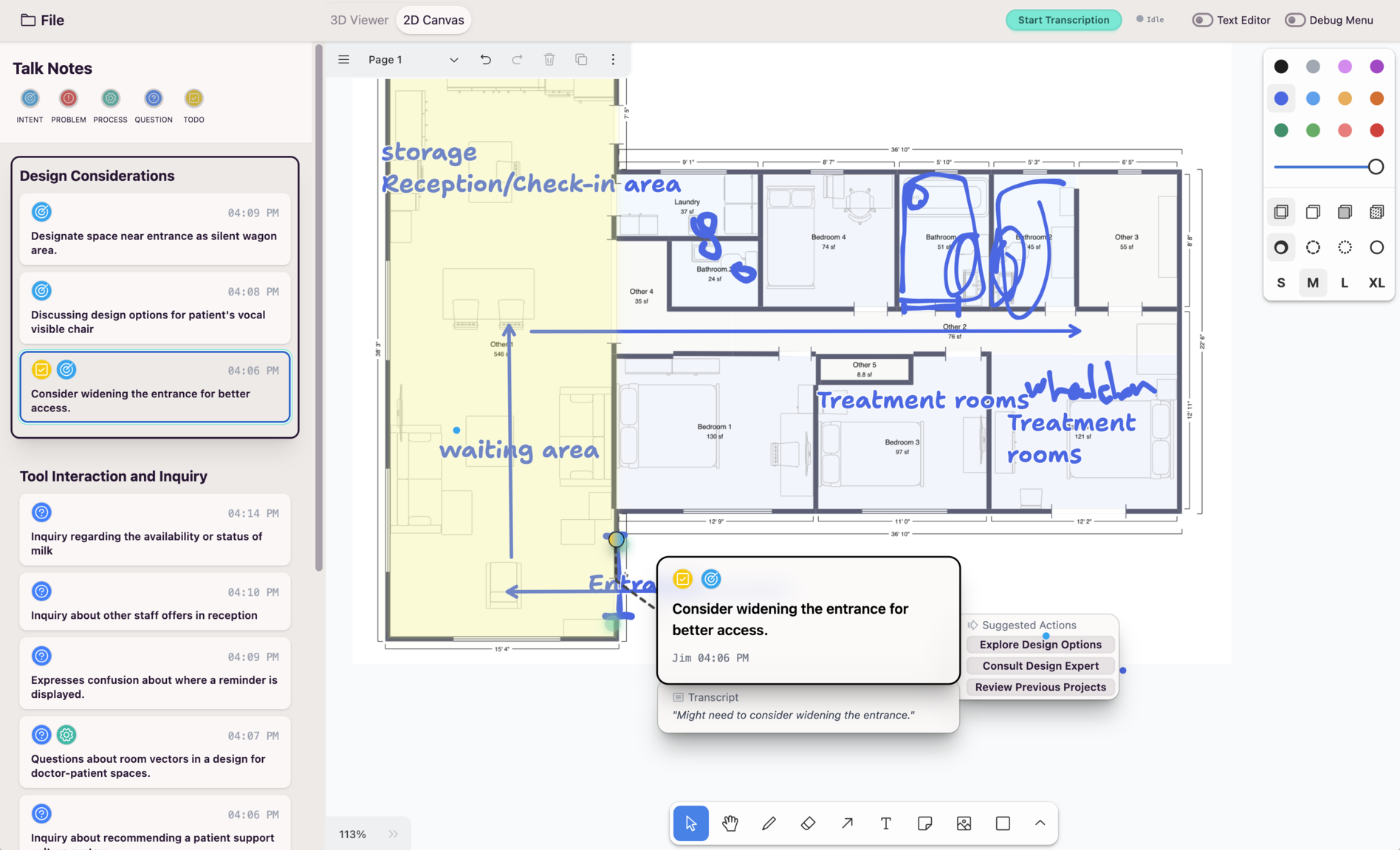}
    \caption{P01 – 2D Annotation Task}
    \label{fig:appendix_task2_P01_2d}
  \end{subfigure}
  \vspace{1em} %
  \begin{subfigure}{0.99\linewidth}
    \centering
    \includegraphics[width=\linewidth]{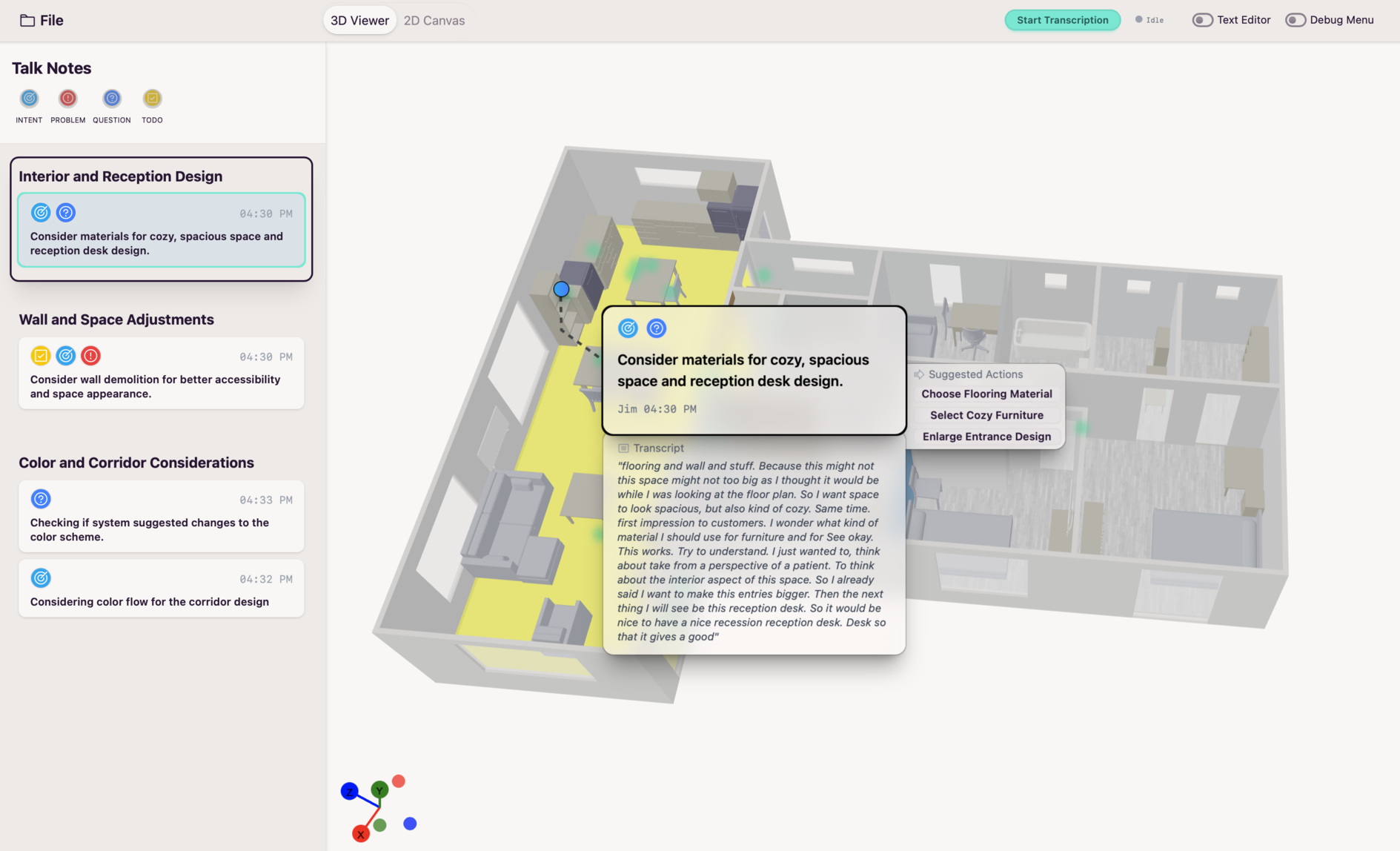}
    \caption{P01 – 3D Review Task}
    \label{fig:appendix_task3_P01}
  \end{subfigure}

  \caption{PointAloud annotation task outcomes P01.}
  \Description{This figure shows two screenshots of the PointAloud system. 
The top subfigure presents a 2D floor plan annotated with colored sketches and linked TalkNotes displayed in the left panel. 
The bottom subfigure presents a 3D model view of the same floor plan, with TalkNotes also visible in the left panel and annotations anchored within the 3D scene. 
Both subfigures illustrate how participants engaged in the annotation and review tasks using PointAloud.}

\end{figure}

\begin{figure}[H]
  \centering
  \begin{subfigure}{0.99\linewidth}
    \centering
    \includegraphics[width=\linewidth]{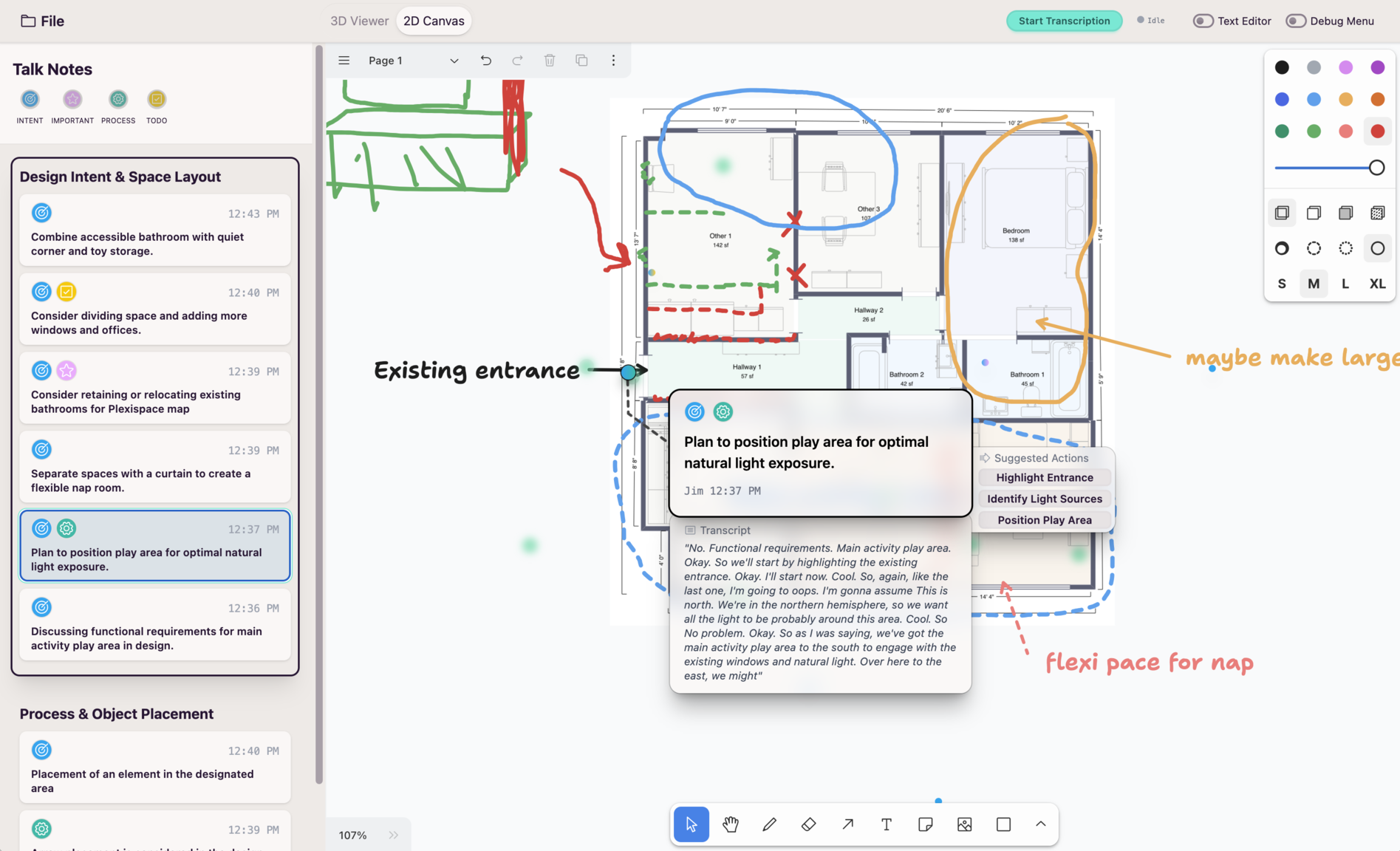}
    \caption{P02 – 2D Annotation Task}
    \label{fig:appendix_task2_P02_2d}
  \end{subfigure}
  \vspace{1em} %
  \begin{subfigure}{0.99\linewidth}
    \centering
    \includegraphics[width=\linewidth]{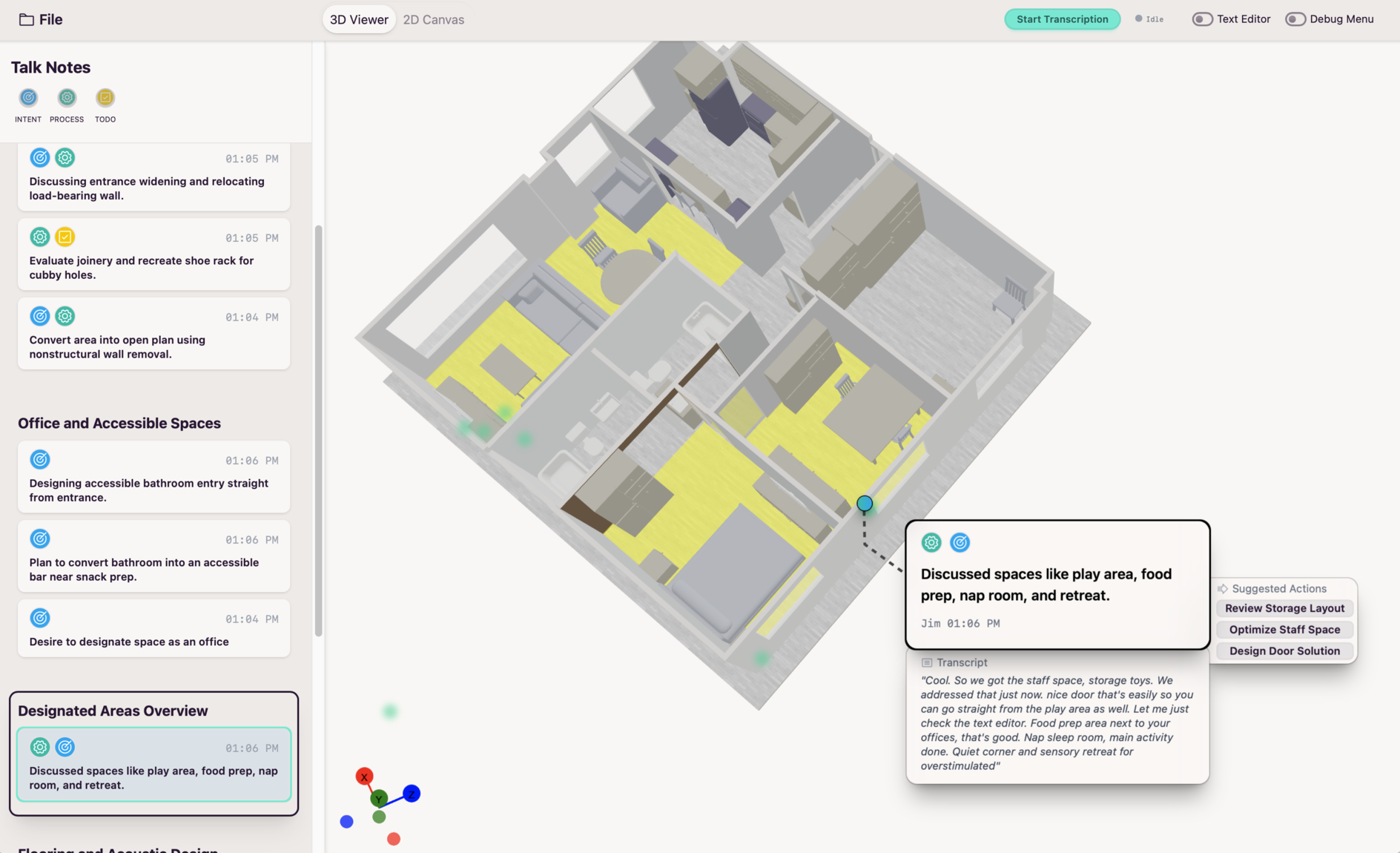}
    \caption{P02 – 3D Review Task}
    \label{fig:appendix_task3_P02}
  \end{subfigure}

  \caption{PointAloud annotation task outcomes P02.}
  \Description{This figure shows two screenshots of the PointAloud system. 
The top subfigure presents a 2D floor plan annotated with colored sketches and linked TalkNotes displayed in the left panel. 
The bottom subfigure presents a 3D model view of the same floor plan, with TalkNotes also visible in the left panel and annotations anchored within the 3D scene. 
Both subfigures illustrate how participants engaged in the annotation and review tasks using PointAloud.}
\end{figure}

\begin{figure}[H]
  \centering
  \begin{subfigure}{0.99\linewidth}
    \centering
    \includegraphics[width=\linewidth]{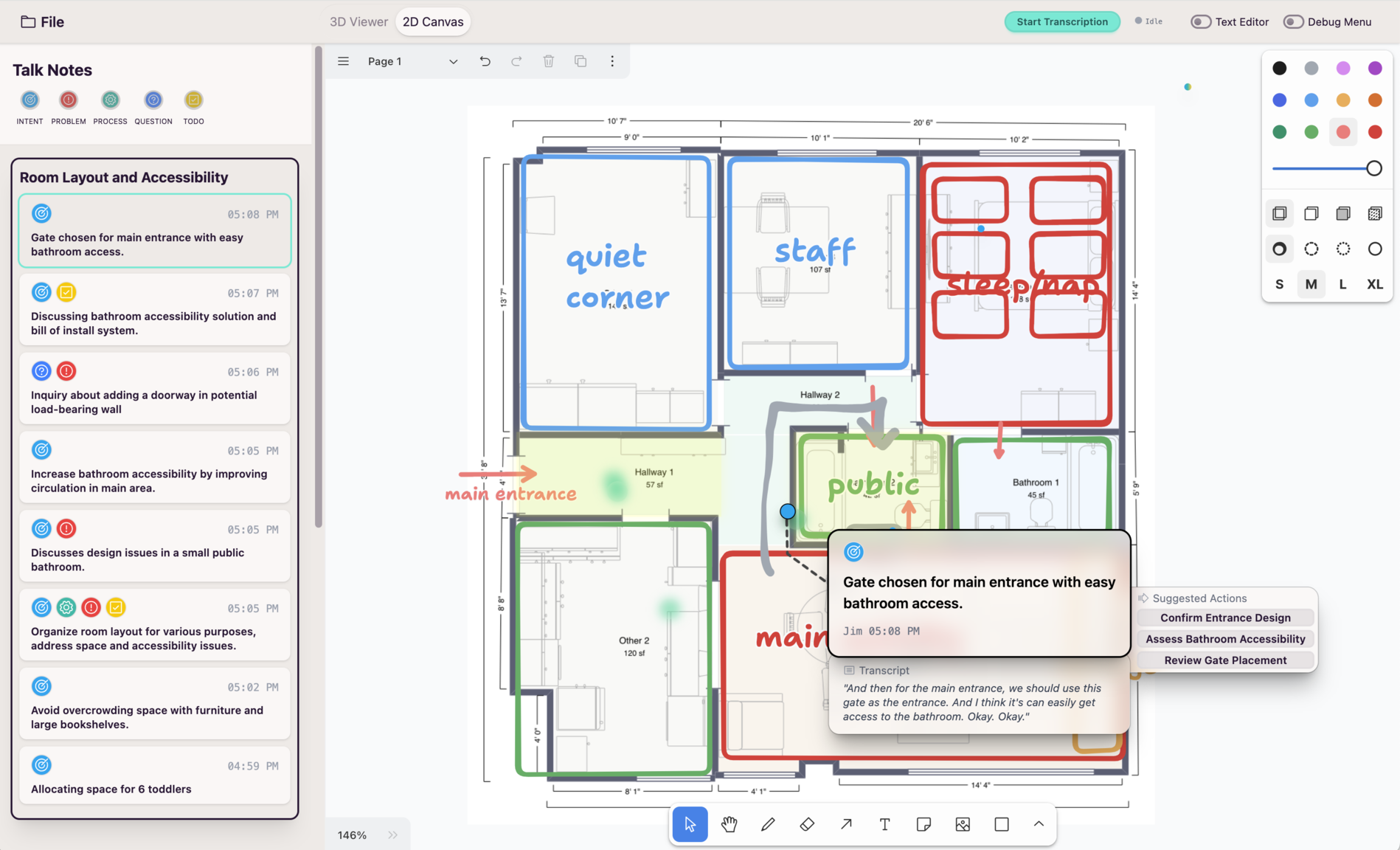}
    \caption{P03 – 2D Annotation Task}
    \label{fig:appendix_task2_P03_2d}
  \end{subfigure}
  \vspace{1em}
  \begin{subfigure}{0.99\linewidth}
    \centering
    \includegraphics[width=\linewidth]{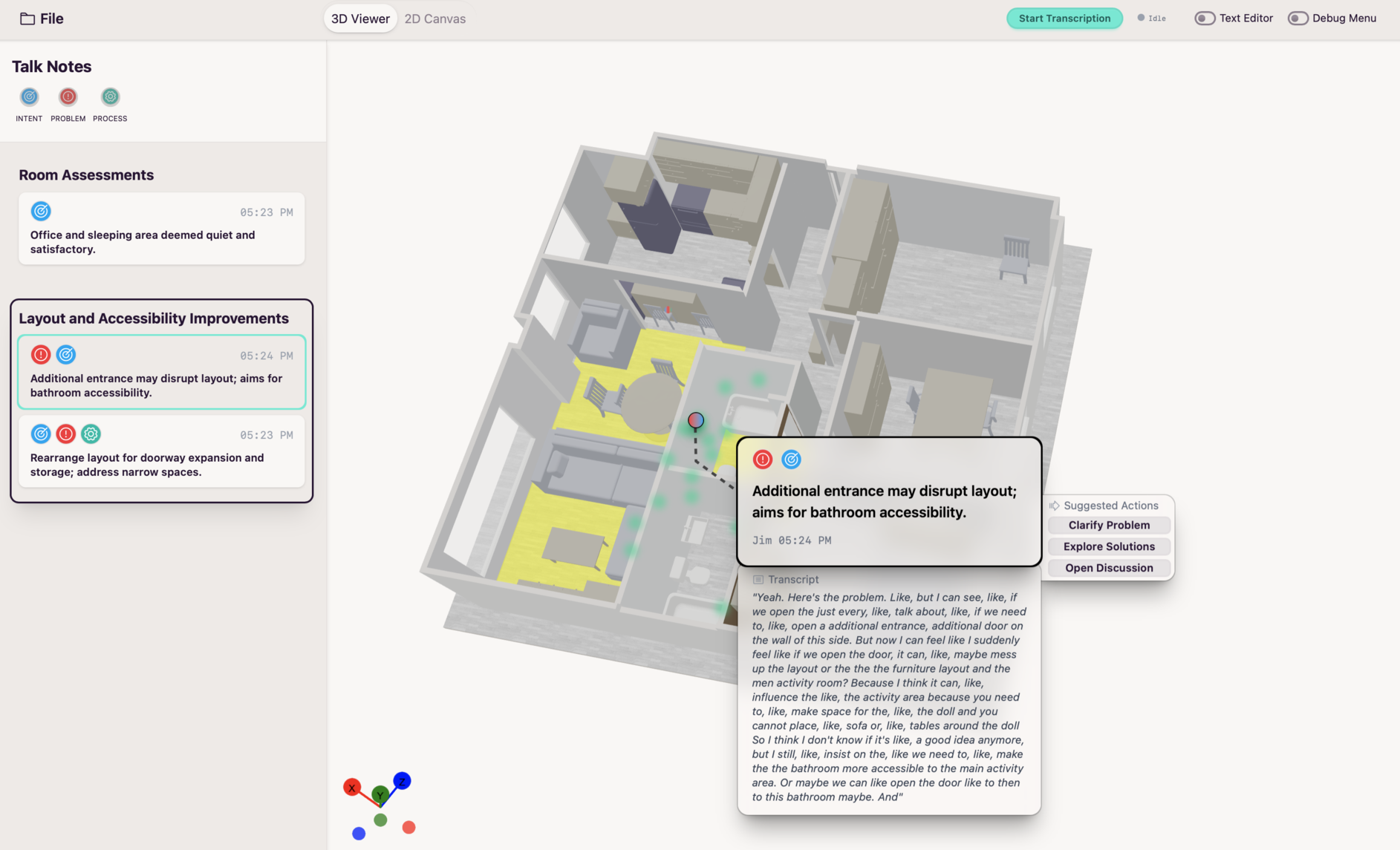}
    \caption{P03 – 3D Review Task}
    \label{fig:appendix_task3_P03}
  \end{subfigure}
  \caption{PointAloud annotation task outcomes P03.}
  \Description{This figure shows two screenshots of the PointAloud system. 
The top subfigure presents a 2D floor plan annotated with colored sketches and linked TalkNotes displayed in the left panel. 
The bottom subfigure presents a 3D model view of the same floor plan, with TalkNotes also visible in the left panel and annotations anchored within the 3D scene. 
Both subfigures illustrate how participants engaged in the annotation and review tasks using PointAloud.}
\end{figure}

\begin{figure}[H]
  \centering
  \begin{subfigure}{0.99\linewidth}
    \centering
    \includegraphics[width=\linewidth]{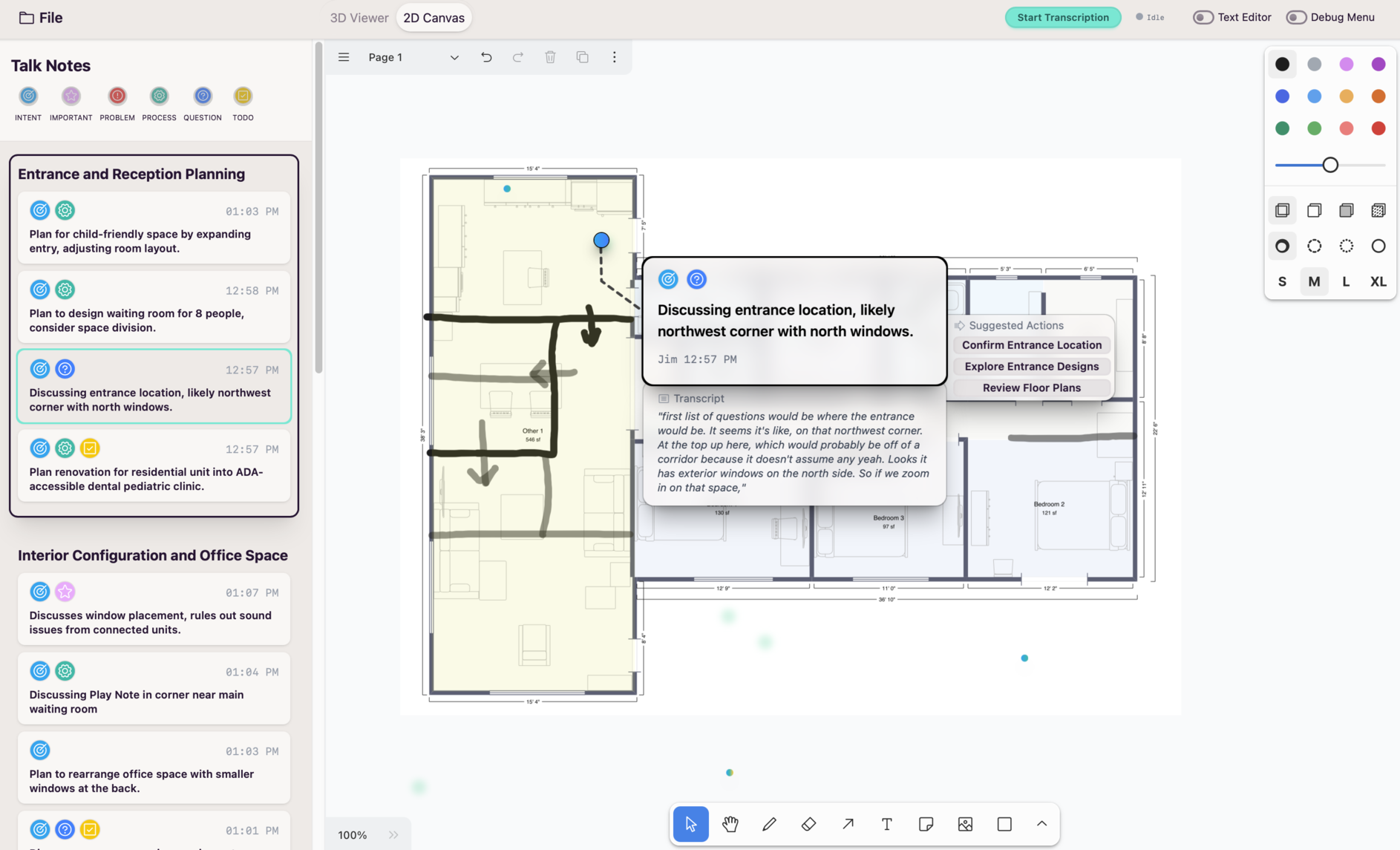}
    \caption{P04 – 2D Annotation Task}
    \label{fig:appendix_task2_P04_2d}
  \end{subfigure}
  \vspace{1em}
  \begin{subfigure}{0.99\linewidth}
    \centering
    \includegraphics[width=\linewidth]{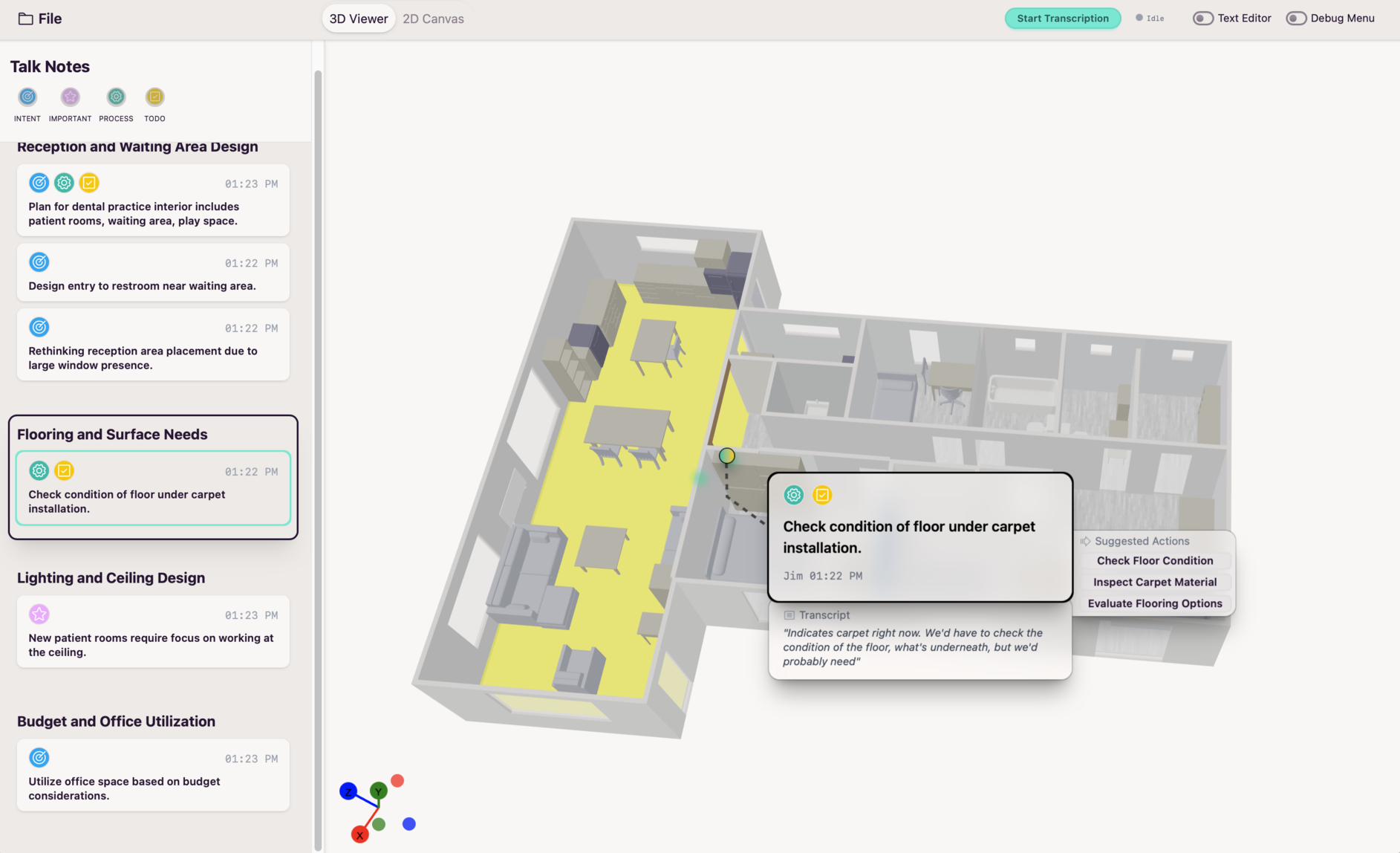}
    \caption{P04 – 3D Review Task}
    \label{fig:appendix_task3_P04}
  \end{subfigure}
  \caption{PointAloud annotation task outcomes P04.}
  \Description{This figure shows two screenshots of the PointAloud system. 
The top subfigure presents a 2D floor plan annotated with colored sketches and linked TalkNotes displayed in the left panel. 
The bottom subfigure presents a 3D model view of the same floor plan, with TalkNotes also visible in the left panel and annotations anchored within the 3D scene. 
Both subfigures illustrate how participants engaged in the annotation and review tasks using PointAloud.}
\end{figure}

\begin{figure}[H]
  \centering
  \begin{subfigure}{0.99\linewidth}
    \centering
    \includegraphics[width=\linewidth]{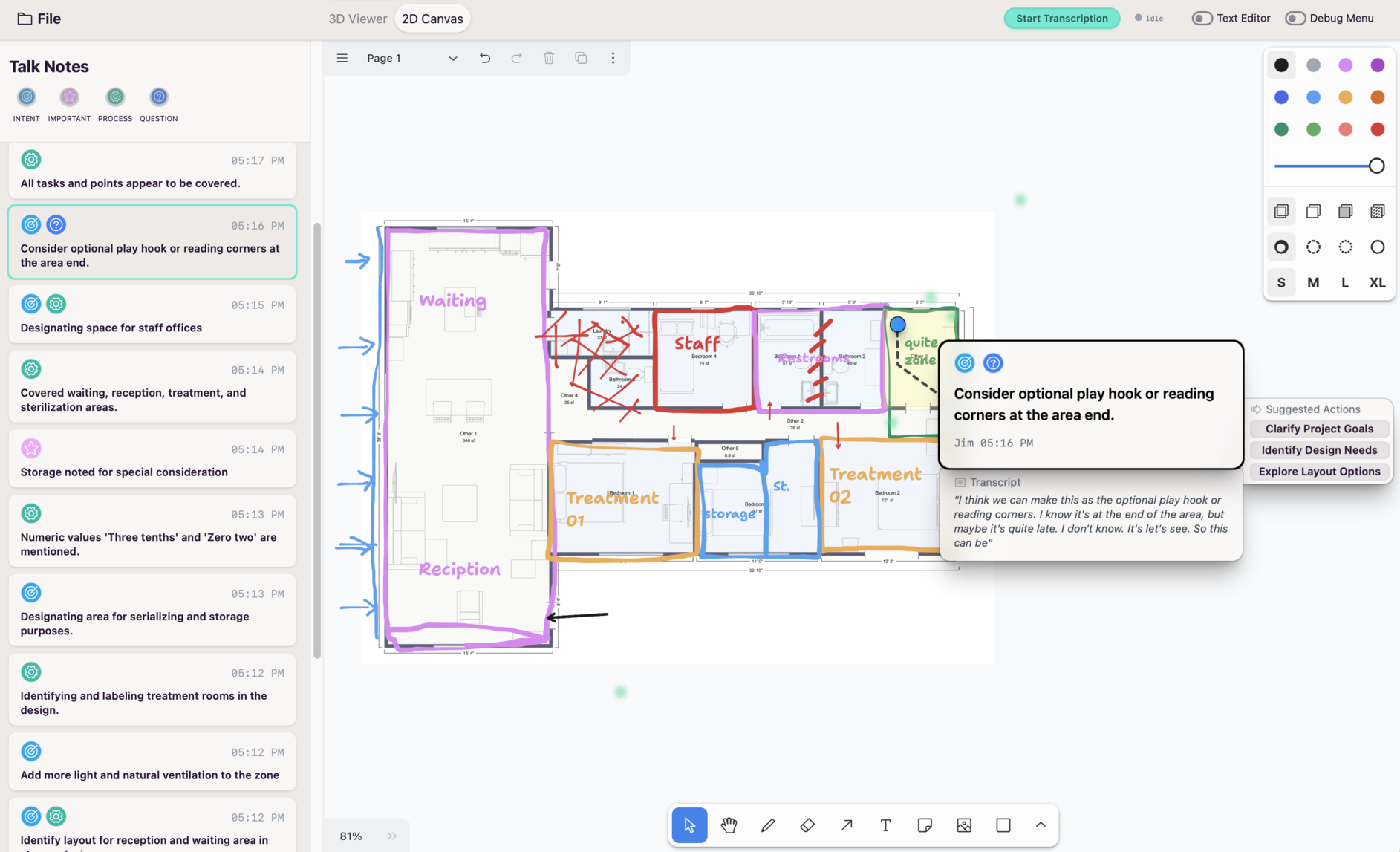}
    \caption{P05 – 2D Annotation Task}
    \label{fig:appendix_task2_P05_2d}
  \end{subfigure}
  \vspace{1em}
  \begin{subfigure}{0.99\linewidth}
    \centering
    \includegraphics[width=\linewidth]{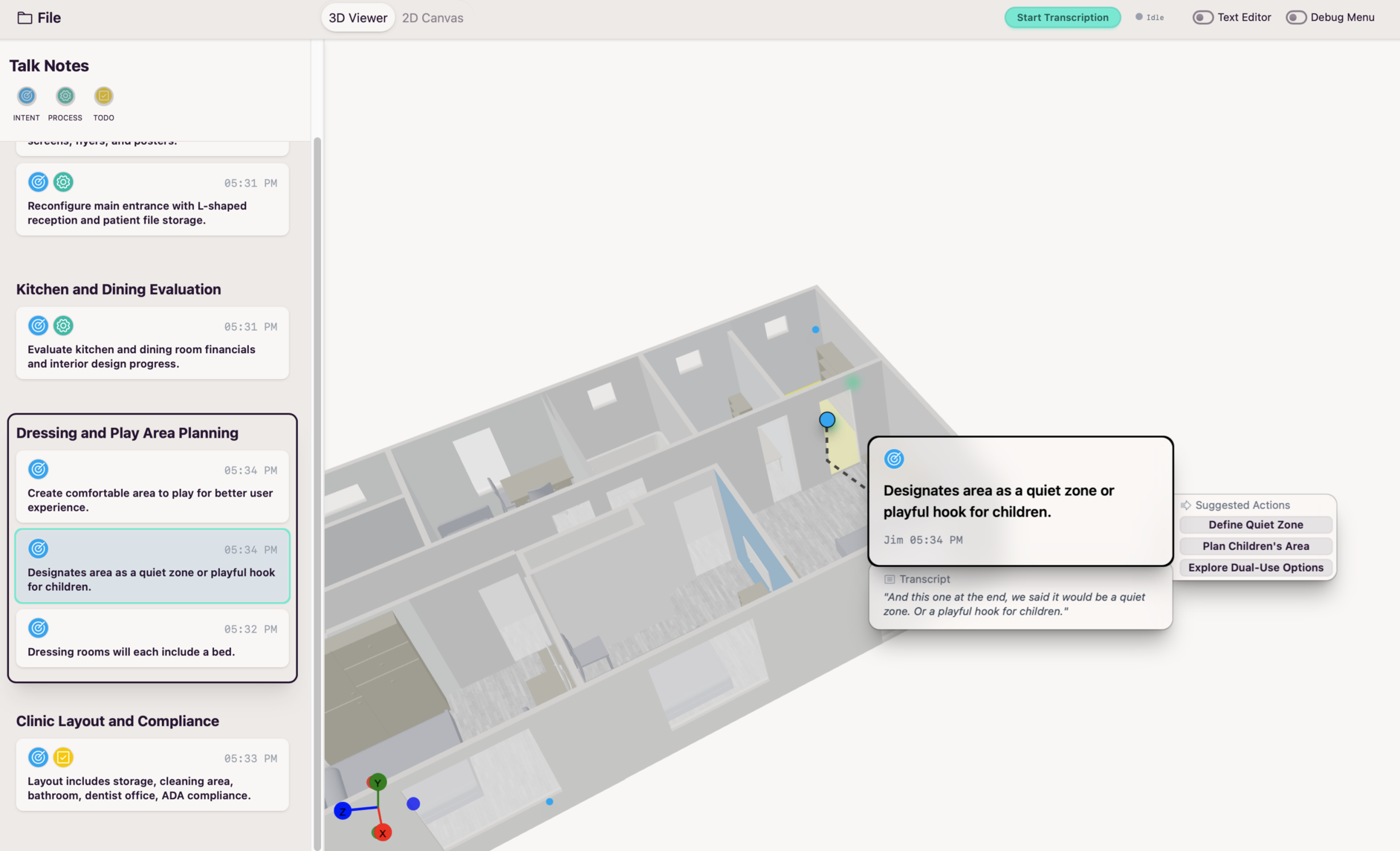}
    \caption{P05 – 3D Review Task}
    \label{fig:appendix_task3_P05}
  \end{subfigure}
  \caption{PointAloud annotation task outcomes P05.}
  \Description{This figure shows two screenshots of the PointAloud system. 
The top subfigure presents a 2D floor plan annotated with colored sketches and linked TalkNotes displayed in the left panel. 
The bottom subfigure presents a 3D model view of the same floor plan, with TalkNotes also visible in the left panel and annotations anchored within the 3D scene. 
Both subfigures illustrate how participants engaged in the annotation and review tasks using PointAloud.}
\end{figure}

\begin{figure}[H]
  \centering
  \begin{subfigure}{0.99\linewidth}
    \centering
    \includegraphics[width=\linewidth]{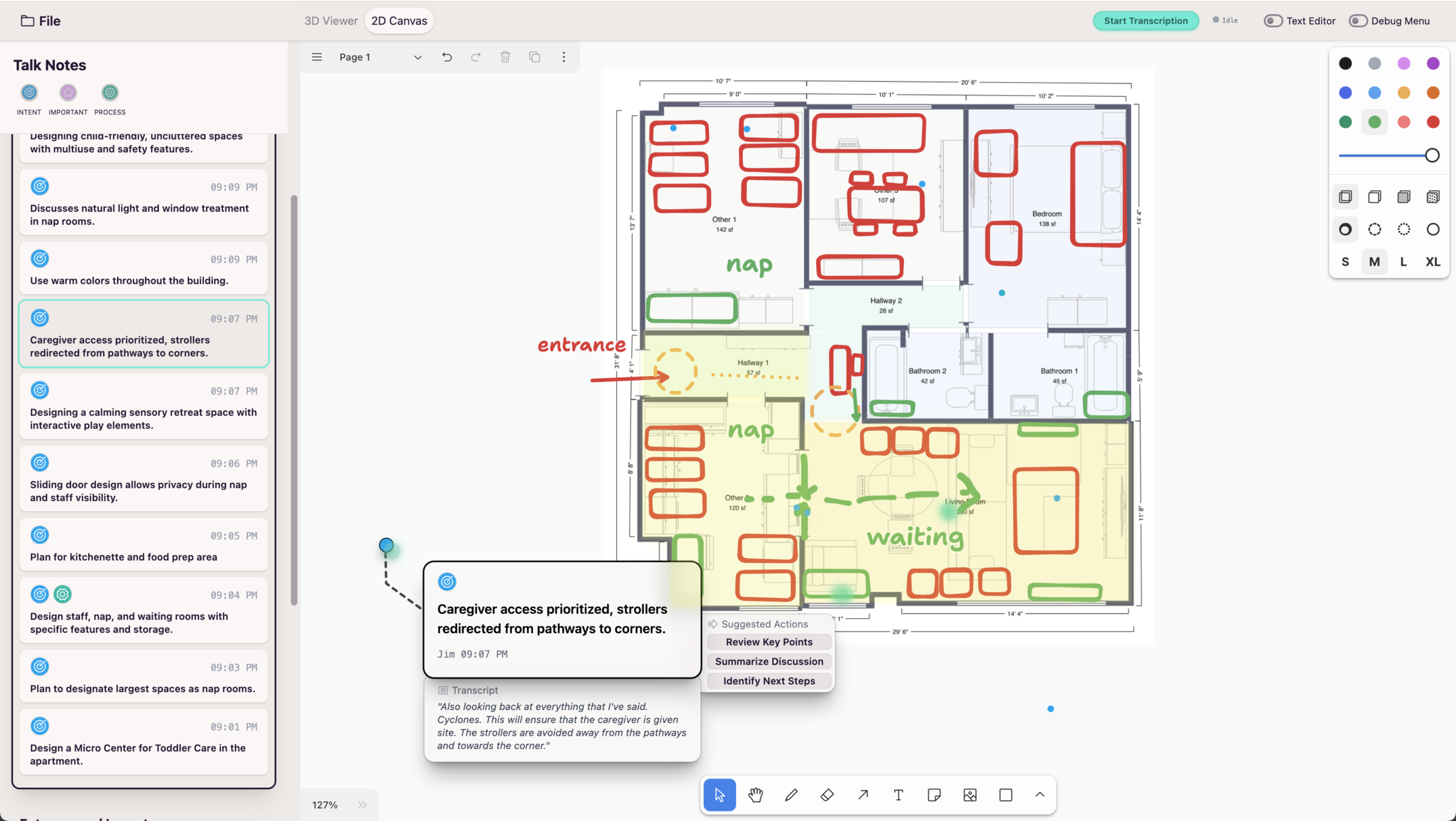}
    \caption{P06 – 2D Annotation Task}
    \label{fig:appendix_task2_P06_2d}
  \end{subfigure}
  \vspace{1em}
  \begin{subfigure}{0.99\linewidth}
    \centering
    \includegraphics[width=\linewidth]{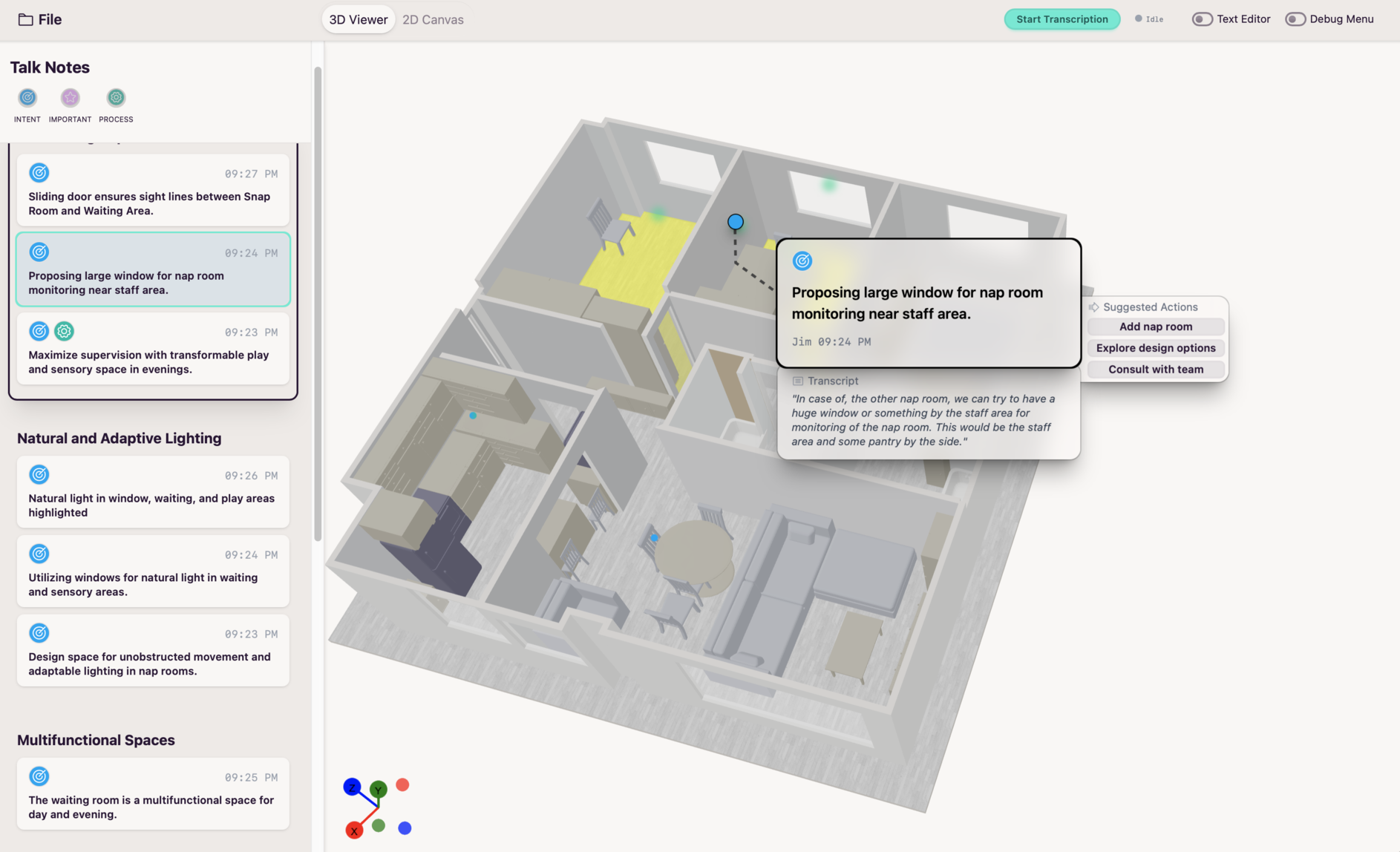}
    \caption{P06 – 3D Review Task}
    \label{fig:appendix_task3_P06}
  \end{subfigure}
  \caption{PointAloud annotation task outcomes P06.}
  \Description{This figure shows two screenshots of the PointAloud system. 
The top subfigure presents a 2D floor plan annotated with colored sketches and linked TalkNotes displayed in the left panel. 
The bottom subfigure presents a 3D model view of the same floor plan, with TalkNotes also visible in the left panel and annotations anchored within the 3D scene. 
Both subfigures illustrate how participants engaged in the annotation and review tasks using PointAloud.}
\end{figure}

\begin{figure}[H]
  \centering
  \begin{subfigure}{0.99\linewidth}
    \centering
    \includegraphics[width=\linewidth]{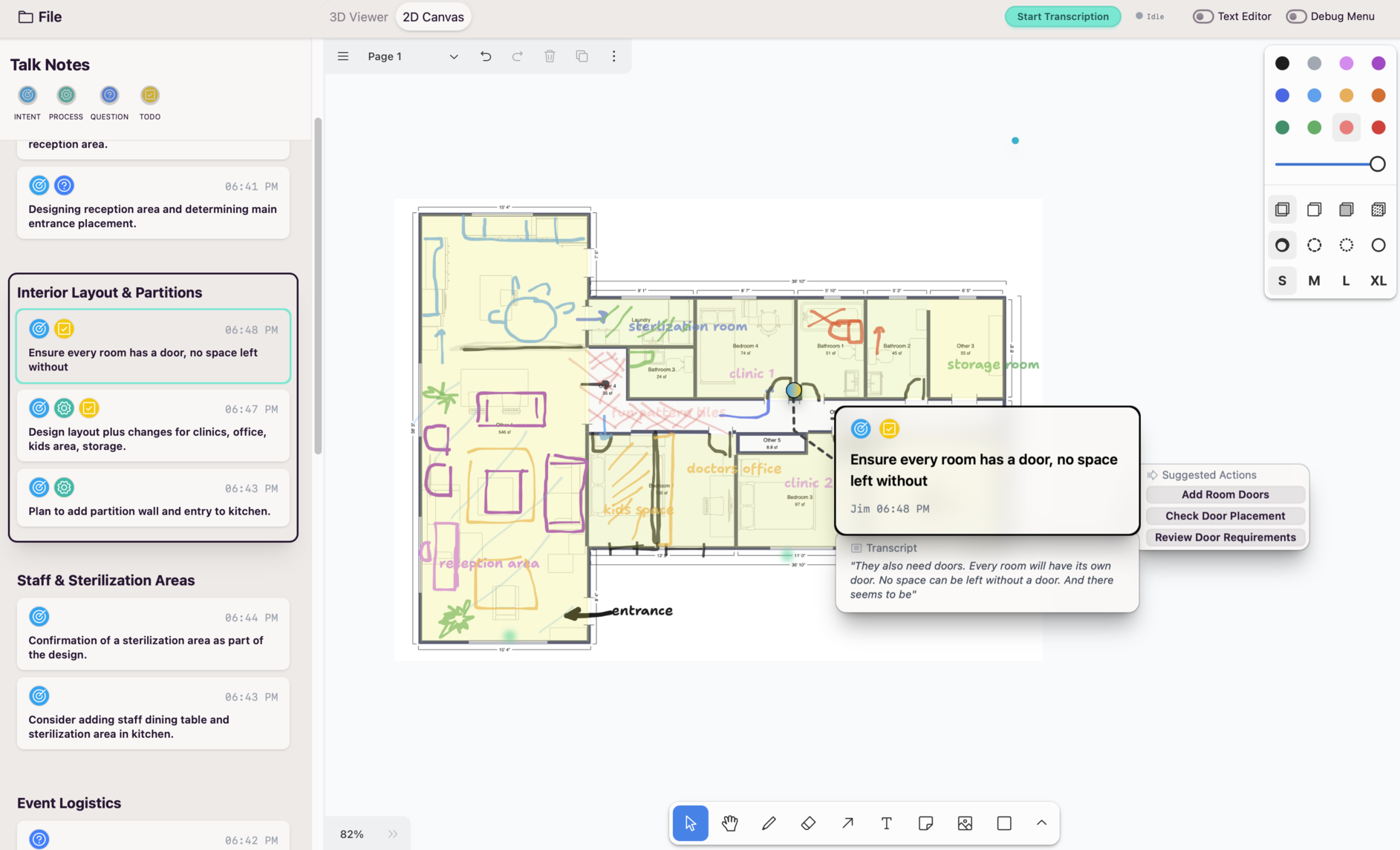}
    \caption{P07 – 2D Annotation Task}
    \label{fig:appendix_task2_P07_2d}
  \end{subfigure}
  \vspace{1em}
  \begin{subfigure}{0.99\linewidth}
    \centering
    \includegraphics[width=\linewidth]{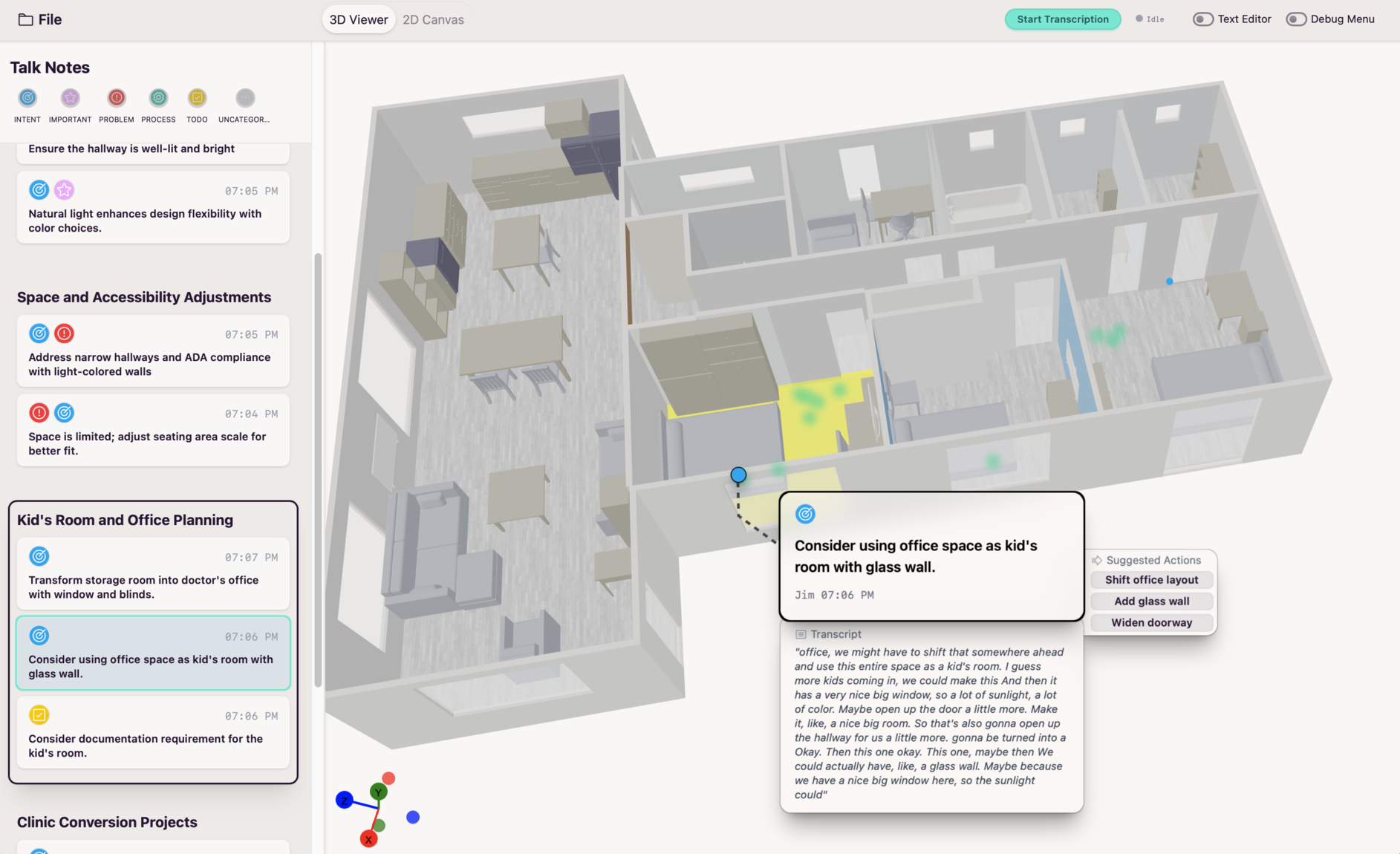}
    \caption{P07 – 3D Review Task}
    \label{fig:appendix_task3_P07}
  \end{subfigure}
  \caption{PointAloud annotation task outcomes P07.}
  \Description{This figure shows two screenshots of the PointAloud system. 
The top subfigure presents a 2D floor plan annotated with colored sketches and linked TalkNotes displayed in the left panel. 
The bottom subfigure presents a 3D model view of the same floor plan, with TalkNotes also visible in the left panel and annotations anchored within the 3D scene. 
Both subfigures illustrate how participants engaged in the annotation and review tasks using PointAloud.}
\end{figure}

\begin{figure}[H]
  \centering
  \begin{subfigure}{0.99\linewidth}
    \centering
    \includegraphics[width=\linewidth]{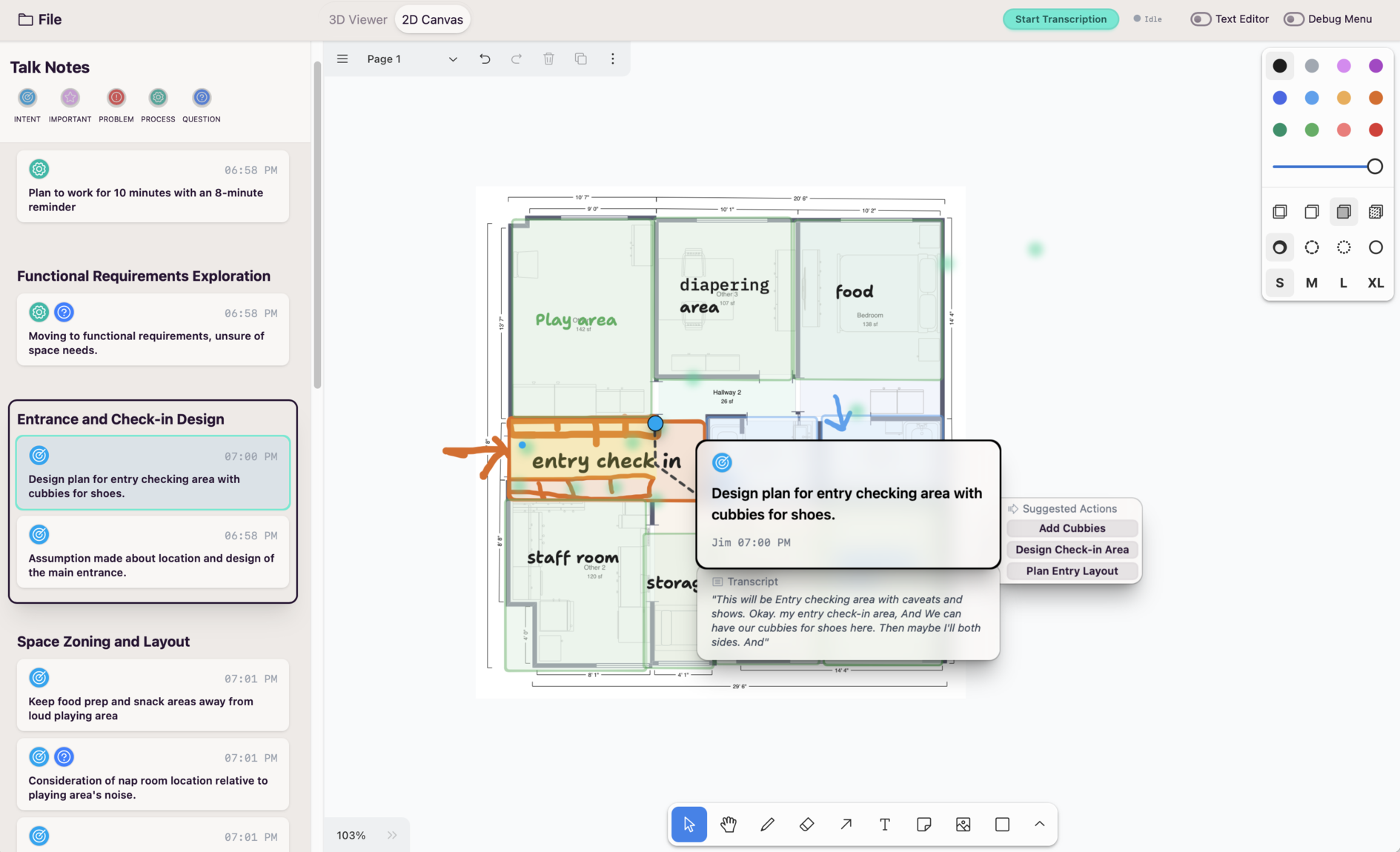}
    \caption{P08 – 2D Annotation Task}
    \label{fig:appendix_task2_P08_2d}
  \end{subfigure}
  \vspace{1em}
  \begin{subfigure}{0.99\linewidth}
    \centering
    \includegraphics[width=\linewidth]{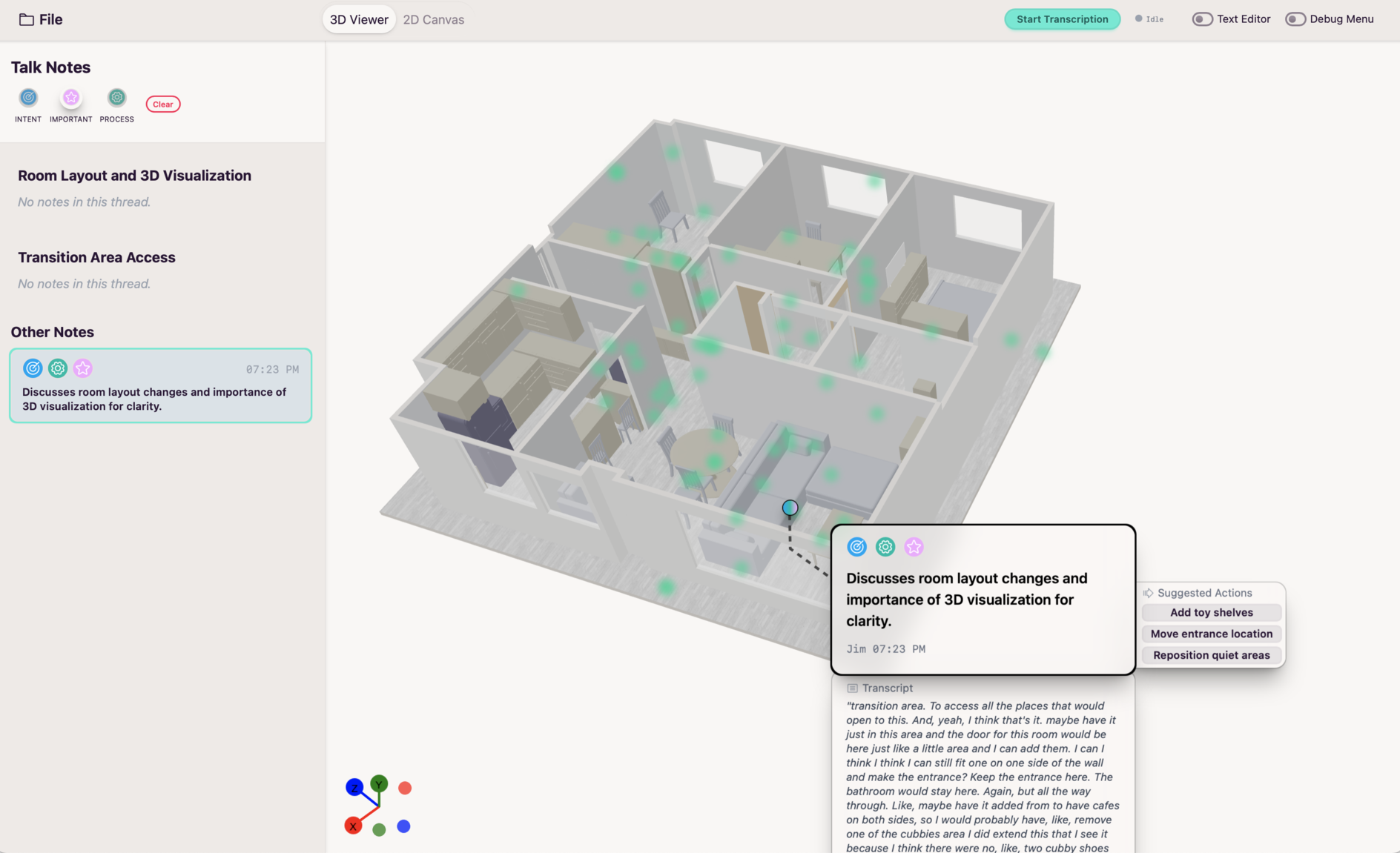}
    \caption{P08 – 3D Review Task}
    \label{fig:appendix_task3_P08}
  \end{subfigure}
  \caption{PointAloud annotation task outcomes P08.}
  \Description{This figure shows two screenshots of the PointAloud system. 
The top subfigure presents a 2D floor plan annotated with colored sketches and linked TalkNotes displayed in the left panel. 
The bottom subfigure presents a 3D model view of the same floor plan, with TalkNotes also visible in the left panel and annotations anchored within the 3D scene. 
Both subfigures illustrate how participants engaged in the annotation and review tasks using PointAloud.}
\end{figure}

\begin{figure}[H]
  \centering
  \begin{subfigure}{0.99\linewidth}
    \centering
    \includegraphics[width=\linewidth]{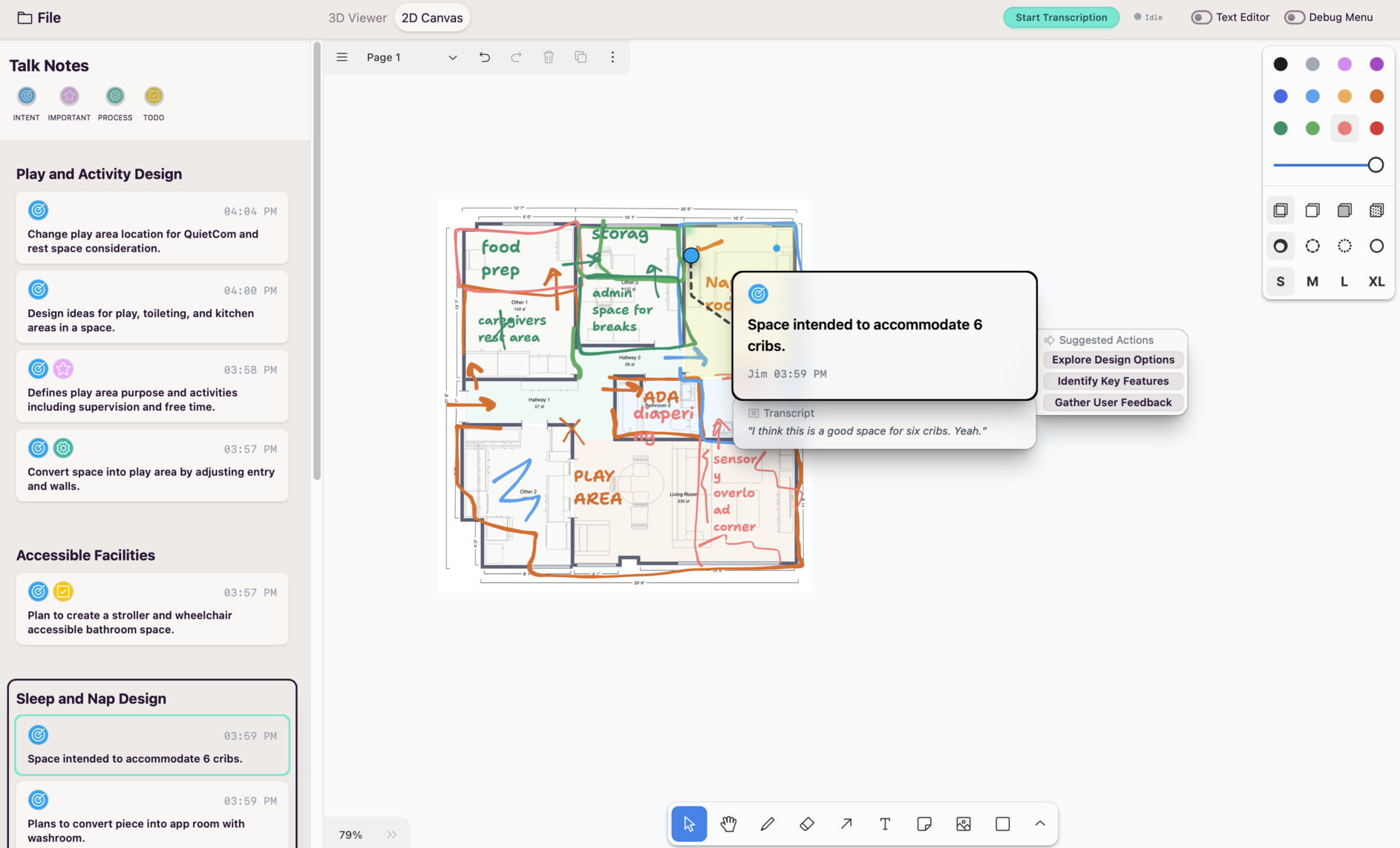}
    \caption{P09 – 2D Annotation Task}
    \label{fig:appendix_task2_P09_2d}
  \end{subfigure}
  \vspace{1em}
  \begin{subfigure}{0.99\linewidth}
    \centering
    \includegraphics[width=\linewidth]{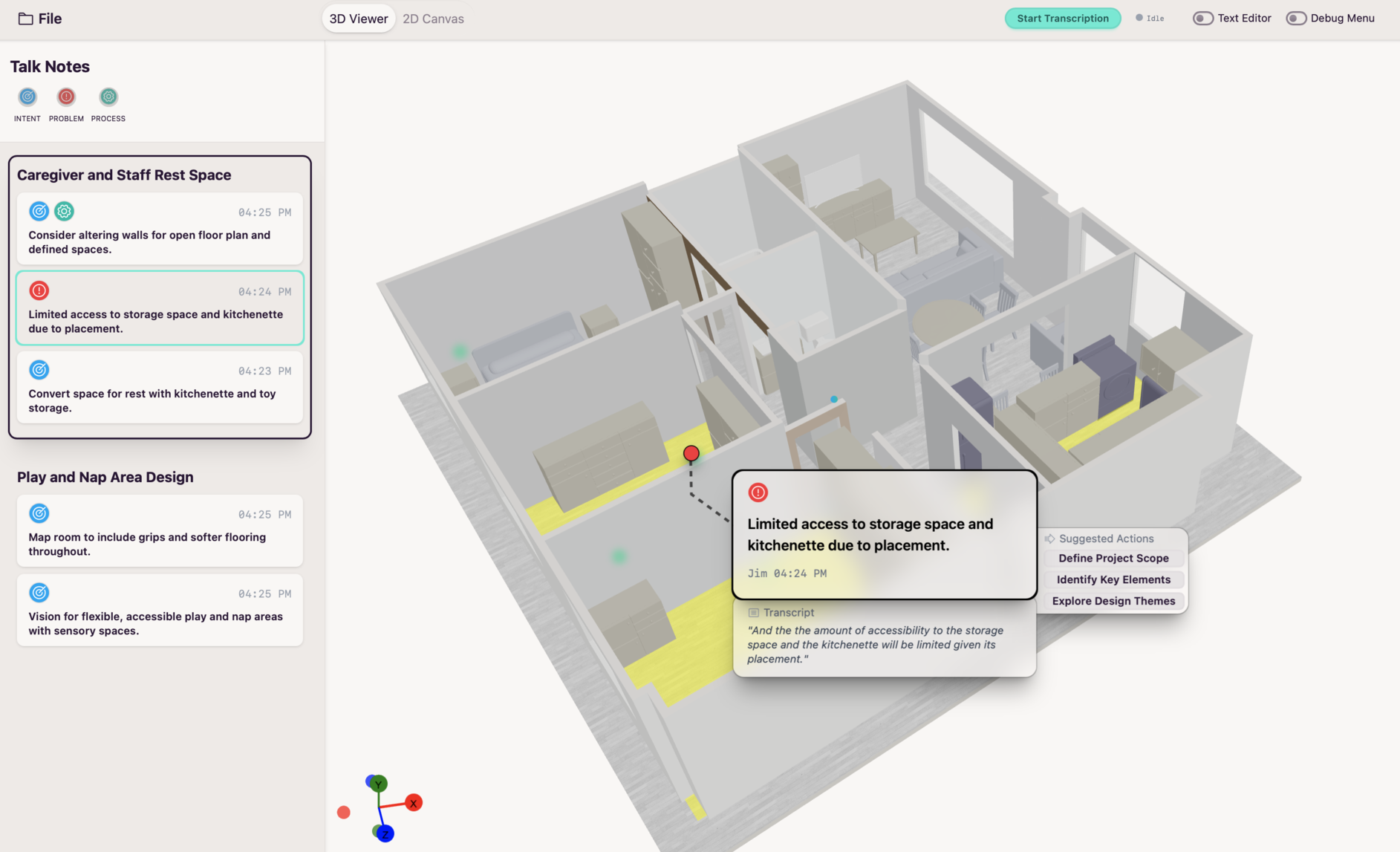}
    \caption{P09 – 3D Review Task}
    \label{fig:appendix_task3_P09}
  \end{subfigure}
  \caption{PointAloud annotation task outcomes P09.}
  \Description{This figure shows two screenshots of the PointAloud system. 
The top subfigure presents a 2D floor plan annotated with colored sketches and linked TalkNotes displayed in the left panel. 
The bottom subfigure presents a 3D model view of the same floor plan, with TalkNotes also visible in the left panel and annotations anchored within the 3D scene. 
Both subfigures illustrate how participants engaged in the annotation and review tasks using PointAloud.}
\end{figure}

\begin{figure}[H]
  \centering
  \begin{subfigure}{0.99\linewidth}
    \centering
    \includegraphics[width=\linewidth]{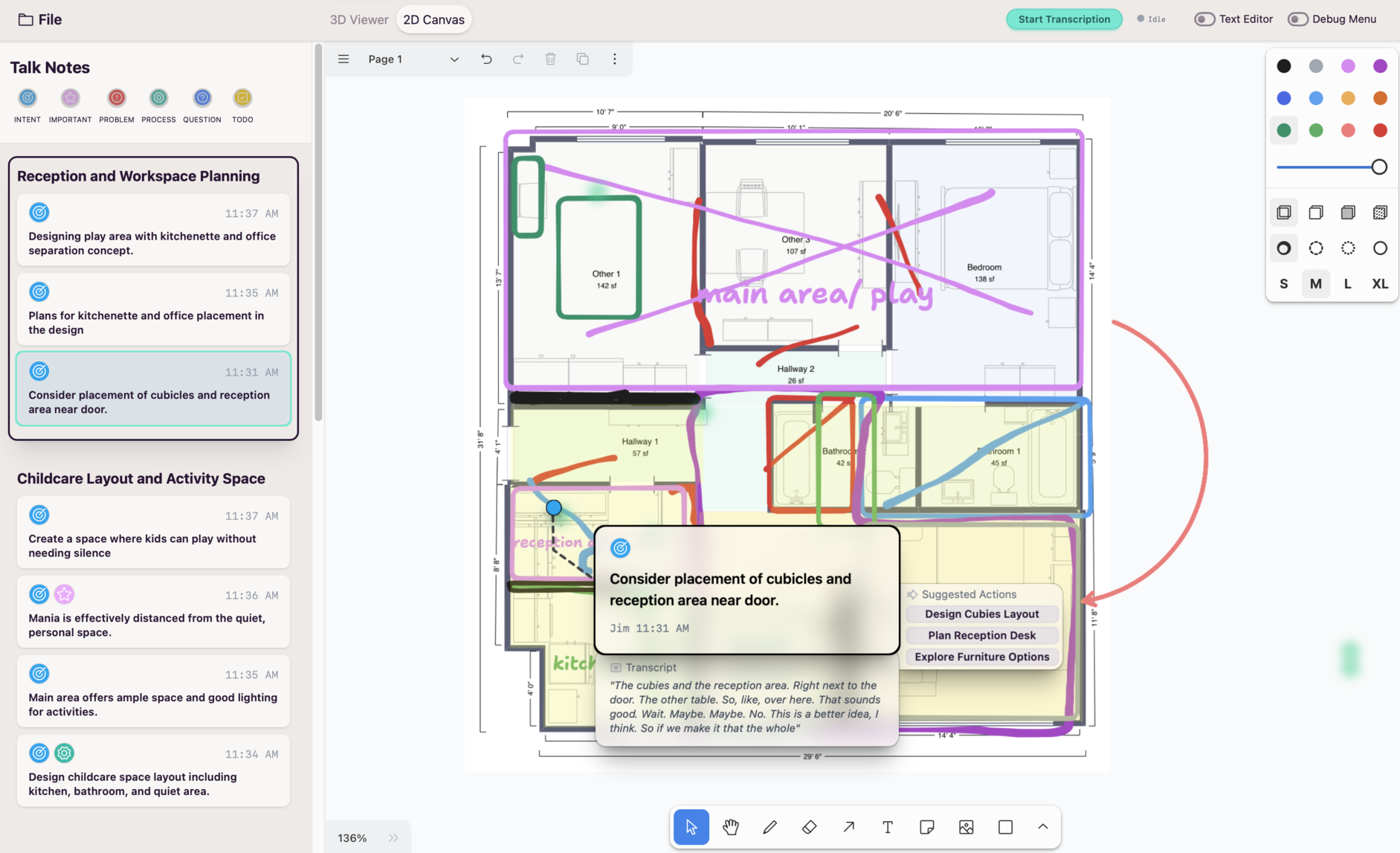}
    \caption{P10 – 2D Annotation Task}
    \label{fig:appendix_task2_P10_2d}
  \end{subfigure}
  \vspace{1em}
  \begin{subfigure}{0.99\linewidth}
    \centering
    \includegraphics[width=\linewidth]{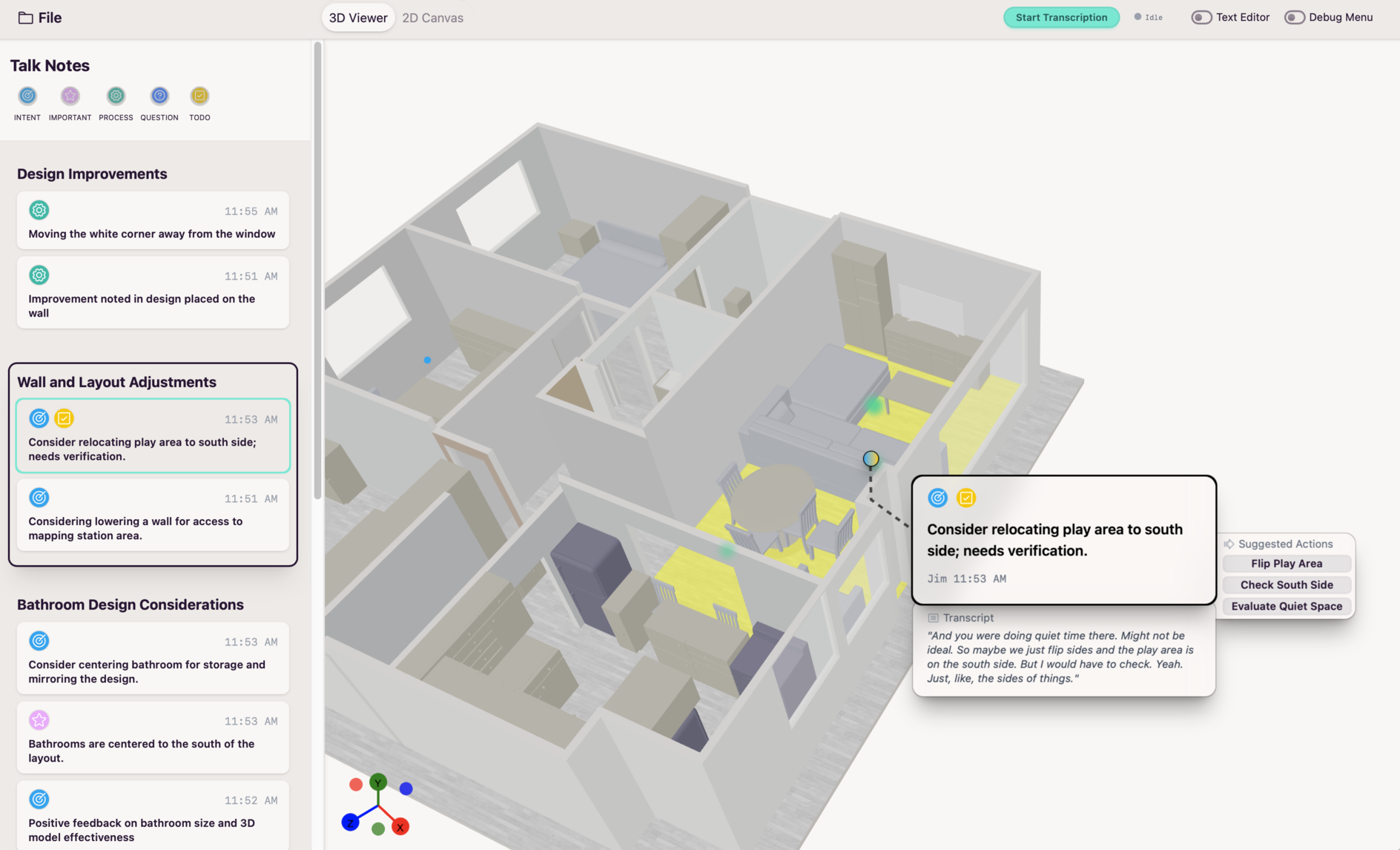}
    \caption{P10 – 3D Review Task}
    \label{fig:appendix_task3_P10}
  \end{subfigure}
  \caption{PointAloud annotation task outcomes P10.}
  \Description{This figure shows two screenshots of the PointAloud system. 
The top subfigure presents a 2D floor plan annotated with colored sketches and linked TalkNotes displayed in the left panel. 
The bottom subfigure presents a 3D model view of the same floor plan, with TalkNotes also visible in the left panel and annotations anchored within the 3D scene. 
Both subfigures illustrate how participants engaged in the annotation and review tasks using PointAloud.}
\end{figure}

\begin{figure}[H]
  \centering
  \begin{subfigure}{0.99\linewidth}
    \centering
    \includegraphics[width=\linewidth]{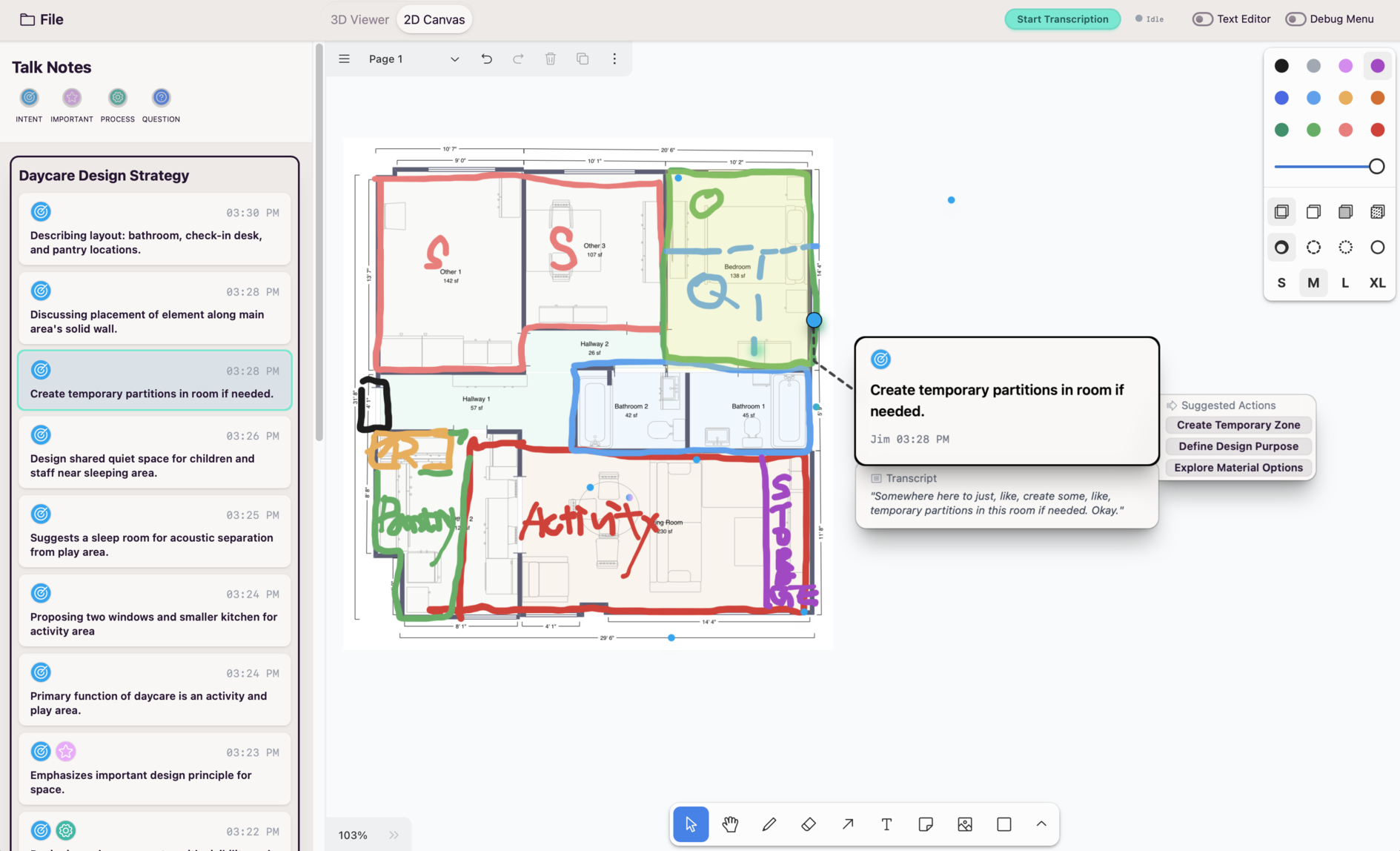}
    \caption{P11 – 2D Annotation Task}
    \label{fig:appendix_task2_P11_2d}
  \end{subfigure}
  \vspace{1em}
  \begin{subfigure}{0.99\linewidth}
    \centering
    \includegraphics[width=\linewidth]{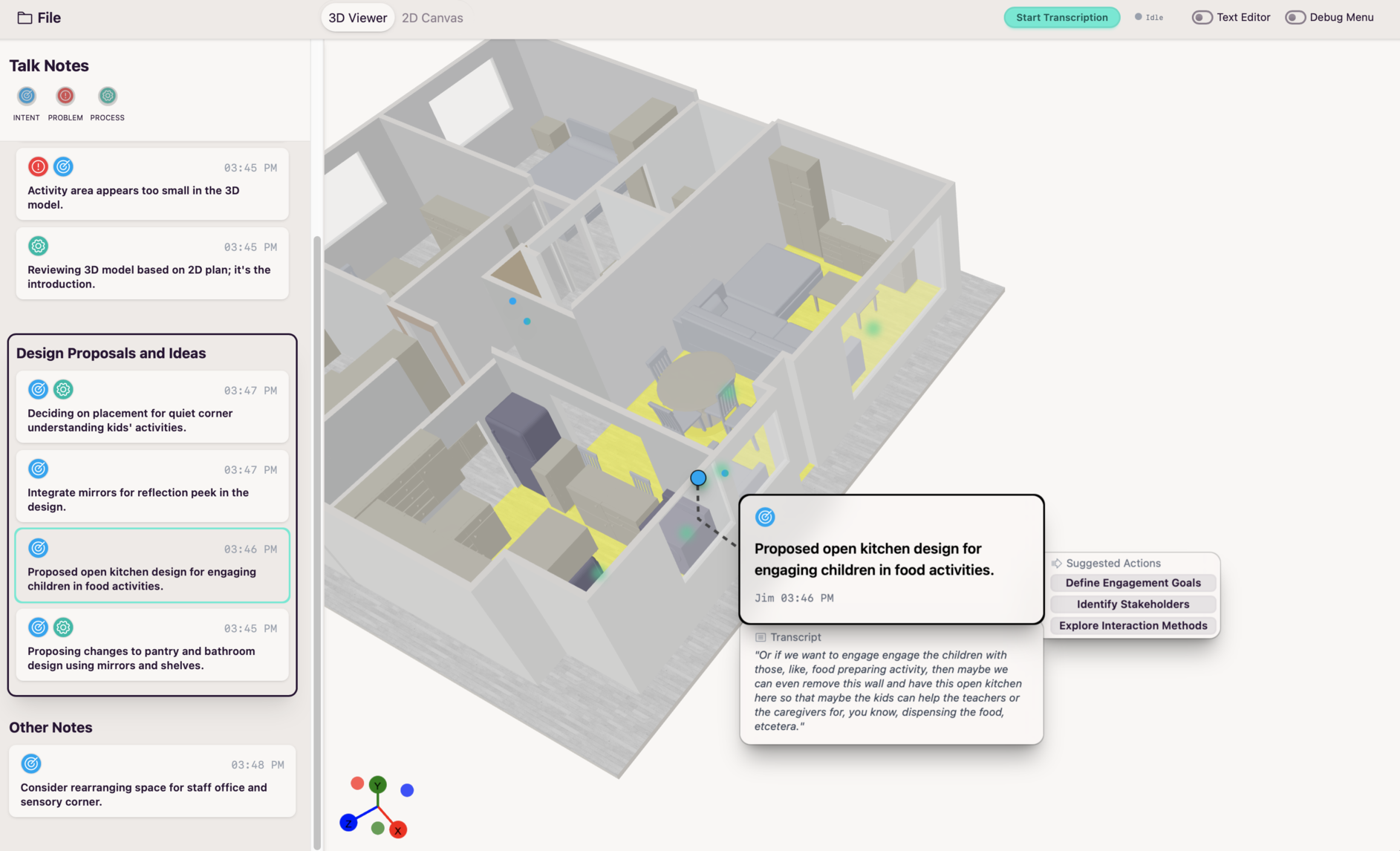}
    \caption{P11 – 3D Review Task}
    \label{fig:appendix_task3_P11}
  \end{subfigure}
  \caption{PointAloud annotation task outcomes P11.}
  \Description{This figure shows two screenshots of the PointAloud system. 
The top subfigure presents a 2D floor plan annotated with colored sketches and linked TalkNotes displayed in the left panel. 
The bottom subfigure presents a 3D model view of the same floor plan, with TalkNotes also visible in the left panel and annotations anchored within the 3D scene. 
Both subfigures illustrate how participants engaged in the annotation and review tasks using PointAloud.}
\end{figure}

\begin{figure}[H]
  \centering
  \begin{subfigure}{0.99\linewidth}
    \centering
    \includegraphics[width=\linewidth]{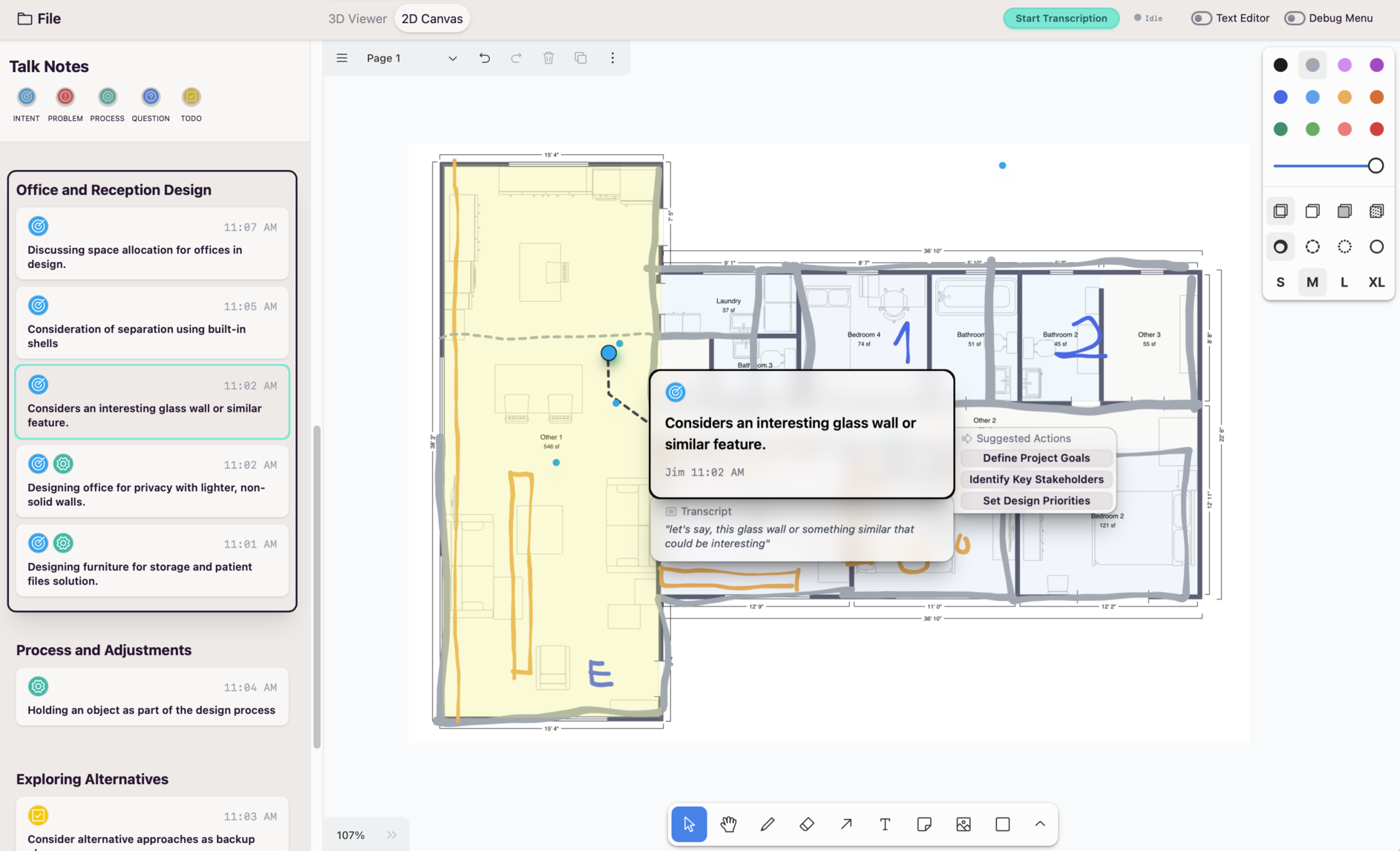}
    \caption{P12 – 2D Annotation Task}
    \label{fig:appendix_task2_P12_2d}
  \end{subfigure}
  \vspace{1em}
  \begin{subfigure}{0.99\linewidth}
    \centering
    \includegraphics[width=\linewidth]{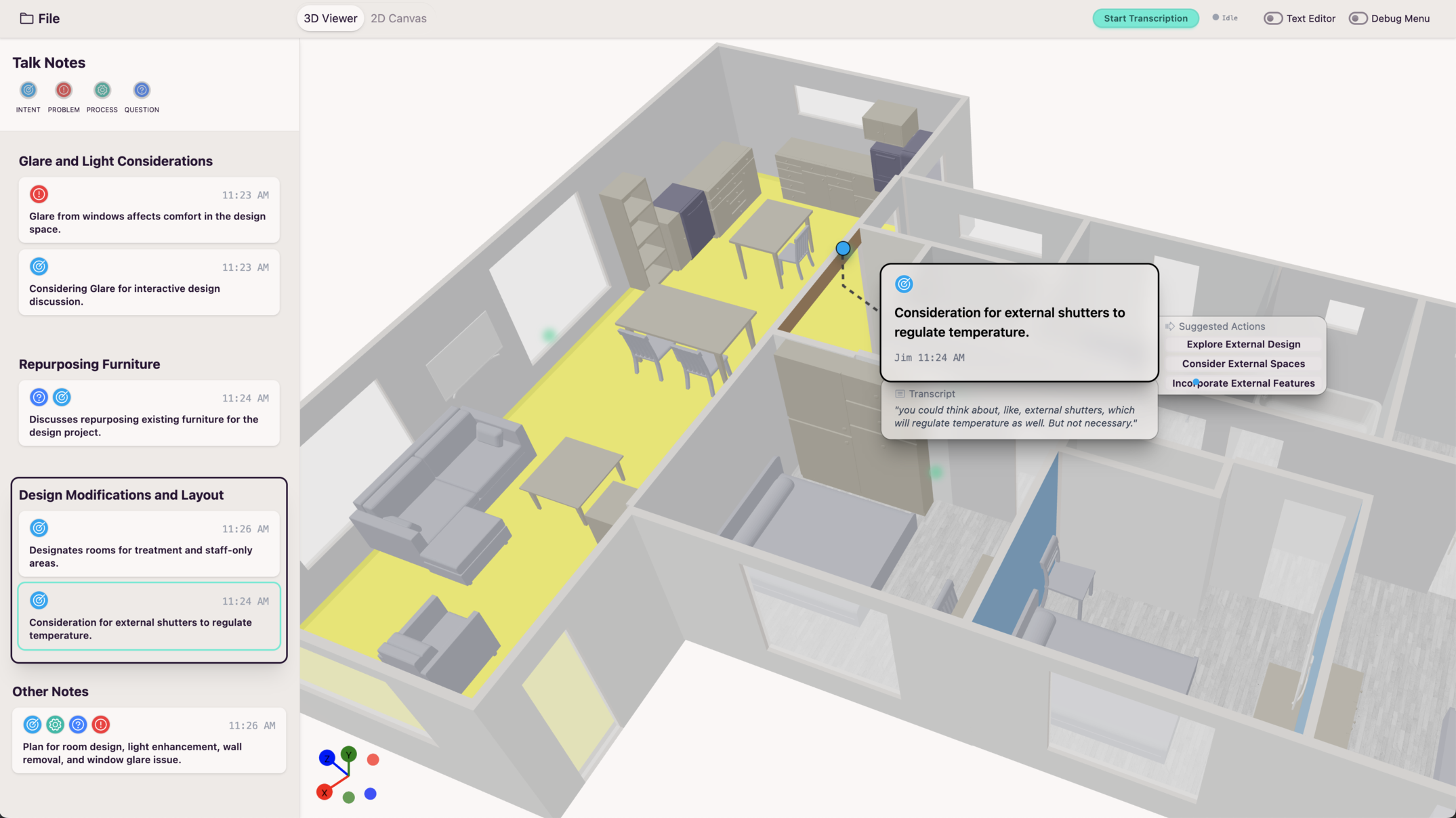}
    \caption{P12 – 3D Review Task}
    \label{fig:appendix_task3_P12}
  \end{subfigure}
  \caption{PointAloud annotation task outcomes P12.}
  \Description{This figure shows two screenshots of the PointAloud system. 
The top subfigure presents a 2D floor plan annotated with colored sketches and linked TalkNotes displayed in the left panel. 
The bottom subfigure presents a 3D model view of the same floor plan, with TalkNotes also visible in the left panel and annotations anchored within the 3D scene. 
Both subfigures illustrate how participants engaged in the annotation and review tasks using PointAloud.}
\end{figure}

\end{document}